\documentclass[twocolumn,tighten,times]{aastex631} 

\usepackage{CJK}
\usepackage{xspace}
\usepackage{booktabs}
\usepackage{lmodern}
\usepackage{slantsc}
\usepackage{hyperref}
\usepackage{amsmath}
\usepackage{bm}

\newcommand{\chisq}{$\chi^2$\xspace}
\newcommand{\chisqone}{$\chi^2_{W1}$\xspace}
\newcommand{\chisqtwo}{$\chi^2_{W2}$\xspace}
\newcommand{\chisqalt}{$\chi^2_{\mathrm{alt}}$\xspace}
\newcommand{\chisqflat}{$\chi^2_{\mathrm{flat}}$\xspace}
\newcommand{\pwd}{$P_{\mathrm{WD}}$\xspace}
\newcommand{\teff}{$T_{\mathrm{eff}}$\xspace}

\newcommand{\msun}{$M_{\odot}$\xspace}

\received{April 18, 2024}
\revised{June 21, 2024}
\accepted{June 24, 2024}

\submitjournal{ApJ}

\shorttitle{Empirical 3.4-$\mu$m Variability of White Dwarfs}
\shortauthors{Guidry et al.}

\begin{document}
\begin{CJK*}{UTF8}{gbsn}
\title{Using 3.4-$\bm{\mu}$m Variability towards White Dwarfs as a Signpost of Remnant Planetary Systems}

\correspondingauthor{Joseph A. Guidry}
\email{jaguidry@bu.edu}

\author[0000-0001-9632-7347]{Joseph A. Guidry}\altaffiliation{NSF Graduate Research Fellow}
\affiliation{Department of Astronomy, Boston University, Boston, MA 02215, USA}
\affiliation{Institute for Astrophysical Research, Boston University, Boston, MA 02215, USA}

\author[0000-0001-5941-2286]{J.~J. Hermes}
\affiliation{Department of Astronomy, Boston University, Boston, MA 02215, USA}
\affiliation{Institute for Astrophysical Research, Boston University, Boston, MA 02215, USA}

\author[0000-0002-8989-0542]{Kishalay De}
\altaffiliation{NASA Einstein Fellow}
\affiliation{MIT-Kavli Institute for Astrophysics and Space Research, Cambridge, MA 02139, USA}

\author[0009-0002-6065-3292]{Lou Baya Ould Rouis}
\affiliation{Department of Astronomy, Boston University, Boston, MA 02215, USA}
\affiliation{Institute for Astrophysical Research, Boston University, Boston, MA 02215, USA}

\author[0009-0008-4954-8807]{Brison B. Ewing}
\affiliation{Department of Astronomy, Boston University, Boston, MA 02215, USA}

\author[0000-0003-1970-4684]{B.~C. Kaiser}
\affiliation{Department of Physics and Astronomy, University of North Carolina at Chapel Hill, Chapel Hill, NC 27599, USA}

\begin{abstract}

Roughly 2\% of white dwarfs harbor planetary debris disks detectable via infrared excesses, but only a few percent of these disks show a gaseous component, distinguished by their double-peaked emission at the near-infrared calcium triplet. Previous studies found most debris disks around white dwarfs are variable at 3.4 and 4.5\,$\mu$m, but they analyzed only a few of the now 21 published disks showing calcium emission. To test if most published calcium emission disks exhibit large-amplitude stochastic variability in the near-infrared, we use light curves generated from the unWISE images at 3.4\,$\mu$m that are corrected for proper motion to characterize the near-infrared variability of these disks against samples of disks without calcium emission, highly variable cataclysmic variables, and 3215 isolated white dwarfs. We find most calcium emission disks are extremely variable: 6/11 with sufficient signal-to-noise show high-amplitude variability in their 3.4-$\mu$m light curves. These results lend further credence to the notion that disks showing gaseous debris in emission are the most collisionally active. Under the assumption that 3.4-$\mu$m variability is characteristic of white dwarfs with dusty debris disks, we generate a catalog of 104 high-confidence near-infrared variable white dwarfs, 84 of which are published as variable for the first time. We do near-infrared spectroscopic follow-up of seven new candidate 3.4-$\mu$m variables, confirming at least one new remnant planetary system, and posit that empirical near-infrared variability can be a discovery engine for debris disks showing gaseous emission.

\end{abstract}


\section{Introduction} \label{sec:intro}

More than 30\% of all white dwarf stars are likely accreting close-in circumstellar debris disks \citep{Zuckerman2003,Zuckerman2010,Koester2014,OuldRouis2024ESSV} composed of planetary bodies disrupted by tidal forces \citep{Jura2003}. These debris disks are most commonly inferred from the observation of metals polluting the otherwise pure hydrogen or pure helium white dwarf photosphere. Mass fractions of these pollutants are largely consistent with a rocky composition similar to bulk Earth \citep{Xu2014,XuBonsor2021}. Metal pollution is interpreted as a signpost for the presence of a remnant planetary system bound to the white dwarf \citep[see][and references therein]{Veras2021}.

Dust liberated from the disruption of these rocky bodies via ensuing grind-down from planetesimal collisions (e.g., \citealt{Malamud2021,Brouwers2022}) can be detected as infrared emission in excess to the white dwarf photosphere. Roughly 2\% of white dwarfs exhibit a detectable infrared excesses indicative of warm, dusty debris disks \citep{Rocchetto2015,Wilson2019,Rebassa-Mansergas2019}. This smaller detection rate could be a consequence of needing near-face-on viewing geometries if these disks tend to be flat. Regardless, all white dwarfs with dusty disks show photospheric metal pollution (see \citealt{Farihi2016} and references therein). These infrared excesses peak at mid-infrared wavelengths ($\gtrsim 5\,\mu$m) and are challenging to observe from the ground; white dwarfs with infrared excesses peaking in the near-infrared ($\lesssim 5\,\mu$m) are more likely to host to an unresolved, cool stellar or brown dwarf companion (e.g, \citealt{Farihi2005,Girven2011,Dennihy2020b}).

At the innermost radii of the debris disk, the temperatures should be sufficient to sublimate the debris into gaseous states. This gas can be viewed in absorption \citep[e.g.,][]{Debes2012b,Xu2016,Steele2021} and as broad, double-peaked emission, mainly at the calcium triplet \citep{Gaensicke2006,Gaensicke2007,Gaensicke2008,Gaensicke2011,Farihi2012,Melis2012,Wilson2014,Melis2020,Dennihy2020b,GentileFusillo2021gas}. The latter class of ``gaseous" debris disk systems are observationally rare: \citet{Manser2020} calculate the occurrence rate of white dwarfs showing emission at the Ca~II triplet to be $\approx$0.1\%. Notably, all 21 published gas disk systems are metal polluted \citep{GentileFusillo2021gas} and most show infrared excesses (see \citealt{Farihi2016} and references therein; \citealt{GentileFusillo2021gas}). Many of the white dwarfs hosting gaseous disks show heavily polluted photospheres, though this circumstellar gas so far does not appear to enhance accretion rates via drag effects \citep{Rogers2024a}. This class of disks showing gaseous calcium is distinct from cases like that of WD\,J0914+1914, which harbors a gas-only, dustless disk fueled by the evaporation of a volatile rich atmosphere of a giant planet, and not a disrupted rocky body \citep{Gaensicke2019}.

Throughout the remainder of this paper we adopt a nomenclature of disks ``showing calcium (Ca~II) gas in emission" and ``non-calcium emission (non-Ca~II)" disks to distinguish the dusty debris disks that have previously been observed to show a gaseous component in emission at the calcium~II triplet in the near-infrared, since gaseous and dusty debris are believed to coexist in all of these planetary debris disks.

\begin{figure*}[t!]
\includegraphics[width=\textwidth]{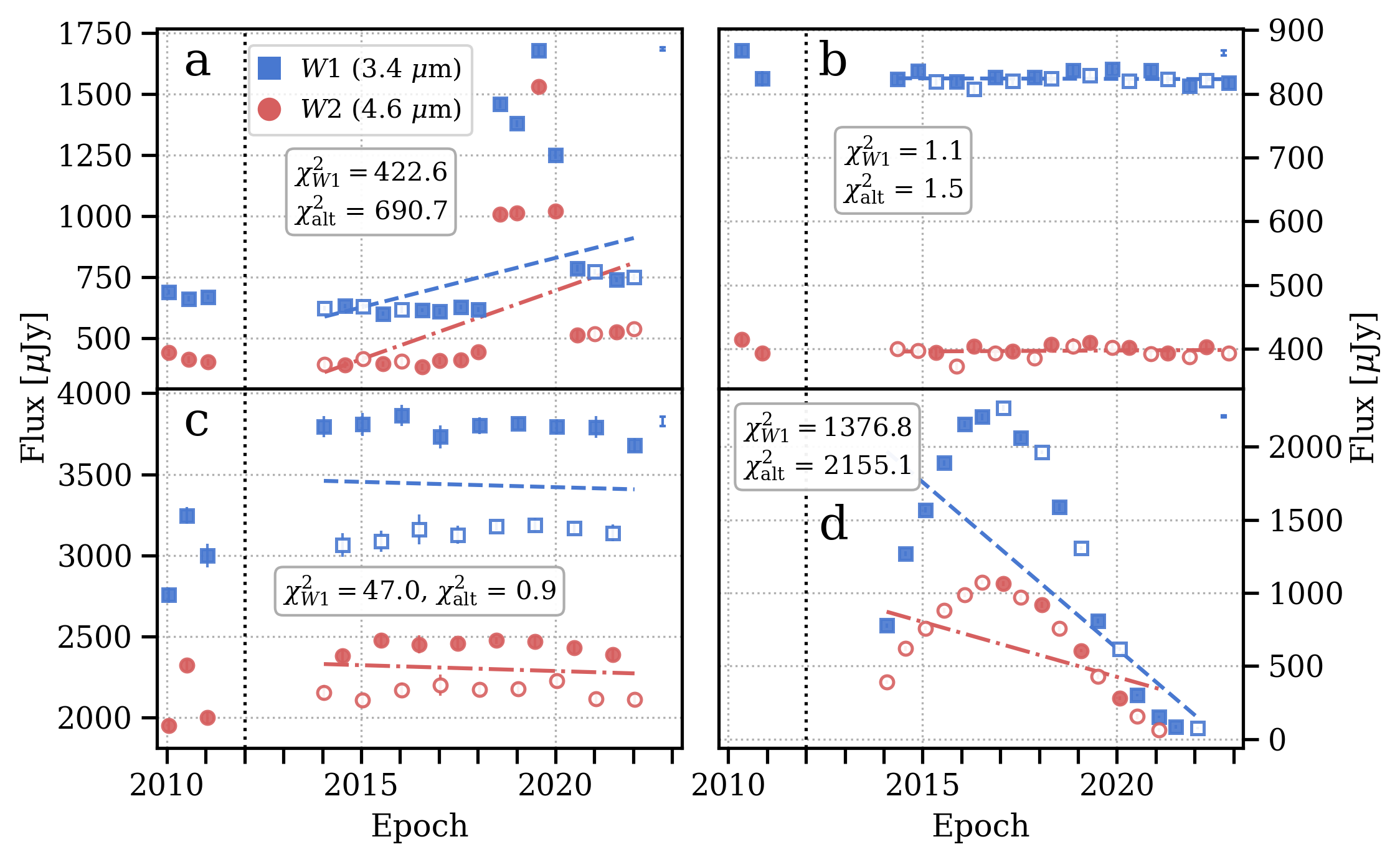}
\caption{WISE light curves for $W1$ (3.4\,$\mu$m, blue squares) and $W2$ (4.6\,$\mu$m, red circles) representing the four most common trends seen in our sample. {\bf (a)} Significant, stochastic variability, such as the outbursting Ca~II gas disk towards the previously studied WD\,0145+234 \citep{Swan2021}. {\bf (b)} A flat light curve representing the expectation for an isolated, unchanging white dwarf (LAWD\,52). {\bf (c)} A light curve showing a strong alternating pattern in both the $W1$ and $W2$ bands caused by the different scanning angles of the PSF from each 6-month WISE observing season, which tends to be exacerbated when there are nearby stars to the target (PG\,1411+219). The solid- and open-faced points indicate the two clusters used to calculate \chisqalt (see text). {\bf (d)} A secular brightening or dimming trend caused by the proper motion of the white dwarf (GJ\,3770) moving into (or away from) another star. The dotted black lines at 2012 demarcate the transition from cryogenic to NEOWISE mission data; we only analyze the NEOWISE epochs for our variability rankings. Blue dashed and red dashed-dotted lines show the fits used to calculate \chisqone and \chisqtwo, respectively (see Section~\ref{sec:methods} for definitions of the annotated \chisq values). Uncertainties are shown but are generally smaller than each symbol. The average $W1$ uncertainties are plotted in the upper right corner of each panel.
\label{fig:4_modes}}
\end{figure*}

Among a sample of 35 non-Ca~II dusty white dwarfs observed by the Wide-field Infrared Survey Explorer (WISE), \citet{Swan2019} found that the majority exhibit significant photometric variability in the near-infrared at 3.4\,$\mu$m. Ensemble variability was again observed by \citet{Swan2020} among a sample of over 40 white dwarf debris disks with Spitzer at 3.4\,$\mu$m and 4.5\,$\mu$m.Within this sample, a few Ca~II gas disk systems have shown exaggerated, high-amplitude variability \citep{Swan2020}. This includes factor-of-two brightenings in the near-infrared seen towards WD\,0145+234 \citep[][Figure~\ref{fig:4_modes} Panel a]{Wang2019}, interpreted as an expansion of the emitting area of the dust disk due to collisions \citep{Swan2021}, which in turn should liberate gas. It also includes dimmings, like those observed with Spitzer towards WD\,J0959$-$0200 \citep{Xu&Jura2014}, SDSS\,J1228+1040 \citep{Xu2018b}, and GD\,56 \citep{Farihi2018}. \citet{Rogers2020} did not observe ensemble variability in the $J$-, $H$-, and $K$-bands over a three-year baseline among dusty white dwarfs.

We explore here the 3.4-$\mu$m variability of all 21 published Ca~II gas disk systems in ensemble for the first time. We aim to test whether they exhibit stochastic variability when bright enough to be confidently detected in WISE. We generate light curves using a new aperture photometry pipeline that corrects for the proper motions of targets as it operates on the unWISE images from the NEOWISE era (2013--2023). We are motivated to test if empirical 3.4-$\mu$m variability can be used as an engine for discovering more disks showing Ca~II emission around white dwarfs.

In Section~\ref{sec:observations} we describe our light curve generation pipeline and follow-up observations. In Section~\ref{sec:methods} we detail our variability ranking and validation method. In Section~\ref{sec:results} we report our catalog of 3.4-$\mu$m variables and characterize the Ca~II disks relative to non-Ca~II disks. We end by discussing the consequences of our work in Section~\ref{sec:discussion}. Our catalog of 3.4-$\mu$m variables is published in Appendix~\ref{sec:var_tables}.


\begin{figure}[t!]
\includegraphics[width=0.48\textwidth]{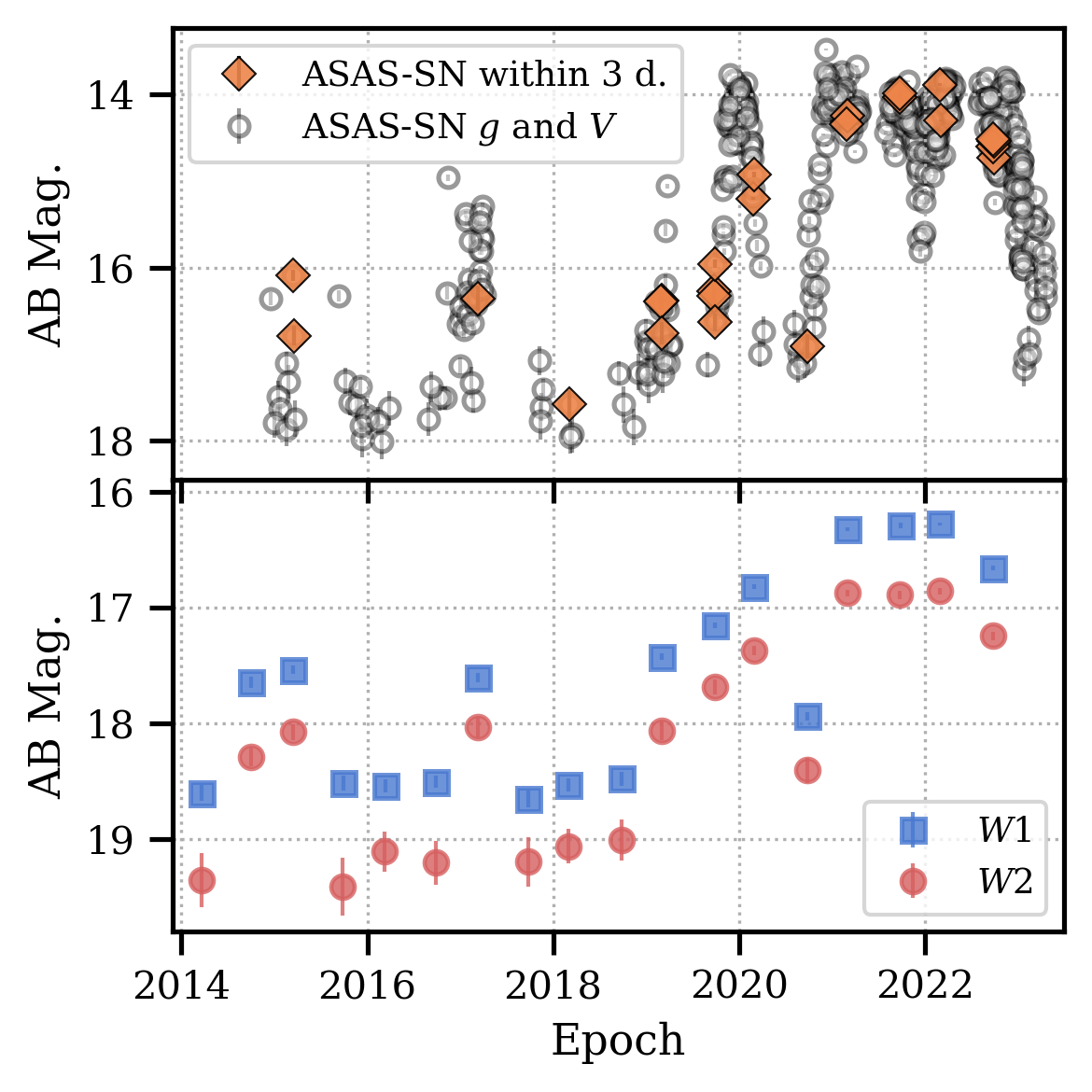}
\caption{{\bf Top:} ASAS-SN light curve \citep{Shappee2014,Koachanek2017,Hart2023} of the cataclysmic variable KR Aur (Gaia DR3 3436435910858051200). Observations shown as orange diamonds were taken to within 3\,days of the WISE epochs, while the remaining open-faced points show the full $g$- and $V$-band light curve. {\bf Bottom:} Our WISE light curve of KR Aur showing a correlation between the optical and near-infrared variability. 
\label{fig:CV_LCs}}
\end{figure}

\section{Observations} \label{sec:observations}

\subsection{New unWISE Light Curve Extractions}\label{sec:pipeline}

Our study utilizes images from the unWISE individual-image catalog\footnote{\url{https://catalog.unwise.me/catalogs.html}} captured by the all-sky survey of the Wide-field Infrared Survey Explorer \citep[WISE,][]{Wright2010}, including the initial cryogenic mission and the ongoing NEOWISE mission \citep{Mainzer2011,Mainzer2014}. The unWISE catalog features coadded images of typically 12$\times$7.7\,s exposures \citep{Meisner2018b}, boosting photometry reliability to magnitudes of $W1 \sim 18$\,mag \citep{Schlafly2019}. The images are 2048$\times$2048 pixels at a scale of 2.75$\arcsec$\ per pixel \citep{Lang2014}, although the typical image resolution from the PSF is roughly 6.1\arcsec\ and 6.3\arcsec\ in the $W1$ (3.4\,$\mu$m) and $W2$ (4.6\,$\mu$m) channels, respectively. For a thorough description of the unWISE catalog see \citet{Schlafly2019}, and for a review of the image processing and generation see \citet{Lang2014} and \cite{Meisner2017a,Meisner2017b}. We use the unWISE images from the 2023 NEOWISE Data Release.

Our pipeline operates on the single coadds for each epoch, treating the $W1$ and $W2$ images separately. Another version of this pipeline uses difference imaging for transient work in WISE \citep{Tran2024,Masterson2024}. Here instead we perform simple aperture photometry, using a fixed circular 2-pixel-radius aperture (equivalently 5.5\arcsec) centered on the coordinates of the target to match the WISE PSF. We also draw an annulus with an inner radius of 5 pixels and width of 3 pixels for background subtraction. Importantly, we propagate the coordinates of each target using the Gaia DR3 measured right ascension, declination, and proper motions to the epoch of the WISE observation, rather than doing source detection from epoch-to-epoch like unTimely \citep{Meisner2023}. Representative light curves are shown in Figure~\ref{fig:4_modes}. The proper motions of field stars are not included, nor are the parallaxes of any sources. We also generate light curves extracted by PSF photometry on the unWISE images using the same procedures. 

We estimate uncertainties on the photometry using the uncertainty images provided by WISE and further processed by unWISE that map uncertainties to the pixel-level \citep{Lang2014}. Objects varying on timescales of seconds generate greater uncertainties due to the larger image-to-image deviations. A full detailed account of this pipeline will be presented in a forthcoming publication, K.~De et al. (in prep).

\subsection{Input White Dwarf Catalog}

We construct a sample for our study that includes all white dwarfs currently known to harbor gaseous debris disks and various control samples to assess the variability of the Ca~II gas disks. The 21 Ca~II gas disk systems are sourced from Table 3 of \citet{GentileFusillo2021gas}. Our first control sample is one of 45 white dwarfs with known infrared excesses due to dusty debris disks that have not shown calcium triplet emission: 28 from the Montreal White Dwarf Database \citep{MWDD2017}, seven systems not in MWDD from \cite{Rogers2020}, and 11 more unique systems from \cite{Wang2023}. We would go on to find two additional known dusty disks in the Gaia-catWISE input described below after ranking all systems by their variability: KPD\,0420+5203 (Gaia DR3 271992414775824640) \citep{Hoard2013,Barber2016} and MCT\,0420$-$7310 (Gaia DR3 4653404070862114176) \citep{Hoard2013,Dennihy2017}, yielding an input of 47 objects.

Our second control is a sample of 14 known cataclysmic variables (CVs) taken from the American Association of Variable Star Observers\footnote{\url{https://targettool.aavso.org/TargetTool/default/index}} with Gaia G-band magnitudes of $13 \leq G \leq 18$\,mag, three recently discovered polars \citep{Guidry2021,Rodriguez2023a} and one AM CVn \citep{Rodriguez2023b} from the Zwicky Transient Facility. The CV sub-sample is meant to be a test to the pipeline, as CVs are known to show both optical and IR outbursts \citep[e.g.,][]{Petrosky2021}; correlated optical and infrared variability of KR Aur, a known CV, is shown in Figure~\ref{fig:CV_LCs}.

We build a third sample from a cross-match between the Gaia eDR3 catalog of white dwarfs \citep{GentileFusillo2021gaia} and the catWISE catalog of WISE-detected sources \citep{Eisenhardt2020}. We intend to search for highly variable white dwarfs within this sample to identify candidate gas disks, as well as contextualize the variability of the known debris disks against that of isolated white dwarfs. We first cut the Gaia white dwarf catalog to 100\,pc and impose \pwd$>0.75$ to define a sample of high-confidence, isolated white dwarfs, eliminating most binary sources like CVs or unresolved WD+MS, post-common-envelope binaries. We require all sources be brighter than $G_{\rm RP} < 17.0$\,mag, and cross-match with catWISE, requiring the $W2$ detections to have a signal-to-noise greater than 5.0 and the $W1$ magnitudes be within 1.5\,mag of the Gaia $RP$ magnitude. This yields a final source catalog of 3215 white dwarfs, including multiple duplicates from the above Ca~II and non-Ca~II disks and CV sub-samples. Aggregating all these samples, we build an input catalog of 3299 white dwarfs.

\subsection{Near-Infrared Spectroscopic Follow-up}\label{sec:followup}

\subsubsection{LDT+DeVeny}

Our primary motivation for ranking white dwarfs by their WISE variability is to build a catalog of candidate remnant planetary systems, especially those with the highest potential to host collisionally active debris disks. We began a preliminary follow-up campaign to test if these candidates show emission at the calcium triplet using the DeVeny Spectrograph \citep{DeVeny2014} at the 4.3-m Lowell Discovery Telescope (LDT) at Happy Jack, Arizona, USA.

For our LDT+DeVeny observations we use a 1.0\arcsec\ slit paired with the DV8 grating (831 line/mm) centered at a blaze wavelength of 8000\,\AA\ with the GG 570 order-blocking filter, yielding a resolution of roughly $0.8$\,\AA\ across $7200-8800$\,\AA. We publish observations here from 2023 June 20 and 2024 February 5. We provide a log of our LDT+DeVeny observations in Table~\ref{tab:spectra} in Appendix~\ref{sec:soar_appendix}.

We reduced our spectra using the {\tt PypeIt} reduction pipeline \citep{pypeit:joss_arXiv}, performing a telluric correction directly on our observations using a third-order polynomial model to estimate the continuum.

\subsubsection{SOAR+Goodman Follow-up}

We also followed up some of our southern candidates using the red camera of the Goodman Spectrograph \citep{Clemens2004} mounted on the 4.1-m Southern Astrophysical Research Telescope (SOAR) at Cerro Pach\'{o}n, Chile on 2023 August 13 and 2023 August 18. Building on the observations of Ca~II gas disks done by \citet{Dennihy2018,Dennihy2020b}, we use the 1200 line/mm grating with the 1.0\arcsec\ slit and GG495 order-blocking filter, giving us spectral coverage from roughly $7570-8750\,$\AA\ with a resolution of roughly $0.6$\,\AA. Alongside the LDT+DeVeny spectra, a log of our SOAR observations can be found in Table~\ref{tab:spectra} in Appendix~\ref{sec:soar_appendix}.

We also reduce our SOAR spectra using {\tt PypeIt}, again performing a telluric correction on our spectra anchored to a third-order polynomial over the region observed.


\section{Variability Detection and Validation}\label{sec:methods}

 We first process each WISE light curve through quality filtering before assessing it for variability, handling the $W1$ and $W2$ observations independently. We reject erroneous measurements (zero flux detections) and require both the $W1$ and $W2$ flux uncertainties be $>1\,\mu$Jy. We further perform a $3\,\sigma$ clip to the flux uncertainties to exclude spuriously small values that would skew a \chisq calculation. Finally we require the signal-to-noise ratio of each epoch be S/N$>$3.0. We then convert the WISE Vega magnitudes to fluxes in Janskys using the published conversions: $F_{\nu,0,W1} = 309.540$\,Jy, $F_{\nu,0,W2} = 171.787$\,Jy \citep{Jarrett2011,Wright2010}. All analysis of the WISE light curves is hereafter conducted in flux space. 
 
Presuming an isolated white dwarf should be constant in infrared flux, we assess the degree of variability for each object by the goodness of linear (first-order polynomial) fits to each light curve. Keeping the $W1$ and $W2$ light curves independent, we fit lines to the NEOWISE-only ($2013-2023$) portion of each light curve, allowing the slopes and offsets to vary. Calibrations for the first few cryogenic epochs collected from $2009-2011$ can cause flux offsets from the NEOWISE data \citep{Meisner2017a}. For every object we record the reduced \chisq of each fit to each band, hereafter referred to as \chisqone and \chisqtwo, respectively. We do the same to for an additional series of linear fits where the slope is fixed to be zero. We designate these ``flat" line fits to the $W1$ light curves as \chisqflat.

A common artifact seen in the light curves from WISE is an alternating set of fluxes due to the rotation of the point-spread function (PSF) from the different scanning angles of the spacecraft from epoch-to-epoch \citep{Eisenhardt2020}. Panel c of Figure~\ref{fig:4_modes} shows an example of this variation, alternating with the typical source revisit timescale of roughly 6 months. This causes the source to exhibit both large \chisqone\ and \chisqflat\ while not necessarily being intrinsically astrophysically variable.

\begin{figure*}[ht!]
\includegraphics[width=\textwidth]{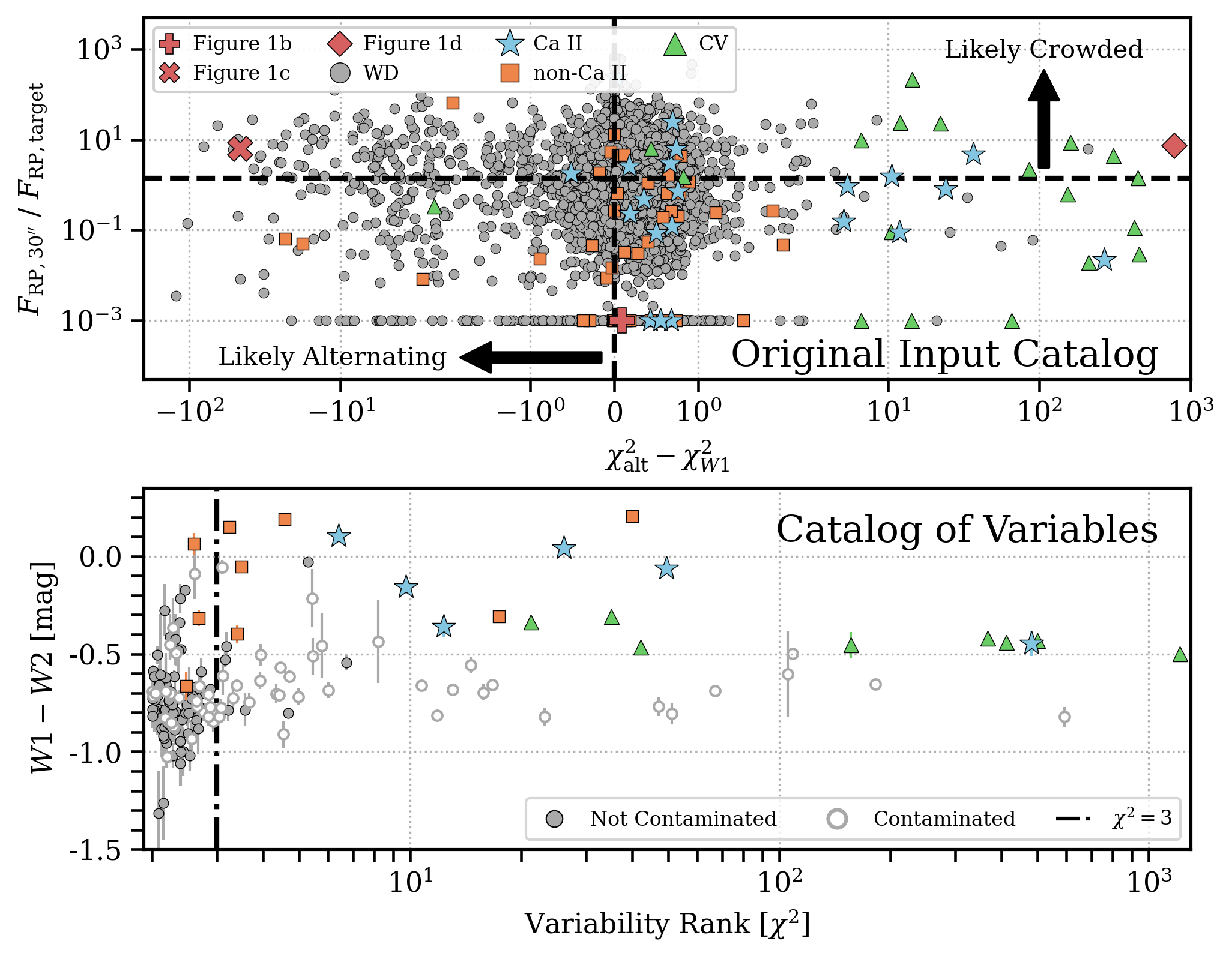}
\caption{{\bf Top:} Excess Gaia $RP$ flux vs. the difference between the ``alternating" \chisq\ (see Panel c of Figure~\ref{fig:4_modes}) and the \chisqone fits. Sources with $\chi^2_{\mathrm{alt}} - \chi^2_{W1} < 0$ are likely to have high \chisqone\ caused by the alternating pattern due to the biannual variation in the WISE PSF scanning angle, rather than astrophysical variability. Sources with 1.415 times more flux in the Gaia $RP$ channel in a 30\arcsec\ cone than their own flux are frequently found in crowded fields, leading to false-positive variability detections due to source crowding of the large WISE PSF and proper motions like that in Panel d of Figure~\ref{fig:4_modes} due to proper motions. We inflate y-axis values by $10^{-3}$ so that sources with zero nearby flux are on-axis. {\bf Bottom:} Our catalog of 163 3.4-$\mu$m variables ranked by variability. Catalog inclusion requires $\chi^2_{\mathrm{alt}} - \chi^2_{W1} > 0$, $G_{\rm RP}$ excess flux $<1.415\times$ the target flux, and \chisqone or \chisqflat $>$ 2.0 (see text). The x-axis values are the maximum value of \chisqone or \chisqflat for each target. Objects with strong apparent contamination from visual inspection with WISE View and/or still show a seasonal alternating pattern (see the Flag column in Table~\ref{tab:final_catalog}) are shown as open-faced (only found for Gaia-catWISE white dwarfs). The vertical dash-dotted line at $\chi^2=3$ marks our definition for high-amplitude variability. In all panels, we distinguish known disks with calcium emission as cyan stars, white dwarfs with dusty disks without calcium emission as orange squares, and known CVs as green triangles.
\label{fig:color_vs_chisq}}
\end{figure*}

We define a second goodness-of-fit metric that analyzes each ``high'' and ``low'' season independently. We use a one-dimensional $k$-means clustering algorithm\footnote{\url{https://github.com/dstein64/kmeans1d}} to sort each light curve temporally, dividing the timestamps by the median separation of the whole light curve, taking the modulo of twice that median revisit timescale to convert the light curve points into two clusters in phase space. The cluster with the higher average flux is classified as the ``high" cluster and the other the ``low.'' We fit an additional pair of lines to each cluster for each filter. We calculate an effective reduced \chisq for these ``high-and-low'' seasons as $\chi^2_{\mathrm{alt}} = (\chi^2_{\mathrm{high}}+\chi^2_{\mathrm{low}})/\sqrt{2}$. The comparison of $\chi^2_{\mathrm{alt}} - \chi^2_{W1}$ is shown in the top panel of Figure~\ref{fig:color_vs_chisq}. We de-prioritize objects with $\chi^2_{\mathrm{alt}} - \chi^2_{W1} < 0$, as they are more likely to have flux changes due to the WISE scanning angle rather than some intrinsic variability, as exhibited in Panel c of Figure~\ref{fig:4_modes}.

We further attempted to quantify contamination by comparing the flux of our target in the Gaia $G_{\rm RP}$ bandpass to the sum of all flux from objects within a 30\arcsec\ radius of the target. We find that the visually classified contaminated objects suffer when the nearby sources sum to $>1.415\times$ the $G_{\rm RP}$ flux of the target. We selected the factor-of-1.415 threshold from a 1-d $k$-means sorting of the total $G_{\rm RP}$ fluxes from the 30\arcsec\ cones into two clusters. The centroid of the cluster of lowest fluxes was 1.415. The top panel of Figure~\ref{fig:color_vs_chisq} shows objects marked as ``likely crowded" based on this analysis, and we discontinue consideration of these crowded objects. We refer to this parameter as ``$G_{\rm RP}$ Excess" in the tables in Appendix~\ref{sec:var_tables}.

Our quality filtering removes 883 likely alternating sources (those with \chisqalt\,$ - $\,\chisqone\,$< 0$) and removes all likely crowded sources (827 total) through our cut on the Gaia $G_{\rm RP}$ fluxes, leaving 1590 objects in total, or about 48\% of the original sample. Although non-Gaussian, if these 1590 objects were normally distributed in \chisqone-space, the $3\,\sigma$ limit would be \chisqone$=2.3$.

We build a catalog variable white dwarfs as seen by WISE by ranking our sample by \chisqone, since signal-to-noise ratio is on average higher in the $W1$ band. We report \chisqtwo values in the tables of Appendix~\ref{sec:var_tables}, but do not incorporate them into our variability analysis. Having removed likely alternating and crowded sources, we impose a third cut to yield a final sample, keeping sources only with a \chisqone$>2.0$ or \chisqflat$>2.0$ to include sources that are showing potentially real, secular variations. Hereafter we abbreviate this ambiguity to simply \chisq. A cut at $\chi^2 > 2.0$ keeps the top 89th percentile of the ``non-alternating" and ``uncrowded" sample. This final catalog, shown in the bottom panel of Figure~\ref{fig:color_vs_chisq} and published in Table~\ref{tab:final_catalog}, contains a total of 163 unique sources. 

We assess the fidelity of the unWISE images for all these candidate variables and special samples (all the known gas disks, known infrared excess systems, known CVs). The unWISE images have a relatively large ($>$6\arcsec) PSF full-width-at-half-maximum \citep{Wright2010}, making source contamination another major issue \citep{Dennihy2020a}. We visually vetted our catalog of variables using the WISE View tool\footnote{\url{http://byw.tools/wiseview}} to inspect each unWISE image of each candidate centered on its J2000 coordinates. We look in particular for the proper motion of targets bringing them towards or away from a nearby field source in a manner correlated with the computed light curve. We mark these objects as contaminated in Figure~\ref{fig:color_vs_chisq} and Table~\ref{tab:final_catalog}. Panel d of Figure~\ref{fig:4_modes} shows an example of a secular brightening of a white dwarf moving into a bright background source. We also denote objects that still show a seasonal alternating pattern as contaminated in Figure~\ref{fig:color_vs_chisq} and Table~\ref{tab:final_catalog}.

\begin{deluxetable*}{rcccccc}[ht!]
\tabletypesize{\scriptsize}
\tablewidth{0pt} 
\tablecaption{Demographic Evolution of our Sample from the Original Input to our Catalog of Variables.\label{tab:sample_breakdown}}
\tablehead{
\colhead{Sample} & \colhead{Original Input} & \colhead{Pass Filtering} & \colhead{S/N\,$>10$} & \colhead{Variable (\chisq$>2$)} & \colhead{Uncontaminated} & \colhead{Highly Variable (\chisq$>3$)}}
\startdata 
Disks with Ca~II Emission & 21 & 13 & 11 & 6 & 6 & 6\\
Disks without Ca~II Emission & 47 & 29 & 17 &  9 & 9 & 6\\
Cataclysmic Variables & 18 & 8 & 8 & 8 & 8 & 8\\
Gaia-catWISE & 3213 & 1540 & 900 & 140 & 81 & 10\\
\tableline
Total & 3299 & 1590 & 936 & 163 & 104 & 30\\
\enddata
\tablecomments{Description of columns: From left to right we record the original number of our input, the amount that pass our quality filtering, the amount that have signal-to-noise$>$10 in our $W1$ light curves, the amount we find variable (\chisq$>$2), the amount of variables that do not appear contaminated upon closer examination, and finally the amount within our catalog that show high-amplitude variability (\chisq$>$3).}
\end{deluxetable*}

\section{Variability Analysis}\label{sec:results}

Here we report the results of our variability ranking. In Section~\ref{sec:catalog} we breakdown our catalog of 3.4-$\mu$m variables. We reserve Section~\ref{sec:gas_vs_dust} for analyzing the Ca~II and non-Ca~II debris disk systems in our catalog. We showcase the results of our near-infrared spectroscopic follow-up in Section~\ref{sec:spectra_results}.

\subsection{Catalog of 3.4-$\mu$m Variables}\label{sec:catalog}

We build our catalog of candidate variables by selecting sources with either \chisqone$>2.0$ or \chisqflat$>2.0$ (top 11\% most variable), which includes 163 unique white dwarfs. We break down in Table~\ref{tab:sample_breakdown} the demographics of our catalog relative to the original inputs. Our catalog is tabulated in full in Table~\ref{tab:final_catalog}. Among these 163 we find 104 uncontaminated sources that we believe to be high-confidence variables at 3.4\,$\mu$m, 84 of which are published as near-infrared variables for the first time based on our literature search.

The known CVs are by far the most variable in our catalog: all eight are within the top 90th or higher percentile of our variable catalog, and all have \chisq$>20$. None of these eight show evidence for having contaminated photometry. The Ca~II gas disk systems also show exceptionally large-amplitude variability. We defer characterization of their variability and that of the dusty disks to Section~\ref{sec:gas_vs_dust}.

We search for optical variability within our catalog to offer some characterization. We report the {\sc excess\_flux\_factor} (abbreviated as EFF in the tables of Appendix~\ref{sec:var_tables}) for all these 3.4-$\mu$m candidate variables in Table~\ref{tab:final_catalog}, a metric that reliably quantifies empirical variability from Gaia photometry \citep{GentileFusillo2021gaia}. We also use {\tt astroquery} \citep{Astroquery_Ginsburg2019} to download any 20-second- and 2-min-cadence PDCSAP light curves captured by the Transiting Exoplanet Survey Satellite (TESS, \citealt{Ricker2015}) of our candidate variables. We report the maximum period and associated amplitude from the 2-min light curves if it is significant to a false-alarm threshold of 0.1\% in Table~\ref{tab:final_catalog}. We use the tool \texttt{TESS\_localize} to confirm with $>$99\% likelihood that the signal is coming from our often-crowded white dwarfs \citep{Higgins2023}.

Beyond the eight previously known CVs, we recover 10 additional known variable white dwarfs in our final sample. This includes the known AM CVn G\,61$-$29 (Gaia DR3 3938156295111047680, \citealt{Nather1981}), eight pulsating white dwarfs (see Table~\ref{tab:final_catalog}), and one spotted magnetic white dwarf \citep[WD\,1953$-$011, Gaia DR3 4235280071072332672,][]{Brinkworth2005,Valyavin2011}. 
Like the CV KR Aur shown in Figure~\ref{fig:CV_LCs}, these recoveries further confirm that our method returns real astrophysical variables. 

At optical wavelengths, 34 of the 163 sources in our catalog are flagged in Gaia DR3 as photometrically variable \citep{Eyer2023}. Within these 34 are all eight CVs, one infrared excess system that is a known ZZ Ceti (Gaia DR3 1641326979142898048), and the gas disk system WD\,0145+234. While WD\,0145+234 has both a negative {\sc varindex} \citep{Guidry2021} and an {\sc excess\_flux\_error}$<$1 \citep{GentileFusillo2021gaia}, suggesting it is not variable over long baselines, the flagging by \citet{Eyer2023} indicates it may show optical variability on some timescale. 

Most of the white dwarfs in our catalog of 3.4-$\mu$m variables do not show coherent periodic variability in the optical as seen by TESS. We find only 13 of our 163 sources in our catalog to show a significant period intrinsic to white dwarf as determined by \texttt{TESS\_localize}. We stress these fractions cannot be compared exactly, given the differences in cadence and baselines. Seven of these 13 are known pulsating white dwarfs, including KX Dra \citep[Gaia DR3 1641326979142898048,][]{Vauclair2000} and WD\,1150$-$153 \citep[Gaia DR3 3571559292842744960,][]{Gianninas2006}, a known non-Ca~II gas debris disk, and four are known CVs. One of these 13 objects is Gaia DR3 3150770626615542784, which shows a significant periodicity at $200.31\,$s at an amplitude of 0.69\% over its 20-second light curves in Sectors $44-46$. \citet{Vincent2020} targeted this object because of its position in the ZZ Ceti instability but did not observe it to vary. We believe these TESS observations confirm it as a new pulsating white dwarf. Three of these 13 appear to show contaminated WISE light curves (see below). We highlight all of these TESS variables in Table~\ref{tab:final_catalog} by including the period and associated amplitude of variability. Lastly, we reject that WD\,0420+520 is a candidate eclipsing binary \citep{Prsa2022}. Our \texttt{TESS\_localize} analysis points towards Gaia DR3 271992483495298688 as the source of the apparent variability.

\begin{figure*}[t!]
\includegraphics[width=\textwidth]{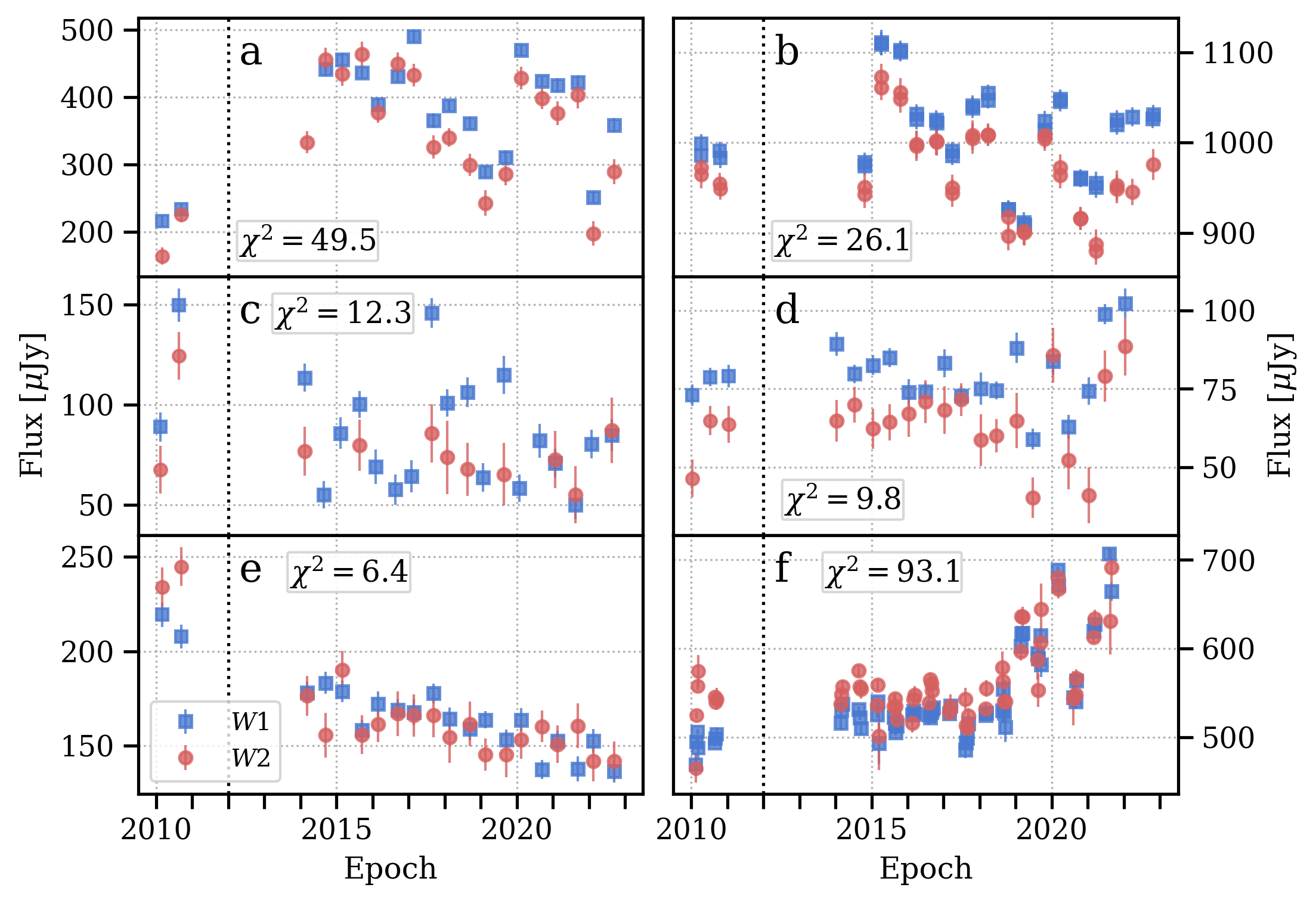}
\caption{WISE light curves of the white dwarfs with known Ca~II gas disks in our variability catalog \citep[all discovered in][unless noted otherwise]{Melis2020}: a LAMOST\,J0510+2315, b WD\,0842+572, c SDSS\,J0347+1624, d WD\,1622+587 \citep[and][]{Dennihy2020b}, e WD\,J0529$-$3401 \citep{GentileFusillo2021gas}. Each shows extreme variability: all have $\chi^2 > 6$ and peak-to-peak flux changes in $W1$\,$>20$\%. None of these systems have had their infrared light curves characterized before. We show Gaia\,J0611$-$6931 \citep[and][]{Dennihy2020b} in Panel F. Despite its exclusion from our catalog for possible crowding, we do not see evidence for obvious contamination in WISE View. Its disk may be undergoing a real brightening. WD\,0145+234 (Figure~\ref{fig:4_modes} Panel A) is the only Ca~II gas disk in our variable catalog not shown here.
\label{fig:6_gas_LCs}}
\end{figure*}

Based on visual inspection of the unWISE images using WISE View, we flag 35 of the 163 variables in our catalog as possibly contaminated by a nearby infrared source (21\%). These 35 (which are made up of sources only from our Gaia-catWISE crossmatch and includes no debris disk systems) appear to show variability correlated with the proper motion of the white dwarf or a nearby field star as seen in WISE View. We still publish these ``contaminated" objects as candidate 3.4-$\mu$m variables, but flag them in Table~\ref{tab:final_catalog} due to the subjectivity of this exercise.

Finally, we note our catalog of variables is still subject to further crowding and contamination due to the large WISE PSF. At least 31 of 163 objects still probably show an alternating pattern in their light curves (seven of which also have contaminated photometry). This is likely in part due to our clustering and sorting technique sometimes failing to identify the ``high" and ``low" seasons generated by the alternating PSF scanning angle of WISE. This technique particularly fails to filter out potentially false-positive variability due to multiple observations being taken during the same epoch. For example, WD\,0028$-$474 (Gaia DR3 4978793541987799040) and WD\,2118$-$388 (Gaia DR3 6583325635088476544) have two flux measurements per epoch for most epochs. In each case, the two concurrent measurements lie on two distinct flux baselines, forming ``high" and ``low" tracks at all epochs, rather than alternating from epoch-to-epoch like in Panel b of Figure~\ref{fig:4_modes}. Neither object shows evidence for rapid variability in TESS. It is unclear what is responsible for the apparent variability seen towards these two systems. Subsequent NEOWISE observing fields do tend to overlap by about 3\arcmin\ at all ecliptic latitudes \citep{Meisner2019}, generating multiple observations at the same epoch with possible differences in flux due to the PSF scanning angle varying as the spacecraft scans from field-to-field \citep{Schlafly2019}. These two objects are not located at extreme ecliptic latitudes (both lie between $-55^{\circ} < \beta < -35^{\circ}$). Both the non-contaminated and contaminated subsets of our catalog of variables appear to be distributed across all ecliptic latitudes.

In the tables in the Appendix we include the {\tt score} statistic calculated in the Gaia DR3 {\tt allwise\_neighbourhood}\footnote{\url{https://gaia.aip.de/metadata/gaiadr3/allwise_neighbourhood/}} table \citep{Marrese2017,Marrese2019,Marrese2022} as an additional Figure of Merit (FoM) for users to vet the sources in our catalog. We do not use the FoM to flag objects due to its incompleteness (41/163 objects in our catalog are not included in the Gaia-AllWISE neighborhood crossmatch analysis). \citet{Dennihy2020a} found white dwarf targets with S/N~$>$~10 in $W1$ and FoM~$>$~4 are unlikely to be confused with nearby field sources.

To grow the number of known white dwarf debris disks showing a gaseous component, we recommend prioritizing near-infrared spectroscopic follow-up like that in Section~\ref{sec:spectra_results} of the 104 uncontaminated high-confidence variables in our catalog, especially the 84 new discoveries among these, down-weighting the sources we flag as contaminated. Since we are interested in generating a sample of candidate strongly variable debris disks, we have not aimed for completeness in our catalog.

\subsection{Characterizing the Variability of Debris Disks with Detected Gaseous Calcium~II Emission}\label{sec:gas_vs_dust}

In addition to the CVs, the Ca~II gas disks appear among the most variable at 3.4\,$\mu$m. All six out of six gas disks in our catalog of variables have \chisq~$>6$ (Figure~\ref{fig:6_gas_LCs}, Figure~\ref{fig:4_modes}a). We publish five out of these six as near-infrared variables for the first time (WD\,0145+234 was characterized by \citealt{Swan2021}).

The six Ca~II gas disks in our variable catalog do not include cases like Gaia\,J0611$-$6931 (Panel F of Figure~\ref{fig:6_gas_LCs}). Gaia\,J0611$-$6931, which might be accreting a disrupted water-rich body \citep{Rogers2024b}, is excluded due to contamination from field crowding, but otherwise does not show obvious evidence for contamination in WISE View, suggesting the apparent brightening towards it could be real and inherent to its debris disk. Likewise, Gaia\,J2100+2122 and SDSS\,J2133+2428 show high-amplitude variability (\chisq~$>6$, see Table~\ref{tab:known_gas_disks}) and are excluded for having an excess of $G_{\rm RP}$ flux due to nearby field stars. Both also do not show evidence for contamination in WISE View.

There are seven Ca~II gas disk systems that pass our quality filtering and do not exhibit significant variability (\chisq~$<2$). Six of these (WD\,0842+231, SDSS\,J0959$-$0200, WD\,1041+092, SDSS\,J1228+1040, HE\,1349$-$2305, SDSS\,J1617+1620) were reported to exhibit significant variability in at least one Spitzer IRAC channel ($3.4\,\mu$m or $4.5\,\mu$m) by \citet{Swan2020}, albeit over significantly different light curve baselines ranging from hundreds of days up to about 10 years, and with better sensitivity than provided by WISE. The near-infrared variability towards WD\,J1930$-$5028 (Gaia DR3 6646693887514073984), the remaining system, has not been assessed due to the recency of the discovery of calcium triplet emission from its debris disk \citep{GentileFusillo2021gas}. Two of these six systems in \citet{Swan2020} have median S/N$<$$10$ in our $W1$ light curves, whereas all six variable gas disks in our variables catalog have S/N$>$$10$ in $W1$. This suggests signal-to-noise in our light curves impacts our determination of variability in some cases, especially when considering the reduced sensitivity in NEOWISE compared to Spitzer+IRAC. When we have robust detections (S/N$>$$10$) we find 6/11 gas disks show variability.

Our catalog of variables includes nine white dwarfs with known infrared excesses due to dusty debris disks with no gaseous emission at the calcium triplet yet detected. While all nine appear to show genuine astrophysical variability, it tends to be at a lower level compared to debris disks with gaseous Ca~II emission: only six show \chisq$>3$ (Figure~\ref{fig:color_vs_chisq}). The most extreme of these is GD\,56 (Gaia DR3 3251748915515143296), which has \chisq\,$=40.0$. We show the light curve of GD\,56 in Panel~9 of Figure~\ref{fig:all_dust_lcs_1}, where the $W1$ and $W2$ fluxes linearly decay until about the year 2020, after which they stabilizes to constant values in both channels. The decaying near-infrared flux towards GD\,56 was first reported by \citet{Farihi2018}. Later analysis by \citet{Swan2021} found this trend to be consistent with the $t^{-1}$ decay expected from a collisional cascade \citep{Dominik2003}. We report near-infrared variability towards the non-Ca~II debris disks of MCT\,0420$-$7310, KPD\,0420+5203, and US\,2966 for the first time.

There are 20 non-Ca~II disk systems that pass our quality filtering but do not show significant variability (\chisq$< 2$). 11 of these 20 systems are included in the survey by \citet{Swan2020}, who report significant variability in Spitzer IRAC photometry towards eight. Only four of these eight are reliably bright enough (S/N$>10$) for us to detect variability, helping explain the why some are not recovered. A total of 8/20 of these non-variable (\chisq$<2$) non-Ca~II disks show S/N$>$$10$ in our $W1$ light curves, yielding a variability rate of 9/17.

We find Ca~II (55\%) and non-Ca~II (53\%) disks are variable at roughly the same rates, except disks with observed Ca~II gas in emission nearly always show larger flux variations when they are detected to vary. This variability is observed across decade-long baselines with roughly 6~month sampling in our $W1$ NEOWISE light curves. We do not find commensurate optical variability towards any of these systems in TESS, save the known pulsating white dwarfs EC\,11507$-$519 (0.8\% pulsation amplitude in $V$-band, \citealt{Gianninas2006}; Figure~\ref{fig:all_dust_lcs_2} Panel 25) and KX Dra (0.5\% pulsation amplitude in TESS, Table~\ref{tab:final_catalog}; Figure~\ref{fig:all_dust_lcs_2} Panel 35) that also harbor dusty disks. Only KX Dra shows a Gaia {\sc excess\_flux\_error} consistent with optical variability. This underlines that the variability we detect towards these white dwarfs in NEOWISE is intrinsic to their dusty disks. 

In Appendix~\ref{sec:light_curves} we show light curves for all 21 published Ca~II gas disk systems in Figure~\ref{fig:all_gas_lcs} and the 47 non-Ca~II disks across Figure~\ref{fig:all_dust_lcs_1} and Figure~\ref{fig:all_dust_lcs_2}. We report our variability parameters for the Ca~II disks in Table~\ref{tab:known_gas_disks} and for the non-Ca~II disks in Table~\ref{tab:known_dust_disks}.

\begin{figure}[t!]
\includegraphics[width=0.48\textwidth]{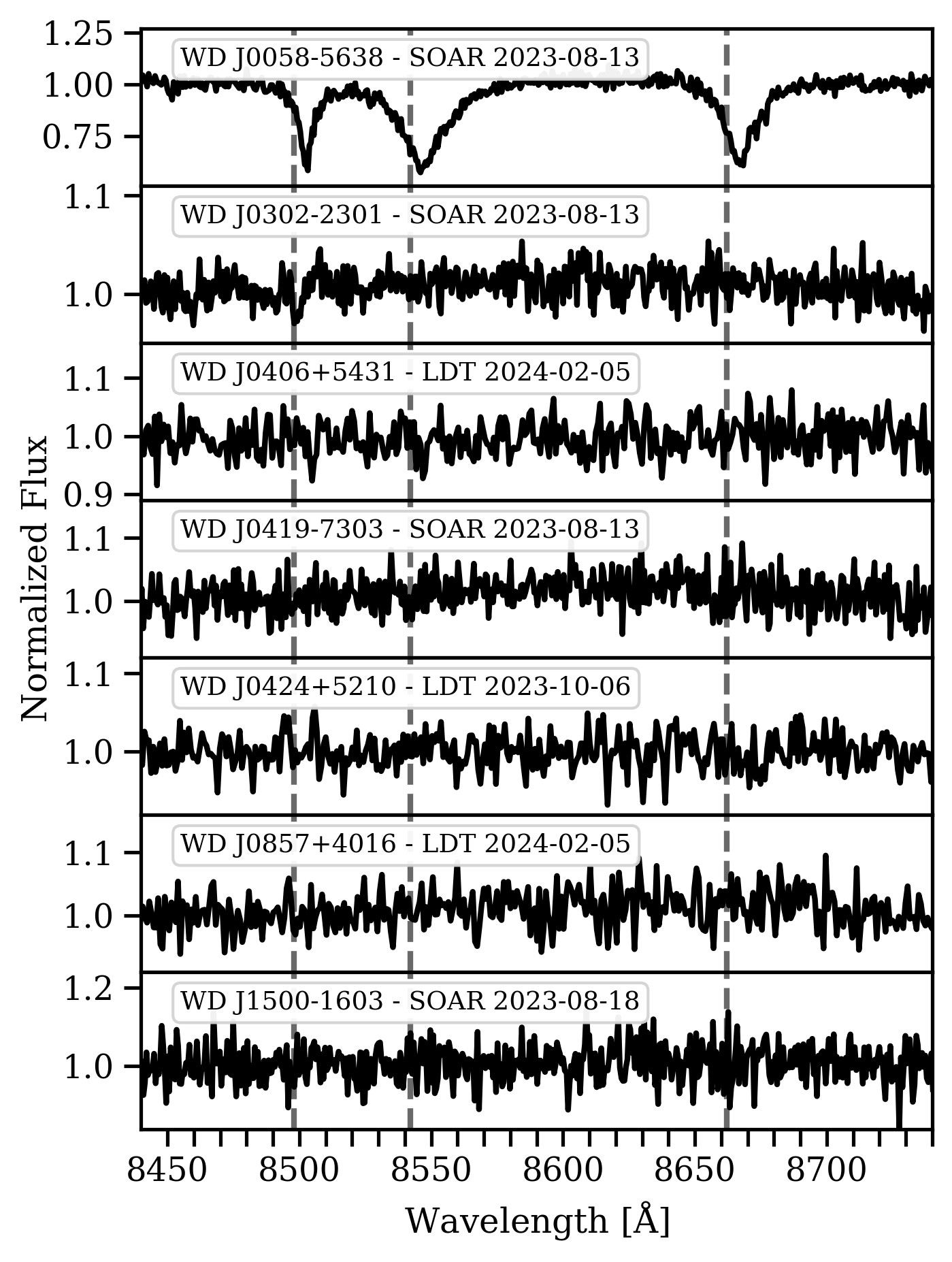}
\caption{
Near-infrared spectroscopic follow-up of seven sources from our catalog centered on the calcium triplet region (rest wavelengths marked as dashed grey vertical lines). While we do not detect Ca~II gas in emission, we confirm one new remnant planetary system: we observe broad Ca~II absorption towards WD\,J0058$-$5638 (top panel). Sources are marked with the observatory and UT date of observation. 
\label{fig:spectra}}
\end{figure}

\subsection{Preliminary Searches for Circumstellar Calcium~II Emission}\label{sec:spectra_results}

We followed up seven objects within our catalog of variables using LDT+DeVeny and SOAR+Goodman as described in Section~\ref{sec:followup}, the results of which we show in Figure~\ref{fig:spectra}. While we fail to discover new systems showing emission at the calcium triplet, we notably confirm one remnant planetary system, marked by Ca~II absorption. The six others show only continua in our wavelength range covering at least $7600-8750\,$\AA.

WD\,J0058$-$5638 (Gaia DR3 4907742035447425024) shows broad absorption at each line of the triplet likely of photospheric origin (S/N$=38.5$). We prioritized WD\,J0058$-$5638 in our follow-up based on its high degree of variability (\chisq$=12.8$) and red catWISE color (Figure~\ref{fig:color_vs_chisq}). While no higher-resolution spectrum of WD\,J0058$-$5638 is published, \citet{Vincent2024} find its Gaia XP spectra indicate it is spectral type DZ with a probability of 1.0. Photometric fits using He-atmosphere models find it is a 9970\,K, 0.65\,\msun\ white dwarf \citep{GentileFusillo2021gaia}.

Also included in our follow-up are MCT\,0420$-$7310 (Gaia DR3 4653404070862114176) and KPD\,0420+5203 (Gaia DR3 271992414775824640), two systems known to harbor dusty disks. Both show only continua at the calcium triplet region. Our spectra of MCT\,0420$-$7310 (S/N$=$38.9) and KPD\,0420+5203 (S/N$=$47.6) are the first published in this region, and suggest their debris disks do not show gaseous calcium in emission at our detection limits.

\section{Discussion and Conclusions}\label{sec:discussion}

We produce a catalog of 163 unique white dwarfs that appear to show astrophysical variability at $3.4\,\mu$m in $W1$ NEOWISE light curves with coverage from 2013--2023 sampled roughly every six months. We generate these light curves using a novel aperture photometry pipeline operating on the unWISE images that corrects for the proper motion of the target. We pass an original input of 3299 white dwarfs, including all 21 published white dwarf Ca~II gas disks and most published non-Ca~II disks, through quality filtering to eliminate known systematics in effort to include only high-fidelity sources. We further vet the of images of the sources in our catalog of 163 variables, finding 104 are high-confidence variables at 3.4\,$\mu$m, 84 of which are published as variable for the first time.

We find that among sources with uncontaminated photometry and sufficient signal-to-noise (median S/N$>$$10$ in our $W1$ light curves), 6/11 Ca~II gas debris disks and 9/17 of non-Ca~II gas disks are variable over baselines of about $9-10$ years at $\approx$$6$-month sampling. Due to the substantially dramatic variability shown by the Ca~II gas disks, we hypothesize that targeted searches for white dwarfs exhibiting variability at wavelengths $\lambda \gtrsim 3.4\,\mu$m represents a promising avenue for discovering more white dwarf debris disks showing gaseous emission. We encourage spectroscopic follow-up of the catalog in Table~\ref{tab:final_catalog} to search for emission at the calcium~II triplet indicative of a gaseous debris disk. This is especially motivated by our discovery of one new remnant planetary system, evidenced by calcium triplet absorption towards the white dwarf (Figure~\ref{fig:spectra}), though we have not yet discovered a new gas disk in our preliminary follow-up campaign. 

Our findings here are in agreement with the work from \citet{Swan2019} and \citet{Swan2020}, who found that most debris disks around white dwarfs are variable at 3.4\,$\mu$m, with the Ca~II gas disks showing the highest flux variations compared to their non-Ca~II counterparts. We also tend to see the degree of variability grow with the light curve baseline, as was observed with Spitzer \citep{Swan2020}. Where there is overlap, all the Ca~II and non-Ca~II debris disks both in our catalog of variables are also found to be variable by \citet{Swan2020}. We identify at least five new variable Ca~II gas disks: LAMOST\,J0510+2315, SDSS\,J0347+1624, WD\,1622+587, and WD\,J0529$-$3401 (Figure~\ref{fig:6_gas_LCs}). We report near-infrared variability towards the non-Ca~II debris disks of MCT\,0420$-$7310, KPD\,0420+5203, and US\,2966 for the first time.

Only 10 of our 104 high-confidence 3.4-$\mu$m variables show periodic optical variability in TESS. Notwithstanding the differences in sampling and baselines and the determination of variability (empirically based on anomalous scatter in NEOWISE versus significant periodicities in TESS), this suggests that the variability we detect towards these debris disks systems is not intrinsic to the white dwarf photosphere, but rather arises from the dusty components of the disks. 

By considering both \chisqone and \chisqflat in our variability ranking, we find at least one more debris disk showing a secular linear trend in flux. The disk around WD\,J0529$-$3401 (Panel e of Figure~\ref{fig:6_gas_LCs}) shows a secular decay reminiscent of that towards GD\,56 (Panel 9 of Figure~\ref{fig:all_dust_lcs_1}). Such a decay would be consistent with that expected from a brightening induced by a collisional cascade \citep{Swan2024,Swan2021,Dominik2003}, albeit over a timescale of at least 10~years. Coupled with the absence of optical variations we find, the heightened 3.4-$\mu$m variability of the disks showing Ca~II emission in our catalog relative to the non-Ca~II disks seems to lends further credence to the notion that the photometric variability of these disks is driven by collisions \citep{Farihi2018,Swan2019,Malamud2021}, and that the gaseous disks may be the most the most collisionally active at this moment \citep{Swan2020}.

Lacking in our analysis is an assessment of the variability seen in our $W2$-band (4.6\,$\mu$m) light curves. We only analyze the $W1$ light curves due to their on-average higher signal-to-noise. Still, the most variable debris disks tend to show significant variability in $W2$ as well (Figure~\ref{fig:6_gas_LCs}). SDSS\,J1228+1040, which is not in our catalog of variables and does not appear contaminated, appears to shows more significant flux variations in $W2$. Indeed, \citet{Xu2018b} found SDSS\,J1228+1040 shows greater variability at $4.5\,\mu$m in Spitzer IRAC photometry. We do tabulate calculated \chisqtwo values in Appendix~\ref{sec:var_tables}. We also do not analyze our PSF-photometry generated light curves since they appear to suffer source detection failures at the lowest signal-to-noise levels, but publish them as well and encourage investigation. 

We also plot the WISE cryogenic epochs in our light curves but exclude these fluxes from our variability ranking. Their inclusion would tend to agree with the finding from \citet{Swan2020} that the degree of variability of white dwarf debris disks grows with the light curve baseline (Figure~\ref{fig:6_gas_LCs}). However, there are systematic offsets in flux between the cryogenic and NEOWISE missions \citep{Meisner2017a} that make this an unreliable comparison.

Unlike their hosts' photospheres, the future for the study of remnant planetary systems around white dwarfs is bright in the infrared. The Near-Earth Object Surveyor, the successor to WISE and NEOWISE, is planned to launch as early as 2028 and survey the sky at 4.6 and 8\,$\mu$m \citep{Mainzer2023}. SPHEREx will compliment NEO Surveyor with spectra spanning $0.75-5\,\mu$m covering the entire sky \citep{Dore2018}. Large, deep spectroscopic surveys like DESI and SDSS-V should discover more gaseous debris disks from their emission at the near-infrared calcium triplet, with target selection propelled by the discovery of hundreds of thousands high-probability white dwarf candidates with Gaia \citep{GentileFusillo2021gaia}. JWST is already characterizing gaseous debris disks \citep{Swan2024}, with surely more on the way. We expect that near-infrared variability will aid in efficiently identifying the most compelling targets for the next generation of flagship observatories like JWST.

\section{Acknowledgments}
We thank the anonymous referee for their constructive comments and suggestions that improved this paper.

This material is based upon work supported by the National Aeronautics and Space Administration under Grant No. 80NSSC23K1068 issued through the Science Mission Directorate. Support for this work was in part provided by NASA TESS Cycle 4 grant 80NSSC22K0737 and Cycle 6 grant 80NSSC24K0878. JAG is supported by the National Science Foundation Graduate Research Fellowship Program under Grant No. 2234657. JAG recognizes Siyi Xu (许\CJKfamily{bsmi}偲\CJKfamily{gbsn}艺, Gemini Observatory) for numerous fruitful discussions that improved this study and the interpretation of these results. K. D. was supported by NASA through the NASA Hubble Fellowship grant \#HST-HF2-51477.001 awarded by the Space Telescope Science Institute, which is operated by the Association of Universities for Research in Astronomy, Inc., for NASA, under contract NAS5-26555. This research was improved by discussions at the KITP Program ``White Dwarfsas Probes of the Evolution of Planets, Stars, the Milky Way and the Expanding Universe" supported by National Science Foundation under grant No. NSF PHY-1748958.

This publication makes use of data products from the Wide-field Infrared Survey Explorer, which is a joint project of the University of California, Los Angeles, and the Jet Propulsion Laboratory/California Institute of Technology, funded by the National Aeronautics and Space Administration. This publication also makes use of data products from NEOWISE, which is a project of the Jet Propulsion Laboratory/California Institute of Technology, funded by the Planetary Science Division of the National Aeronautics and Space Administration.

These results make use of the Lowell Discovery Telescope at Lowell Observatory. Lowell is a private, nonprofit institution dedicated to astrophysical research and public appreciation of astronomy and operates the LDT in partnership with Boston University, the University of Maryland, the University of Toledo, Northern Arizona University, and Yale. We particularly thank LDT telescope operators Ishara Nisley and Cecilia Siqueiros.

Our work is based on observations obtained at the Southern Astrophysical Research (SOAR) telescope, which is a joint project of the Minist\'{e}rio da Ci\^{e}ncia, Tecnologia e Inova\c{c}\~{o}es (MCTI/LNA) do Brasil, the US National Science Foundation's NOIRLab, the University of North Carolina at Chapel Hill (UNC), and Michigan State University (MSU). We particularly thank Sean Points and Felipe Navarete for observing on our behalf and Jay Elias for facilitating our observing plan.

This work has made use of data from the European Space Agency (ESA) mission
{\it Gaia} (\url{https://www.cosmos.esa.int/gaia}), processed by the {\it Gaia}
Data Processing and Analysis Consortium (DPAC,
\url{https://www.cosmos.esa.int/web/gaia/dpac/consortium}). Funding for the DPAC
has been provided by national institutions, in particular the institutions
participating in the {\it Gaia} Multilateral Agreement.

This paper includes data collected by the TESS mission. Funding for the TESS mission is provided by the NASA's Science Mission Directorate.

\vspace{5mm}
\facilities{WISE, NEOWISE, Gaia, LDT, SOAR, TESS, ASAS-SN}

\software{Astropy \citep{astropy2013,astropy2018},  
          lmfit \citep{newville_matthew_2014_11813},
          Matplotlib \citep{Matplotlib2007},
          NumPy \citep{numpy2020}
          pandas \citep{pandas2020},
          PypeIt \citep{pypeit:joss_arXiv},
          SciPy \citep{2020SciPy-NMeth},
          seaborn \citep{Waskom2021},
          TOPCAT \citep{Taylor2005_TOPCAT},
          SIMBAD and VizieR (curated by CDS, Strasbourg, France),
          and the NASA Astrophysics Data System (ADS).}

\appendix

\section{Variability Rank Tables}\label{sec:var_tables}

Here we show various tables summarizing our variability ranking and validation for the various inputs in our sample. In Table~\ref{tab:final_catalog} we show these parameters for our catalog of variables, which includes all 104 high-confidence variables and the 59 likely contaminated sources. In Table~\ref{tab:known_gas_disks} we show these parameters for the 21 previously published white dwarf debris disks that show gaseous calcium in emission. In Table~\ref{tab:known_dust_disks} we show these parameters for the 47 disks in our study that do not show gaseous emission.

\movetabledown=1.5cm
\begin{longrotatetable}
\startlongtable
\begin{deluxetable}{rrrccccccccccccccccc}
\tabletypesize{\tiny}
\tablewidth{0pt} 
\tablecaption{Variability Parameters for our Catalog of 163 Variable White Dwarfs \label{tab:final_catalog}}
\tablehead{
\colhead{Gaia DR3 Source ID} &  \colhead{R.A.}& \colhead{Decl.} & \colhead{Sample} & \colhead{Flag\hspace{-0.2cm}} & \colhead{\chisqone} & \colhead{\chisqalt} & \colhead{$\chi^2_{\mathrm{flat}}$} & \colhead{\chisqtwo} & \colhead{S/N} &  \colhead{$G_{\rm RP}$ Excess} & \colhead{$G_{\rm RP}$} & \colhead{$W1$} & \colhead{$W1-W2$} & \colhead{Spectral} & \colhead{\teff} & \colhead{$M_{\rm WD}$}   & \colhead{EFF} & \colhead{FoM} & \colhead{TESS}\\
 & \colhead{(J2000)} & \colhead{(J2000)} & & & & & & & \colhead{$W1$} &  & (mag) & (mag) & (mag) & \colhead{Type} & (K) & (\msun) & & & (hr, \%)}
\startdata 
2312821266302222720 & 0.66720 & $-$34.22766 & GC & * & 1.3 & 1.9 & 2.2 &  & 5.1 & 0.000 & 14.5 & 20.2 & $-$0.3 & DC [1] & 6460 & 0.73 & 1.8 & 9.3 & \\
537563127588521600 & 1.28048 & 73.21927 & GC & * & 2.5 & 3.1 & 2.6 & 1.0 & 43.1 & 0.231 & 14.3 & 17.8 & $-$0.8 & DBZA [2] & 14160 & 0.66 & 1.4 & 7.6 & \\
4978793541987799040 & 7.69668 & $-$47.21010 & GC & 2 & 47.1 & 68.1 & 46.5 & 4.0 & 18.9 & 0.000 & 15.2 & 19.0 & $-$0.8 & DA [3] & 13600 & 0.31 & 0.3 & 9.0 & \\
381396329995329408 & 9.98592 & 42.49150 & GC & * & 2.1 & 2.8 & 2.3 & 0.0 & 9.4 & 0.295 & 16.4 & 19.5 & $-$0.8 & DA [4] & 9990 & 0.55 & 0.0 & 7.6 & \\
414603986339418624 & 12.18956 & 50.12801 & GC & * & 2.4 & 3.7 & 2.3 & 0.9 & 18.5 & 0.361 & 16.1 & 18.7 & $-$0.8 & DA [5] & 10130 & 0.76 & 0.0 &  & \\
4907742035447425024 & 14.56430 & $-$56.63616 & GC & * & 12.0 & 18.9 & 12.8 & 4.9 & 14.3 & 0.560 & 16.9 & 19.1 & $-$0.2 & DZ [5] & 10250 & 0.74 & 0.7 & 8.5 & \\
5026963661794939520 & 17.15014 & $-$32.62872 & IR, GC &  & 2.2 & 2.6 & 2.5 & 0.9 & 28.9 & 1.110 & 15.5 & 18.1 & $-$0.7 & DAZ [3] & 16050 & 0.63 & 0.0 &  & \\
4913589203924379776 & 18.08809 & $-$56.24098 & IR, GC &  & 2.6 & 4.1 & 2.7 & 1.1 & 20.4 & 0.000 & 15.9 & 18.8 & $-$0.3 & DBAZ [6] & 18790 & 0.59 & 0.0 & 8.7 & \\
321856332242222336 & 18.88398 & 37.62655 & CV &  & 137.5 & 203.3 & 156.3 & 12.8 & 23.5 & 0.000 & 17.2 & 18.3 & $-$0.5 & CV [7] &  &  &  & 7.7 & \\
409956900443661312 & 19.72531 & 52.45381 & GC & * & 1.5 & 2.5 & 2.1 & 0.5 & 15.5 & 0.475 & 15.6 & 18.9 & $-$0.7 & DA [3] & 10760 & 0.60 & 0.7 & 5.1 & \\
4704847230628098048 & 21.63848 & $-$65.94308 & GC & * & 2.5 & 4.0 & 2.4 & 0.9 & 19.7 & 0.572 & 15.6 & 19.0 & $-$0.9 & DA [8] & 13400 & 0.76 & 1.2 & 6.7 & \\
2480731779699826176 & 22.59595 & $-$4.79939 & GC & 1 & 1.2 & 1.9 & 3.4 & 0.9 & 12.1 & 0.355 & 15.0 & 19.0 & $-$0.7 & DA [3] & 16280 & 0.57 & 0.0 & 9.0 & \\
4686195252815125888 & 22.74033 & $-$73.83004 & GC & 1 & 1.9 & 2.4 & 14.6 & 1.0 & 28.7 & 0.566 & 16.4 & 18.7 & $-$0.6 & DA [9] & 7820 & 0.52 & 0.0 &  & \\
4912427947846613760 & 23.99450 & $-$54.73145 & GC & * & 2.1 & 2.1 & 2.3 & 1.1 & 27.5 & 0.360 & 15.4 & 18.5 & $-$0.7 & DA [5] & 10310 & 0.61 & 1.3 & 8.0 & \\
572487740053132288 & 25.36576 & 83.58301 & GC & * & 2.1 & 2.4 & 2.7 & 1.6 & 98.0 & 0.013 & 13.3 & 16.8 & $-$0.7 & DA [3] & 19000 & 0.65 & 0.3 & 10.4 & \\
5135466183642594304 & 26.84096 & $-$21.94761 & GC & [a] & 1.4 & 2.1 & 2.4 & 0.9 & 65.6 & 0.000 & 15.2 & 16.9 & $-$0.5 & DA [3] & 11080 & 0.60 & 7.8 & 8.9 & 0.12, 0.5\\
291057843317534464 & 26.97841 & 23.66211 & Ca~II, GC &  & 422.6 & 690.7 & 482.4 & 254.9 & 54.3 & 0.021 & 14.1 & 16.9 & $-$0.4 & DA [10] & 12980 & 0.66 & 0.8 & 5.5 & \\
2574761734934286848 & 28.77645 & 12.14059 & GC & * & 2.2 & 3.4 & 2.1 &  & 6.6 & 0.000 & 17.0 & 19.8 & $-$0.4 & DA [5] & 7810 & 0.61 & 0.0 & 6.9 & \\
4970215770740383616 & 33.45361 & $-$33.75834 & GC & 1 & 1.7 & 1.8 & 10.8 & 0.9 & 25.4 & 0.555 & 15.1 & 18.3 & $-$0.7 & DAZ [1] & 4640 & 0.37 & 27.4 & 1.4 & \\
5063539946887524864 & 37.11329 & $-$32.70939 & GC & * & 2.1 & 2.4 & 2.1 & 1.1 & 34.4 & 0.011 & 14.1 & 17.9 & $-$0.7 & DA [3] & 22280 & 0.63 & 0.3 & 8.0 & \\
4945893096067016960 & 39.19143 & $-$44.87680 & GC & * & 2.1 & 2.6 & 2.0 &  & 6.7 & 0.000 & 16.6 & 20.2 & $-$0.8 & DA [5] & 15690 & 0.77 & 1.9 & 6.2 & \\
548423278811480960 & 40.38260 & 74.71518 & GC & 1 & 1.3 & 2.1 & 3.7 & 0.9 & 15.5 & 0.505 & 15.8 & 19.1 & $-$0.7 & DA [3] & 9130 & 0.60 & 0.0 & 9.3 & \\
5052391208218637440 & 43.82827 & $-$32.11233 & GC & * & 2.2 & 2.5 & 2.2 &  & 6.6 & 0.000 & 17.0 & 19.9 & $-$0.7 & DA [5] & 6770 & 0.83 & 1.5 & 8.2 & \\
5078074837769743744 & 45.65272 & $-$23.03090 & GC & * & 2.5 & 3.5 & 2.5 & 0.0 & 8.4 & 0.105 & 15.9 & 19.6 & $-$0.7 & DA [3] & 22880 & 0.86 & 1.2 & 7.9 & \\
4733604373137759616 & 48.32741 & $-$56.12639 & GC & 1,[b] & 1.3 & 1.7 & 11.9 & 1.0 & 52.4 & 0.636 & 14.6 & 17.9 & $-$0.8 & DA [11] & 10980 & 0.60 & 6.1 &  & 0.22, 0.2\\
4847443748331819136 & 49.20349 & $-$44.62911 & GC & * & 1.8 & 2.6 & 2.5 & 0.8 & 22.4 & 0.200 & 17.0 & 18.8 & $-$0.8 & DA [5] & 5590 & 0.64 & 0.8 &  & \\
5098597840775610112 & 50.37700 & $-$22.45211 & GC & * & 2.7 & 2.8 & 2.7 & 0.0 & 7.8 & 0.038 & 16.7 & 19.7 & $-$0.6 & DA [5] & 5210 & 0.38 & 13.0 & 4.2 & \\
5053839127592396032 & 52.21176 & $-$32.64194 & GC & 1 & 1.2 & 1.3 & 5.0 & 0.9 & 13.0 & 1.399 & 16.6 & 19.2 & $-$0.7 & DC [1] & 4640 & 0.45 & 1.1 & 8.5 & \\
63126196662620416 & 55.84636 & 19.97044 & GC & * & 2.0 & 2.6 & 2.0 & 0.6 & 27.8 & 0.487 & 15.3 & 18.0 & $-$0.6 & DA [12] & 7110 & 0.87 & 0.0 & 8.9 & \\
43629828277884160 & 56.90288 & 16.40270 & Ca~II & * & 12.3 & 17.7 & 12.1 & 0.5 & 11.2 & 0.940 & 16.7 & 19.2 & $-$0.4 & DAZ [13] & 17710 & 0.52 & 0.0 & 6.9 & \\
570086853335107200 & 57.49233 & 82.90492 & GC & * & 2.2 & 3.1 & 2.1 & 0.4 & 12.8 & 0.153 & 16.2 & 19.6 & $-$0.8 & DA [5] & 13430 & 0.61 & 1.2 & 6.0 & \\
5094843764483219712 & 58.24161 & $-$19.93029 & GC & * & 2.2 & 3.0 & 2.1 & 0.5 & 8.8 & 0.000 & 16.5 & 19.6 & $-$1.0 & DA [5] & 10250 & 0.59 & 2.6 & 7.1 & \\
276192136878054656 & 61.52921 & 54.52561 & GC & * & 3.6 & 4.5 & 3.4 & 0.3 & 8.9 & 0.945 & 15.5 & 19.5 & $-$0.8 & DA [3] & 17010 & 0.63 & 0.9 &  & \\
3251748915515143296 & 62.75904 & $-$3.97294 & IR &  & 4.9 & 6.8 & 40.0 & 3.3 & 56.2 & 0.268 & 15.7 & 17.1 & 0.2 & DA [3] & 14900 & 0.61 & 0.1 & 10.7 & \\
229143725086190336 & 64.52301 & 42.18394 & GC & 1 & 1.3 & 2.0 & 3.3 & 1.9 & 23.1 & 0.989 & 16.4 & 18.3 & $-$0.7 & DA [14] & 5050 & 0.49 & 0.2 &  & \\
4653404070862114176 & 64.90760 & $-$73.06236 & IR, GC & * & 2.4 & 3.9 & 3.5 & 1.3 & 59.6 & 0.150 & 15.7 & 17.7 & $-$0.1 & DA [3] & 17790 & 0.58 & 0.0 & 9.6 & \\
232990572675079296 & 65.37243 & 46.13271 & GC & 1 & 1.2 & 1.2 & 2.2 & 1.1 & 41.0 & 0.996 & 14.9 & 17.6 & $-$0.8 & DA [14] & 7810 & 0.59 & 0.0 & 7.0 & \\
271992414775824640 & 66.06547 & 52.16962 & IR, GC & * & 3.1 & 4.9 & 17.4 & 2.2 & 40.1 & 0.876 & 15.2 & 17.6 & $-$0.3 & DA [3] & 22170 & 0.66 & 0.9 & 8.9 & \\
3186021141200137472 & 69.44754 & $-$8.81962 & GC & 2 & 6.8 & 7.4 & 23.1 & 1.3 & 14.2 & 0.073 & 13.2 & 19.1 & $-$0.8 & DQ [15] & 6720 & 0.66 & 1.8 & 9.7 & \\
4867104257483687040 & 71.11395 & $-$35.29586 & GC & * & 1.5 & 2.3 & 2.3 & 0.1 & 10.9 & 0.138 & 15.7 & 19.5 & $-$0.9 & DBA [16] & 13870 & 0.75 & 0.0 & 8.1 & \\
4893995356962398208 & 72.26348 & $-$24.21089 & GC & * & 1.6 & 1.8 & 2.0 & 1.2 & 22.6 & 1.228 & 16.7 & 18.4 & $-$0.6 & DA [11] & 4870 & 0.56 & 0.0 &  & \\
4776747972460494208 & 73.81052 & $-$54.69595 & GC & * & 1.9 & 2.1 & 4.7 & 1.5 & 55.2 & 0.291 & 14.5 & 17.9 & $-$0.8 & DA [17] & 17360 & 0.55 & 0.0 & 5.9 & \\
3212803625248377088 & 76.55461 & $-$4.13535 & CV &  & 21.3 & 31.7 & 20.2 & 5.3 & 26.4 & 0.091 & 16.7 & 18.2 & $-$0.3 & CV [18] &  &  & 51.6 & 10.0 & \\
3415788525598117248 & 77.50894 & 23.26151 & Ca~II, GC & * & 41.9 & 65.9 & 49.5 & 16.0 & 41.8 & 0.796 & 15.2 & 17.4 & $-$0.1 & DAZ [19] & 21250 & 0.73 & 0.0 & 9.0 & \\
4823042355496728832 & 82.30968 & $-$34.01892 & Ca~II & * & 2.0 & 2.2 & 6.4 & 0.8 & 28.5 & 0.230 & 17.6 & 18.4 & 0.1 & DAZ [20] & 17190 & 0.46 & 0.0 &  & \\
4805197659735412224 & 86.36308 & $-$39.59992 & GC & * & 1.4 & 1.5 & 2.2 & 1.0 & 31.4 & 0.085 & 16.3 & 18.4 & $-$0.7 & DA [5] & 5220 & 0.27 & 0.0 & 8.9 & \\
3429193393407592576 & 87.69361 & 26.20574 & GC & * & 2.5 & 3.3 & 2.4 & 0.8 & 9.7 & 0.822 & 15.6 & 19.2 & $-$0.8 & DA [21] & 21210 & 0.57 & 0.0 &  & \\
3218697767783768320 & 87.83117 & $-$0.17253 & GC & * & 2.2 & 2.6 & 2.6 & 1.2 & 59.0 & 0.015 & 14.0 & 17.0 & $-$0.7 & DQP [1] & 6180 & 0.72 & 4.5 & 9.5 & \\
3022568258011201024 & 88.03243 & $-$5.41682 & CV &  & 1216.6 & 1640.3 & 1152.6 & 526.6 & 81.3 & 0.113 & 13.5 & 15.3 & $-$0.5 & CV [22] &  &  &  & 10.1 & \\
4650712810002383104 & 89.51038 & $-$72.48012 & GC & 2 & 2.9 & 5.5 & 2.6 & 1.2 & 14.8 & 1.284 & 16.7 & 19.5 & $-$0.8 & DC [1] & 6850 & 0.83 & 0.0 & 7.1 & \\
1114054838014073984 & 91.81668 & 73.53325 & GC & 1 & 1.1 & 1.5 & 5.5 & 0.8 & 10.1 & 0.049 & 16.6 & 19.6 & $-$0.5 & DA [14] & 6520 & 0.92 & 1.3 & 6.6 & \\
943770757800160384 & 101.90802 & 37.51585 & GC & 1 & 0.2 & 0.2 & 6.0 &  & 6.3 & 0.063 & 12.3 & 19.8 & $-$0.7 & DA [3] & 21390 & 0.71 & 0.2 & 10.5 & \\
950478049312082688 & 104.43818 & 40.12384 & GC & 2 & 2.3 & 2.9 & 2.3 & 0.0 & 11.3 & 0.061 & 16.6 & 19.3 & $-$0.9 & DA [14] & 6580 & 0.46 & 0.6 & 7.8 & \\
5536077746353130240 & 113.40764 & $-$42.89924 & GC & 3 & 1.7 & 2.4 & 15.8 & 1.2 & 14.0 & 0.514 & 14.3 & 19.2 & $-$0.7 & DA [10] & 15020 & 0.70 & 0.3 &  & \\
5295512298775133824 & 115.47008 & $-$57.14576 & GC & * & 2.1 & 2.1 & 2.0 & 2.0 & 24.4 & 0.891 & 15.3 & 18.8 & $-$0.7 & DA [3] & 19540 & 0.47 & 0.0 &  & \\
3150770626615542784 & 117.92301 & 11.34175 & GC & * & 2.1 & 3.1 & 2.1 & 0.7 & 8.0 & 0.690 & 16.4 & 19.6 & $-$0.9 & DA [23] & 11810 & 0.56 & 0.8 & 6.1 & 0.06, 0.7\\
5210507882302442368 & 121.86563 & $-$76.53352 & CV &  & 367.0 & 520.4 & 346.0 & 245.6 & 98.9 & 0.626 & 15.1 & 16.3 & $-$0.4 & CV [24] &  &  & 111.8 & 2.2 & \\
1091051096255456384 & 123.11779 & 62.60623 & CV &  & 499.9 & 956.4 & 489.1 & 288.5 & 92.0 & 0.028 & 14.3 & 15.9 & $-$0.4 & CV [25] &  &  &  & 10.0 & \\
908364559240386944 & 125.00536 & 38.57653 & GC & * & 1.9 & 2.5 & 2.2 &  & 5.9 & 0.084 & 16.2 & 19.9 & $-$0.9 & DA [10] & 7640 & 0.62 & 0.0 & 5.6 & \\
911639111025648512 & 130.03196 & 40.25104 & GC & [c] & 1.4 & 2.2 & 2.1 & 1.2 & 17.1 & 0.000 & 15.6 & 18.8 & $-$0.5 & DA [3] & 11570 & 0.61 & 3.4 &  & 0.14, 0.9\\
5639391810273308416 & 130.38422 & $-$32.94248 & GC & 1 & 51.2 & 76.7 & 49.9 & 6.9 & 20.6 & 0.089 & 11.7 & 18.4 & $-$0.8 & DA [3] & 9080 & 0.47 & 0.9 & 7.3 & \\
5734737438536674432 & 130.70190 & $-$13.78700 & GC & * & 2.5 & 2.9 & 2.5 & 1.1 & 41.3 & 0.218 & 15.1 & 17.7 & $-$0.2 & DZ [3] & 5520 & 0.75 & 4.4 & 10.3 & \\
5522446550952428288 & 131.23160 & $-$44.14232 & GC & 1 & 2.0 & 2.3 & 2.8 & 1.8 & 35.8 & 0.386 & 15.2 & 17.7 & $-$0.8 & DA [5] & 9190 & 0.42 & 0.0 &  & \\
910341618586847232 & 131.35268 & 38.03232 & GC & * & 2.6 & 3.0 & 2.5 & 0.6 & 19.4 & 0.044 & 15.7 & 18.6 & $-$0.7 & DA [3] & 8000 & 0.58 & 0.0 & 8.0 & \\
1037258898615762048 & 131.51030 & 57.05795 & Ca~II & * & 25.2 & 37.1 & 26.1 & 9.0 & 95.4 & 0.090 & 16.9 & 16.4 & 0.0 & DA [3] & 15410 & 0.57 & 0.1 & 8.0 & \\
5220278009486912896 & 131.94610 & $-$73.21382 & GC & 1 & 0.9 & 1.9 & 2.2 & 0.5 & 26.4 & 0.443 & 15.4 & 18.9 & $-$1.0 & DA [3] & 17030 & 0.54 & 0.5 &  & \\
720353602808918016 & 134.37693 & 40.27029 & GC & * & 2.0 & 2.3 & 2.3 & 1.1 & 15.2 & 0.248 & 15.0 & 18.9 & $-$1.0 & DA [3] & 22600 & 0.61 & 0.4 & 9.1 & \\
5622626147732121856 & 134.79668 & $-$36.79189 & GC & 1 & 3.0 & 4.0 & 2.9 & 1.4 & 17.6 & 0.364 & 15.4 & 18.8 & $-$0.8 & DA [26] & 9610 & 0.58 & 0.4 & 2.5 & \\
712888090655562624 & 134.81103 & 32.95338 & GC & 2 & 2.8 & 3.4 & 2.8 & 0.9 & 17.0 & 0.068 & 15.0 & 18.7 & $-$0.7 & DQ [3] & 10370 & 1.05 & 2.4 & 8.9 & \\
711744456827031680 & 135.67776 & 31.34542 & GC & 3 & 2.7 & 4.0 & 2.9 & 0.9 & 15.3 & 0.000 & 15.1 & 18.6 & $-$0.8 & DA [3] & 9840 & 0.74 & 0.1 &  & \\
1123700235048742016 & 136.38352 & 73.24860 & GC & 2 & 2.2 & 2.5 & 2.1 & 0.2 & 9.4 & 1.059 & 16.1 & 19.8 & $-$0.5 & DC [14] & 5130 & 0.57 & 0.7 & 9.7 & \\
5635344954642200576 & 136.66879 & $-$30.20627 & GC & 3 & 3.2 & 3.6 & 3.9 & 0.6 & 19.3 & 0.335 & 15.7 & 18.6 & $-$0.6 & DA [5] & 9580 & 0.59 & 0.0 & 1.8 & \\
692134843040270080 & 136.89278 & 27.65095 & GC & * & 2.2 & 2.7 & 2.1 & 1.6 & 28.1 & 0.953 & 16.2 & 18.1 & $-$0.8 & DA [21] & 14660 & 0.81 & 0.0 &  & \\
591040864898749312 & 139.01121 & 10.18630 & GC & 2 & 2.3 & 2.7 & 2.2 & 1.2 & 15.9 & 0.258 & 15.6 & 18.9 & $-$0.9 & DQZ [27] & 9460 & 0.78 & 0.0 & 8.9 & \\
3843957354387777152 & 140.73361 & 1.05282 & GC & 1 & 1.1 & 1.4 & 4.3 & 1.6 & 34.6 & 0.269 & 16.1 & 17.9 & $-$0.7 & DAZ [28] & 6060 & 0.57 & 0.0 &  & \\
5669427512997660800 & 153.00767 & $-$18.72591 & GC & * & 6.7 & 7.6 & 6.4 & 0.3 & 9.1 & 0.000 & 15.0 & 19.5 & $-$0.5 & DZ [2] & 6520 & 0.75 & 0.7 & 9.8 & \\
3830623164560911872 & 154.02864 & $-$1.32143 & GC & 2 & 1.5 & 1.7 & 2.6 & 1.0 & 8.1 & 0.677 & 15.0 & 19.6 & $-$0.1 & DA [3] & 7760 & 0.29 & 0.4 & 10.2 & \\
5446683430923603840 & 155.40966 & $-$34.80381 & GC & * & 2.1 & 3.2 & 2.0 & 0.0 & 6.3 & 0.559 & 16.3 & 20.0 & $-$1.3 & DA [26] & 14210 & 0.57 & 1.2 & 6.1 & \\
3882547566823086848 & 157.22873 & 11.09741 & GC & 1 & 1.5 & 1.8 & 2.3 & 0.9 & 19.7 & 1.252 & 15.7 & 18.5 & $-$0.5 & DA [3] & 14710 & 0.62 & 0.2 &  & \\
1076941716370493696 & 159.24298 & 71.18298 & GC & 1 & 3.7 & 5.9 & 3.9 & 0.0 & 6.8 & 1.202 & 15.9 & 20.2 & $-$0.5 & DC [29] & 4690 & 0.68 & 0.2 & 2.2 & \\
5394487930327131904 & 160.94445 & $-$39.11038 & GC & * & 2.0 & 2.9 & 1.9 & 0.0 & 10.8 & 0.717 & 15.4 & 19.4 & $-$0.7 & DA [5] & 25710 & 0.63 & 2.1 &  & \\
3555923756458320000 & 162.83398 & $-$18.17569 & GC & * & 2.2 & 3.0 & 2.0 &  & 6.9 & 0.601 & 16.9 & 19.9 & $-$0.7 & DA [5] & 9150 & 0.60 & 0.0 & 4.6 & \\
3763445409285757824 & 164.39636 & $-$7.52310 & GC & 3 & 8.1 & 13.2 & 105.4 & 1.2 & 13.4 & 0.196 & 14.0 & 18.9 & $-$0.6 & DC [30] & 7520 & 0.87 & 0.5 &  & \\
861050512312844672 & 166.92737 & 59.97490 & GC & * & 2.6 & 2.9 & 2.5 & 1.1 & 52.6 & 0.003 & 14.0 & 17.7 & $-$0.7 & DA [3] & 18310 & 0.69 & 0.0 & 10.1 & \\
3810099989754827136 & 168.90400 & 0.55475 & GC & * & 2.5 & 2.9 & 2.4 & 0.4 & 10.4 & 0.000 & 16.7 & 19.1 & $-$0.8 & DA [3] & 5050 & 0.39 & 0.0 & 9.0 & \\
3817262208497857024 & 169.16826 & 6.45074 & GC & * & 2.5 & 2.7 & 2.6 & 0.2 & 8.9 & 0.000 & 16.2 & 19.5 & $-$0.7 & DA [3] & 6260 & 0.54 & 0.0 & 8.6 & \\
3790040465258127616 & 169.56276 & $-$3.23488 & GC & 1 & 4.5 & 7.2 & 13.1 & 0.0 & 6.9 & 0.000 & 15.2 & 19.9 & $-$0.7 & DQ [30] & 9900 & 0.70 & 0.0 &  & \\
3540018805367474816 & 175.48576 & $-$23.34921 & GC & * & 2.2 & 2.4 & 2.1 &  & 7.1 & 0.433 & 16.5 & 19.8 & $-$0.7 & DA [3] & 13410 & 0.60 & 0.5 &  & \\
1074721664954807040 & 175.91038 & 71.68904 & CV &  & 411.4 & 623.4 & 405.7 & 352.3 & 135.4 & 0.019 & 14.4 & 15.2 & $-$0.4 & CV [31] &  &  &  & 11.2 & 3.97, 3.0\\
4019789359821201536 & 176.99441 & 28.53223 & IR & * & 1.9 & 3.1 & 2.6 & 0.9 & 10.1 & 0.245 & 17.5 & 19.4 & 0.1 & DA [3] & 11990 & 0.58 & 0.3 & 8.3 & \\
3571559292842744960 & 178.31358 & $-$15.61014 & IR, GC & [d] & 4.6 & 6.6 & 4.3 & 3.2 & 40.5 & 0.046 & 16.0 & 17.7 & 0.2 & DA [3] & 11770 & 0.62 & 0.1 & 9.4 & \\
1129251428738666880 & 178.67848 & 79.24235 & GC & 2 & 2.8 & 3.0 & 2.8 & 0.2 & 13.1 & 0.000 & 15.8 & 19.6 & $-$0.8 & DA [3] & 10470 & 0.57 & 0.4 & 8.4 & \\
3466326581136347520 & 181.69834 & $-$32.57586 & GC & 2 & 2.2 & 2.3 & 3.1 & 0.5 & 13.5 & 0.054 & 15.8 & 19.1 & $-$0.6 & DA [3] & 19970 & 0.61 & 0.0 &  & \\
3576269871237617024 & 184.79672 & $-$12.89442 & GC & 1 & 3.2 & 5.0 & 8.2 & 1.9 & 25.3 & 0.257 & 16.4 & 18.2 & $-$0.4 & DA [5] & 8310 & 0.58 & 0.0 &  & \\
1543723709905857792 & 188.73446 & 47.62593 & GC & * & 2.7 & 3.0 & 2.7 & 1.4 & 33.4 & 0.000 & 14.6 & 18.2 & $-$0.8 & DA [3] & 15230 & 0.59 & 0.4 & 9.7 & \\
6127333286605955072 & 189.70606 & $-$49.80006 & GC & [e] & 2.2 & 2.6 & 2.3 & 0.8 & 27.6 & 0.764 & 13.8 & 18.1 & $-$0.6 & DA [3] & 11350 & 1.04 & 0.9 & 9.0 & 0.16, 0.1\\
1540371883066128768 & 191.39846 & 42.64019 & GC & 2 & 2.0 & 3.0 & 1.9 &  & 8.2 & 0.058 & 17.0 & 19.9 & $-$0.7 & DA [27] & 9540 & 0.80 & 0.0 &  & \\
3938156295111047680 & 196.42659 & 18.01771 & GC &  & 5.1 & 7.0 & 5.3 & 1.7 & 14.6 & 0.100 & 15.9 & 18.9 & $-$0.0 & CV [32] & 11530 & 0.53 & 10.1 & 8.7 & 0.78, 1.1\\
3743646365965636352 & 198.42570 & 13.68691 & GC & * & 2.3 & 3.6 & 2.3 &  & 5.7 & 0.119 & 16.9 & 20.1 & $-$0.4 & DC [33] & 9710 & 0.91 & 0.0 & 6.6 & \\
3506567328028533120 & 199.08138 & $-$20.12557 & GC & 1 & 1.1 & 1.3 & 4.5 & 0.3 & 9.5 & 0.549 & 16.4 & 19.2 & $-$0.9 & DZ [34] & 5230 & 0.83 & 19.6 &  & \\
3630035787972473600 & 202.55677 & $-$8.57485 & GC & 3 & 108.0 & 163.9 & 108.8 & 14.8 & 45.4 & 0.044 & 12.5 & 17.1 & $-$0.5 & DA [3] & 14560 & 0.59 & 0.3 & 4.2 & \\
1470081444731625728 & 202.79520 & 34.14484 & GC & * & 3.2 & 4.8 & 3.1 & 0.6 & 18.1 & 0.041 & 15.3 & 18.8 & $-$0.8 & DA [3] & 16720 & 0.55 & 0.6 & 8.5 & \\
3714914271705535360 & 205.34082 & 5.01273 & GC & 2 & 1.7 & 1.8 & 2.3 & 1.6 & 17.0 & 0.050 & 15.5 & 18.7 & $-$0.4 & DC [28] & 4020 & 0.44 & 0.0 & 10.0 & \\
1444161214019904768 & 207.63476 & 24.57086 & GC & * & 2.4 & 3.9 & 2.3 & 0.0 & 8.3 & 0.000 & 16.2 & 19.6 & $-$0.9 & DA [23] & 13140 & 0.67 & 0.4 & 7.6 & \\
3644717669817337856 & 210.64444 & $-$4.91747 & GC & 3 & 1.2 & 1.2 & 2.6 & 0.7 & 15.2 & 1.270 & 16.4 & 18.7 & $-$0.9 & DA [5] & 7730 & 0.55 & 0.0 &  & \\
6217739938701063808 & 221.84281 & $-$30.58990 & GC & * & 2.2 & 5.1 & 1.9 &  & 4.0 & 0.365 & 16.9 & 20.4 & $-$0.6 & DA [35] & 7100 & 0.63 & 0.0 & 7.1 & \\
6306842338087328512 & 225.01548 & $-$16.06410 & GC & * & 1.8 & 6.3 & 2.1 &  & 7.8 & 0.103 & 16.9 & 19.6 & $-$0.7 & DA [5] & 7150 & 0.60 & 0.0 & 5.6 & \\
1164185283975493120 & 233.32823 & 7.29944 & GC & * & 1.8 & 2.6 & 2.1 &  & 6.4 & 0.000 & 16.6 & 20.0 & $-$1.3 & DA [23] & 10260 & 0.54 & 0.7 & 5.6 & \\
1641326979142898048 & 235.43704 & 64.89805 & IR, GC & [f] & 1.9 & 2.5 & 3.2 & 1.6 & 65.7 & 0.188 & 15.6 & 17.8 & 0.2 & DAV [36] & 11600 & 0.63 & 10.2 & 10.5 & 0.19, 0.5\\
1598771000065137536 & 237.14875 & 57.14068 & GC & 2 & 2.1 & 2.3 & 2.2 & 0.9 & 16.7 & 0.000 & 16.9 & 19.4 & $-$0.7 & DC [37] & 5910 & 0.75 & 0.9 & 8.2 & \\
1696309737421796352 & 237.84308 & 71.75329 & CV &  & 39.8 & 54.1 & 42.1 & 6.8 & 45.2 & 0.000 & 16.5 & 18.2 & $-$0.5 & CV [38] &  &  &  & 6.2 & 1.63, 0.7\\
4406733207018066944 & 241.47717 & $-$0.46974 & GC & * & 2.3 & 3.4 & 2.2 &  & 8.2 & 0.527 & 16.5 & 19.6 & $-$0.7 & DA [3] & 9350 & 0.59 & 0.1 & 7.7 & \\
1429618420396285952 & 243.31921 & 55.35721 & GC & * & 2.1 & 2.9 & 2.4 & 1.3 & 29.0 & 0.041 & 16.0 & 18.7 & $-$0.3 & DA [3] & 11980 & 0.66 & 0.6 & 8.9 & \\
1623866184737702912 & 245.74852 & 58.67525 & Ca~II & * & 9.8 & 14.8 & 9.2 & 2.9 & 20.9 & 0.152 & 17.0 & 19.2 & $-$0.2 & DB [3] & 18790 & 0.48 & 1.5 &  & \\
1653044367185115264 & 247.16083 & 70.88936 & GC & * & 3.2 & 3.4 & 2.5 &  & 4.9 & 0.000 & 16.9 & 20.9 & $-$0.5 & DA [14] & 4900 & 0.57 & 0.3 & 8.4 & \\
4563152548683139968 & 251.58233 & 19.76768 & GC & 2 & 3.1 & 5.1 & 2.9 & 0.9 & 18.5 & 0.242 & 15.1 & 18.8 & $-$0.8 & DB [3] & 15260 & 0.61 & 0.0 & 9.2 & \\
1705751041906319872 & 251.78541 & 77.21271 & GC & * & 2.4 & 3.9 & 2.4 &  & 9.8 & 0.026 & 16.6 & 20.1 & $-$1.0 & DA [23] & 12740 & 0.57 & 1.1 & 5.5 & \\
4379328051494006784 & 253.15663 & $-$1.23234 & GC & * & 2.0 & 2.6 & 2.1 & 0.2 & 7.8 & 0.468 & 16.6 & 19.7 & $-$0.6 & DA [3] & 5540 & 0.75 & 0.0 & 5.9 & \\
4565048312887877888 & 254.29109 & 21.44685 & GC & 2 & 4.3 & 5.4 & 4.4 & 0.7 & 7.8 & 0.199 & 13.9 & 19.8 & $-$0.7 & DA [3] & 9200 & 0.60 & 0.0 & 10.1 & \\
1638563322306634368 & 258.27492 & 69.52375 & GC & 2 & 67.1 & 68.2 & 59.7 & 0.6 & 62.6 & 0.000 & 13.4 & 17.7 & $-$0.7 & DA [3] & 15730 & 0.61 & 0.0 & 10.4 & \\
1348275102768198912 & 261.87223 & 42.08119 & GC & * & 1.5 & 1.7 & 2.6 & 0.7 & 16.5 & 0.042 & 16.5 & 19.3 & $-$0.8 & DA [5] & 9580 & 0.65 & 0.0 &  & \\
5913195150770419840 & 262.59637 & $-$57.92381 & GC & * & 2.5 & 4.0 & 2.4 &  & 7.5 & 0.911 & 15.4 & 19.6 & $-$1.0 & DBA [39] & 14060 & 0.61 & 0.4 & 6.5 & \\
1706631093589103872 & 263.40091 & 79.82129 & GC & 1 & 1.1 & 1.8 & 2.4 & 1.3 & 28.1 & 0.346 & 16.5 & 18.9 & $-$0.7 & DC [14] & 5760 & 0.84 & 1.0 &  & \\
4557279335526790528 & 265.13949 & 23.34029 & GC & * & 3.1 & 3.5 & 3.2 &  & 6.4 & 0.000 & 16.7 & 20.2 & $-$0.5 & DC [5] & 7170 & 0.75 & 0.0 & 7.7 & \\
5920900901901635968 & 266.90216 & $-$54.60866 & GC & 1 & 1.3 & 2.0 & 4.7 & 1.5 & 28.6 & 1.139 & 15.2 & 18.0 & $-$0.6 & DC [1] & 4350 & 0.47 & 0.8 &  & \\
1417200020676904704 & 269.49775 & 54.68952 & GC & 2 & 2.7 & 3.8 & 2.5 & 1.1 & 10.7 & 0.316 & 16.8 & 19.8 & $-$0.7 & DA [14] & 7670 & 0.56 & 0.0 &  & \\
4470233817461336704 & 271.84740 & 3.95050 & GC & 2 & 2.6 & 2.6 & 2.7 & 0.9 & 33.1 & 0.582 & 14.5 & 17.8 & $-$0.8 & DA [3] & 10220 & 0.64 & 0.1 &  & \\
4497414466452138496 & 274.27702 & 13.47361 & GC & 3 & 5.6 & 5.8 & 5.8 & 0.1 & 9.7 & 0.818 & 15.0 & 19.5 & $-$0.5 & DA [3] & 4870 & 0.48 & 0.3 &  & \\
2158285185808357504 & 275.33333 & 61.01872 & GC & 1 & 1.2 & 1.3 & 4.5 & 1.7 & 6.6 & 0.157 & 14.7 & 20.3 & $-$0.6 & DA [3] & 4820 & 0.46 & 0.0 & 3.5 & \\
2256410856215182464 & 278.82519 & 64.35491 & GC & * & 2.2 & 2.3 & 2.0 & 0.6 & 25.9 & 0.045 & 16.7 & 18.7 & $-$0.7 & DC [1] & 4900 & 0.58 & 1.1 & 9.5 & \\
2156469995191226112 & 279.53127 & 60.36160 & GC & * & 2.9 & 3.2 & 2.8 & 0.6 & 40.7 & 0.024 & 15.6 & 18.4 & $-$0.7 & DA [5] & 8990 & 0.52 & 0.0 &  & \\
2127566548919332608 & 291.61211 & 46.33753 & GC & 2 & 2.3 & 3.0 & 2.2 & 0.8 & 29.1 & 0.826 & 15.6 & 18.6 & $-$0.9 & DA [40] & 8140 & 0.72 & 0.4 & 7.7 & 4.66, 2.4\\
6740493675455190784 & 296.22801 & $-$36.05303 & GC & * & 1.9 & 2.7 & 2.0 & 1.1 & 42.0 & 0.126 & 14.1 & 17.5 & $-$0.7 & DA [26] & 13080 & 0.67 & 0.8 & 9.9 & \\
2264107613704439808 & 296.25519 & 73.27496 & GC & * & 1.3 & 1.3 & 2.2 & 1.5 & 21.7 & 1.387 & 16.5 & 19.0 & $-$0.9 & DA [5] & 8330 & 0.56 & 0.0 &  & \\
2263690864438162944 & 297.69110 & 71.92769 & GC & * & 2.0 & 2.7 & 2.0 & 0.9 & 14.2 & 1.258 & 16.7 & 19.5 & $-$0.8 & DA [3] & 11870 & 0.72 & 0.0 &  & \\
4235280071072332672 & 299.12178 & $-$1.04241 & GC & 1 & 63.3 & 96.7 & 592.3 & 7.8 & 35.8 & 0.519 & 13.3 & 17.6 & $-$0.8 & DAH [1] & 7790 & 0.70 & 12.0 &  & \\
6427468556182061312 & 302.09929 & $-$66.07714 & GC & * & 2.2 & 2.8 & 2.4 & 1.4 & 14.2 & 0.208 & 16.1 & 18.9 & $-$0.2 & DA [17] & 21380 & 0.72 & 1.1 &  & \\
4223980836743666176 & 304.46180 & $-$1.44747 & GC & * & 2.2 & 3.5 & 2.1 &  & 7.2 & 1.211 & 15.7 & 19.7 & $-$0.7 & DA [5] & 16260 & 0.57 & 0.4 & 7.8 & \\
1831553382794173824 & 308.59100 & 25.06382 & GC & 1 & 1.3 & 1.7 & 16.7 & 1.2 & 15.4 & 0.072 & 11.7 & 18.8 & $-$0.7 & DA [37] & 20140 & 0.59 & 0.3 & 10.3 & \\
6471102606408911360 & 309.57045 & $-$53.07371 & GC & * & 2.0 & 2.0 & 2.6 & 1.2 & 28.7 & 0.103 & 14.4 & 18.2 & $-$0.7 & DB [3] & 15720 & 0.40 & 0.1 & 6.7 & \\
6683212727417918848 & 309.67282 & $-$36.82044 & GC & * & 1.9 & 2.4 & 2.3 & 0.9 & 28.4 & 0.215 & 14.8 & 18.0 & $-$0.8 & DA [3] & 9490 & 0.62 & 0.0 & 9.2 & \\
6681773947733560192 & 310.95503 & $-$39.05501 & GC & 1,[g] & 3.1 & 3.7 & 2.9 & 1.5 & 34.1 & 1.187 & 13.7 & 17.6 & $-$0.1 & DA [3] & 11220 & 0.63 & 6.7 & 0.3 & 0.17, 0.3\\
6794613321399870080 & 312.37830 & $-$30.08413 & GC & * & 2.9 & 4.5 & 2.9 & 0.2 & 13.3 & 0.971 & 15.4 & 18.9 & $-$0.8 & DBA [39] & 12030 & 0.70 & 0.0 & 6.5 & \\
2221406155496413952 & 317.46853 & 65.12276 & GC & * & 2.1 & 3.6 & 2.3 & 1.1 & 21.8 & 0.144 & 15.5 & 19.0 & $-$0.8 & DAH [26] & 21400 & 0.88 & 0.0 & 0.4 & \\
2690763205257368832 & 317.94327 & 1.34841 & GC & 2 & 2.2 & 3.0 & 2.6 &  & 8.1 & 0.813 & 15.4 & 19.4 & $-$0.7 & DA [14] & 15730 & 0.58 & 0.3 & 8.2 & \\
6896767366186700416 & 318.03855 & $-$8.82688 & CV &  & 35.1 & 41.7 & 34.5 & 8.4 & 51.2 & 0.000 & 16.4 & 17.3 & $-$0.3 & CV [41] &  &  & 34.3 & 10.3 & 1.54, 7.1\\
6583325635088476544 & 320.52351 & $-$38.64296 & GC & 2 & 182.1 & 272.5 & 180.1 & 16.7 & 24.0 & 0.059 & 15.7 & 18.7 & $-$0.7 & DC [1] & 5100 & 0.47 & 0.0 & 7.9 & \\
2220815923910913920 & 320.63282 & 66.01214 & GC & 2 & 2.0 & 2.4 & 2.0 & 0.9 & 13.7 & 0.434 & 15.8 & 19.4 & $-$0.7 & DB [5] & 11250 & 0.74 & 0.0 &  & \\
1785659488269249408 & 320.77531 & 19.28408 & GC & * & 2.7 & 3.2 & 2.6 &  & 6.8 & 0.511 & 16.7 & 19.9 & $-$0.9 & DA [23] & 9840 & 0.63 & 0.0 & 5.2 & \\
6828110179364684672 & 321.60953 & $-$22.73151 & GC & * & 2.3 & 3.0 & 2.1 & 0.0 & 10.2 & 0.057 & 15.4 & 19.2 & $-$0.9 & DA [3] & 15020 & 0.69 & 0.0 & 8.0 & \\
6830482410058056192 & 325.89003 & $-$19.20734 & GC & * & 2.1 & 3.1 & 2.4 & 0.0 & 7.1 & 0.000 & 16.9 & 19.7 & $-$0.5 & DA [5] & 7460 & 0.53 & 0.0 & 7.0 & \\
2673699059671833984 & 328.42542 & $-$2.24063 & GC & * & 2.1 & 3.4 & 2.1 & 0.3 & 9.9 & 0.081 & 17.0 & 19.4 & $-$0.9 & DA [5] & 6620 & 0.56 & 0.0 & 2.3 & \\
2676566272464334720 & 329.69637 & $-$2.67345 & GC & 2 & 2.2 & 2.9 & 2.1 & 1.3 & 19.1 & 0.000 & 16.5 & 18.6 & $-$0.7 & DC [1] & 4730 & 0.48 & 0.0 & 9.2 & \\
6613289285448236288 & 331.15850 & $-$31.45382 & GC & 1 & 1.8 & 2.4 & 5.4 & 1.6 & 11.5 & 0.068 & 16.2 & 19.1 & $-$0.2 & DA [11] & 4810 & 0.53 & 0.1 & 0.2 & \\
2595728287804350720 & 336.07265 & $-$16.26310 & IR, GC &  & 1.6 & 2.3 & 3.4 & 0.7 & 19.5 & 0.663 & 15.9 & 18.5 & $-$0.4 & DAZ [6] & 9900 & 0.67 & 0.0 & 8.2 & \\
2737918849495239936 & 337.11568 & 17.90141 & GC & * & 2.4 & 3.7 & 2.4 &  & 8.5 & 0.280 & 16.3 & 19.6 & $-$1.0 & DA [23] & 14650 & 0.59 & 0.7 & 7.0 & \\
2623101763649861632 & 337.14432 & $-$6.20543 & GC & * & 2.2 & 3.3 & 2.2 & 1.4 & 12.9 & 0.000 & 16.3 & 19.0 & $-$0.9 & DA [5] & 6940 & 0.54 & 0.6 & 8.1 & \\
2842650153836732928 & 346.57374 & 24.53547 & GC & [h] & 2.2 & 3.4 & 2.1 & 1.2 & 20.3 & 0.000 & 15.3 & 18.5 & $-$0.7 & DA [3] & 11430 & 0.60 & 10.0 & 8.8 & 0.22, 1.7\\
2407167579853954688 & 348.87669 & $-$14.66824 & GC & 2 & 2.9 & 4.2 & 2.7 & 0.5 & 20.5 & 0.079 & 15.3 & 18.4 & $-$0.8 & DA [35] & 8850 & 0.84 & 0.0 & 7.9 & \\
2393546245693497728 & 351.32675 & $-$17.86607 & GC & * & 2.4 & 3.1 & 2.3 &  & 7.6 & 0.000 & 15.6 & 19.6 & $-$1.1 & DA [3] & 22030 & 0.58 & 0.6 & 7.0 & \\
6485572518732377856 & 355.75224 & $-$64.79386 & GC & 1 & 2.0 & 2.9 & 2.0 & 1.8 & 15.4 & 0.019 & 16.5 & 19.0 & $-$0.7 & DC [1] & 5790 & 0.58 & 0.5 & 0.8 & \\
2763719512612656000 & 356.55201 & 11.98061 & GC & * & 2.0 & 3.2 & 2.0 & 0.2 & 10.9 & 0.374 & 16.7 & 19.3 & $-$0.8 & DC [14] & 6210 & 0.78 & 0.0 & 7.7 & \\
6525205175564946816 & 356.55229 & $-$47.24147 & GC & * & 1.4 & 2.1 & 2.0 & 0.7 & 16.3 & 0.458 & 15.5 & 19.0 & $-$0.8 & DA [5] & 12500 & 0.64 & 0.0 & 7.7 & \\
\enddata
\tablecomments{Description of columns: Legend for the Sample column: Ca~II are objects with previously identified Ca~II gas emission (Table~\ref{tab:known_gas_disks}), IR refers to non-Ca~II debris disk systems (Table~\ref{tab:known_dust_disks}), CV are known cataclysmic variables, and GC are objects from our Gaia-catWISE crossmatch. Objects flagged with asterisks are believed to be newly discovered near-infrared variables; they are not known CVs nor are they published in \citet{Swan2019,Swan2020,Swan2021}. We also flag known ZZ Cetis: [a] \citep{Fontaine2003}, [b] \citep{Romero2022}, [c] \citep{Vauclair1997}, [d] \citep{Gianninas2006}, [e] \citep{Kanaan1992}, [f] \citep{Vauclair2000}, [g] \citep{Romero2019}, [h] \citep{Vauclair1987}. We define $G_{\rm RP}$ Excess as the ratio of the summed flux in the Gaia $RP$ bandpass in a 30\arcsec\ cone around the target white dwarf to target white dwarf's $RP$ flux. Flag legend: 1=contaminated as seen in WISE View, 2=false-positive variability due to a seasonal alternating pattern, 3=contaminated both as seen in WISE View and from alternating. S/N$_{W1}$ is the median signal-to-noise ratio of our $W1$ light curves. All magnitudes are reported here on the AB scale. catWISE $W1-W2$ colors are also on the AB scale. References for the tabulated spectral types are as follows: [1] \citet{OBrien2024}, [2] \citet{Coutu2019}, [3] \citet{McCookSion1999}, [4] \citet{Guo2015}, [5] \citet{Vincent2024}, [6] \citet{Koester2005}, [7] \citet{1984IBVS.2483....1M}, [8] \citet{Kilkenny2016}, [9] \citet{Subasavage2022}, [10] \citet{Gianninas2011}, [11] \citet{OBrien2023}, [12] \citet{Limoges2013}, [13] \citet{Dennihy2020b}, [14] \citet{Limoges2015}, [15] \citet{Koester1982}, [16] \citet{Bergeron2021}, [17] \citet{ODonoghue2013}, [18] \citet{1991PVSS...17...38B}, [19] \citet{Melis2020}, [20] \citet{GentileFusillo2021gas}, [21] \citet{Zhang2013}, [22] \citet{1978JBAA...88..292H}, [23] \citet{Kilic2020}, [24] \citet{1974MNRAS.168..235W}, [25] \citet{1990AJ....100..226U}, [26] \citet{Raddi2017}, [27] \citet{Kleinman2013}, [28] \citet{Sayres2012}, [29] \citet{Sion2014}, [30] \citet{Bergeron2001}, [31] \citet{Simon2000}, [32] \citet{1971ApJ...170L..39B}, [33] \citet{Caron2023}, [34] \citet{Bergeron1997}, [35] \citet{KawkaVennes2012}, [36] \citet{Kilic2012}, [37] \citet{McCleery2020}, [38] \citet{2000IBVS.4932....1K}, [39] \citet{Bergeron2011}, [40] \citet{Tremblay2020}, [41] \citet{2009AJ....137.4061H}. \teff and $M_{\rm WD}$ are derived from Gaia photometry assuming a hydrogen-dominated atmosphere \citep{GentileFusillo2021gaia}. EFF={\sc excess\_flux\_error} \citep{GentileFusillo2021gaia}. FoM is the Figure of Merit metric computed from the Gaia-AllWISE neighbor crossmatch analysis \citep{Marrese2017,Marrese2019,Marrese2022}. In the TESS column we report the significant periodicities in TESS that we attributed intrinsic to the white dwarf using \texttt{TESS\_localize} in hours and the associated amplitude in \%.}
\end{deluxetable}
\end{longrotatetable}


\movetabledown=1.0cm
\begin{longrotatetable}
\begin{deluxetable}{crcrrcccccccccccccc}
\tabletypesize{\tiny}
\tablewidth{0pt} 
\tablecaption{Variability Parameters for the 21 Published White Dwarf Debris Disks That Show Calcium~II Gas in Emission \label{tab:known_gas_disks}}
\tablehead{
    \colhead{Figure~\ref{fig:all_gas_lcs}}& 
    \colhead{Gaia DR3 Source ID}& 
    \colhead{Alt. Name}& 
    \colhead{R.A.}& 
    \colhead{Decl.}& 
    \colhead{Flag}&
    \colhead{\chisqone}&
    \colhead{\chisqalt}&
    \colhead{$\chi^2_{\mathrm{flat}}$}&
    \colhead{\chisqtwo}&
    \colhead{S/N}&
    \colhead{$G_{\rm RP}$ Excess}&
    \colhead{$G_{\rm RP}$}&
    \colhead{$W1$}&
    \colhead{$W1-W2$}&
    \colhead{\teff}&
    \colhead{$M_{\rm WD}$}&
    \colhead{EFF}&
    \colhead{FoM}\\
    & &  & \colhead{(J2000)} & \colhead{(J2000)} & & & &  & & \colhead{$W1$} &  &  (mag) & (mag) & (mag) & (K) & (\msun) & &
}
\startdata 
a & 2860923998433585664 & SDSS\,J0006+2858 & 1.64464 & 28.97960 & 1 & 0.3 & 0.3 & 0.3 & 1.0 & 5.3 & 237.861 & 16.5 & 18.1 & $-$0.0 & 21980 & 0.55 & 1.7 & \\
b & 291057843317534464 & WD\,0145+234 & 26.97841 & 23.66211 &  & 422.6 & 690.7 & 482.4 & 254.9 & 54.3 & 0.021 & 14.1 & 16.9 & $-$0.4 & 12980 & 0.66 & 0.8 & 5.5\\
c & 2489275328645218560 & SDSS\,J0234$-$0406 & 38.56461 & $-$4.10258 & 1 & 1.2 & 1.8 & 1.3 & 0.6 & 12.5 & 3.039 & 16.5 & 19.1 & $-$0.5 & 13100 & 0.63 & 0.7 & 8.3\\
d & 43629828277884160 & SDSS\,J0347+1624 & 56.90288 & 16.40270 & * & 12.3 & 17.7 & 12.1 & 0.5 & 11.2 & 0.940 & 16.7 & 19.2 & $-$0.4 & 17710 & 0.52 & 0.0 & 6.9\\
e & 3415788525598117248 & LAMOST\,J0510+2315 & 77.50894 & 23.26151 & * & 41.9 & 65.9 & 49.5 & 16.0 & 41.8 & 0.796 & 15.2 & 17.4 & $-$0.1 & 21250 & 0.73 & 0.0 & 9.0\\
f & 4823042355496728832 & WD\,J0529–3401 & 82.30968 & $-$34.01892 & * & 2.0 & 2.2 & 6.4 & 0.8 & 28.5 & 0.230 & 17.6 & 18.4 & 0.1 & 17190 & 0.46 & 0.0 & \\
g & 5279484614703730944 & Gaia\,J0611−6931 & 92.88208 & $-$69.51726 & 1 & 45.6 & 129.8 & 93.1 & 22.6 & 120.3 & 4.803 & 16.9 & 17.1 & 0.1 & 16140 & 0.54 & 0.0 & \\
h & 3105360521513256832 & Gaia\,J0644−0352 & 101.02179 & $-$3.86845 & 1 & 2.2 & 2.9 & 2.2 & 2.1 & 7.6 & 25.020 & 16.3 & 19.2 & 0.0 & 17380 & 0.56 & 0.0 & 7.2\\
i & 671450448046315520 & SDSS\,J0738+1835 & 114.67736 & 18.58603 & 1 & 1.1 & 1.8 & 1.0 & 0.3 & 5.3 & 6.238 & 17.8 & 20.0 & $-$0.3 & 15740 & 0.72 & 0.1 & \\
j & 689352219629097856 & WD\,0842+231 & 131.41326 & 22.95785 &  & 1.7 & 2.4 & 1.7 & 1.7 & 22.8 & 0.000 & 16.0 & 18.3 & $-$0.2 & 19180 & 0.58 & 0.0 & 3.2\\
k & 1037258898615762048 & WD\,0842+572 & 131.51030 & 57.05795 & * & 25.2 & 37.1 & 26.1 & 9.0 & 95.4 & 0.090 & 16.9 & 16.4 & 0.0 & 15410 & 0.57 & 0.1 & 8.0\\
l & 3829599892897720832 & SDSS\,J0959$-$0200 & 149.76957 & $-$2.01323 &  & 1.0 & 1.5 & 1.0 & 0.0 & 4.7 & 0.000 & 18.2 & 20.2 & $-$0.0 & 11330 & 0.48 & 0.0 & 6.2\\
m & 3869060540584643328 & WD\,1041+092 & 160.92305 & 8.93287 &  & 0.8 & 1.4 & 0.8 & 0.0 & 5.1 & 0.000 & 17.4 & 20.2 & $-$1.0 & 16060 & 0.59 & 0.0 & \\
n & 3904415787947492096 & SDSS\,J1228+1040 & 187.24973 & 10.67585 &  & 1.5 & 2.2 & 2.0 & 1.3 & 15.8 & 0.122 & 16.6 & 18.8 & $-$0.0 & 19310 & 0.62 & 0.4 & 5.6\\
o & 6287259310145684608 & HE\,1349$-$2305 & 208.18385 & $-$23.33483 &  & 0.9 & 1.4 & 1.7 & 1.0 & 13.8 & 0.086 & 16.6 & 18.9 & $-$0.4 & 16540 & 0.53 & 0.0 & \\
p & 4465269178854732160 & SDSS\,J1617+1620 & 244.32100 & 16.33960 &  & 1.4 & 2.2 & 1.7 & 1.0 & 13.2 & 0.718 & 17.2 & 19.2 & $-$0.4 & 13220 & 0.63 & 0.0 & 8.2\\
q & 1623866184737702912 & WD\,1622+587 & 245.74852 & 58.67525 & * & 9.8 & 14.8 & 9.2 & 2.9 & 20.9 & 0.152 & 17.0 & 19.2 & $-$0.2 & 18790 & 0.48 & 1.5 & \\
r & 6646693887514073984 & WD\,J1930–5028 & 292.65688 & $-$50.47136 &  & 1.0 & 1.3 & 1.0 & 0.7 & 12.3 & 0.482 & 17.1 & 19.2 & 0.0 & 13570 & 0.63 & 0.9 & 6.8\\
s & 1837948790953103232 & Gaia\,J2100+2122 & 315.14437 & 21.38247 & 2 & 6.7 & 6.2 & 6.7 & 1.9 & 53.8 & 1.772 & 15.4 & 17.3 & 0.0 & 25030 & 0.65 & 0.0 & 10.2\\
t & 1797494081674032512 & SDSS\,J2133+2428 & 323.46132 & 24.46831 & 1 & 12.9 & 23.5 & 15.1 & 4.2 & 17.2 & 1.536 & 17.5 & 18.9 & $-$0.3 & 26320 & 0.58 & 0.0 & 9.4\\
u & 2600209141285265280 & ATLAS\,J2212–1352 & 333.01198 & $-$13.87777 & 1 & 1.7 & 1.9 & 1.7 & 1.7 & 10.5 & 2.554 & 17.5 & 19.1 & 0.1 & 19360 & 0.53 & 0.0 & 8.7\\
\enddata
\tablecomments{Description of columns: Flag legend: 1=likely crowded ($G_{\rm RP}$ Excess$>$1.415), 2=likely alternating (\chisqalt$<$\chisqone, see Section~\ref{sec:methods}). Asterisks denote new near-infrared variables. We define $G_{\rm RP}$ Excess as the ratio of the summed flux in the Gaia $RP$ bandpass in a 30\arcsec\ cone around the target white dwarf to target white dwarf's $RP$ flux (Section~\ref{sec:methods}). S/N$_{W1}$ is the median signal-to-noise ratio of our $W1$ light curves. All magnitudes are reported here on the AB scale. catWISE $W1-W2$ colors are also on the AB scale. \teff and $M_{\rm WD}$ are derived from Gaia photometry assuming a hydrogen-dominated atmosphere \citep{GentileFusillo2021gaia}. EFF={\sc excess\_flux\_error} \citep{GentileFusillo2021gaia}. FoM is the Figure of Merit metric computed from the Gaia-AllWISE neighbor crossmatch analysis \citep{Marrese2017,Marrese2019,Marrese2022}. We also guide the reader to Table~3 of \citet{GentileFusillo2021gas} for a summary of the gaseous species in emission towards these 21 white dwarfs.}
\end{deluxetable}
\end{longrotatetable}

\movetabledown=1.5cm
\begin{longrotatetable}
\startlongtable
\begin{deluxetable}{crcrrcccccccccccccc}
\tabletypesize{\tiny}
\tablewidth{0pt} 
\tablecaption{Variability Parameters for the White Dwarf Debris Disks That Do Not Show Calcium~II Gas in Emission \label{tab:known_dust_disks}}
\tablehead{
    \colhead{Figure~\ref{fig:all_dust_lcs_1}, \ref{fig:all_dust_lcs_2}}& 
    \colhead{Gaia DR3 Source ID}& 
    \colhead{SIMBAD ID}& 
    \colhead{R.A.}& 
    \colhead{Decl.}& 
    \colhead{Flag}&
    \colhead{\chisqone}&
    \colhead{\chisqalt}&
    \colhead{$\chi^2_{\mathrm{flat}}$}&
    \colhead{\chisqtwo}&
    \colhead{S/N}&
    \colhead{$G_{\rm RP}$ Excess}&
    \colhead{$G_{\rm RP}$}&
    \colhead{$W1$}&
    \colhead{$W1-W2$}&
    \colhead{\teff}&
    \colhead{$M_{\rm WD}$}&
    \colhead{EFF}&
    \colhead{FoM}\\
     &  &  & \colhead{(J2000)} & \colhead{(J2000)} & & & &  & & \colhead{$W1$} &  &  (mag) & (mag) & (mag) & (K) & (\msun) & &
}
\startdata
1 & 2859951106737135488 & PG\,0010+281 & 3.33780 & 28.33884 &  & 1.0 & 1.8 & 1.0 & 1.5 & 15.6 & 0.200 & 15.9 & 18.9 & $-$0.5 & 25150 & 0.52 & 0.0 & 5.6\\
2 & 5026963661794939520 & Ton\,S\,193 & 17.15014 & $-$32.62872 &  & 2.2 & 2.6 & 2.5 & 0.9 & 28.9 & 1.110 & 15.5 & 18.1 & $-$0.7 & 16050 & 0.63 & 0.0 & \\
3 & 4913589203924379776 & JL\,234 & 18.08809 & $-$56.24098 &  & 2.6 & 4.1 & 2.7 & 1.1 & 20.4 & 0.000 & 15.9 & 18.8 & $-$0.3 & 18790 & 0.59 & 0.0 & 8.7\\
4 & 95297185335797120 & Wolf\,88 & 27.23665 & 19.04097 & 2 & 5.2 & 4.3 & 4.9 & 1.8 & 27.8 & 0.022 & 15.5 & 18.0 & 0.0 & 11690 & 0.77 & 0.0 & 10.2\\
5 & 2518936059258271232 & GALEX\,J015550.2+043131 & 28.95924 & 4.52516 &  & 1.5 & 2.2 & 1.4 &  & 5.7 & 0.000 & 16.9 & 20.0 & $-$0.8 & 14110 & 0.55 & 0.0 & 6.0\\
6 & 547501815051141248 & EGGR\,474 & 42.96365 & 73.69309 & 1 & 0.6 & 0.7 & 0.6 &  & 6.3 & 4.494 & 17.0 & 20.2 & $-$0.6 & 8600 & 0.81 & 0.0 & 3.1\\
7 & 5187830356195791488 & GD\,40 & 45.72127 & $-$1.14272 &  & 1.2 & 1.7 & 1.2 & 1.3 & 22.2 & 0.000 & 15.5 & 18.4 & $-$0.2 & 14090 & 0.60 & 0.0 & 9.1\\
8 & 13611477211053824 & HS\,0307+0746 & 47.53816 & 7.95891 &  & 1.1 & 1.6 & 1.0 & 0.0 & 8.3 & 0.000 & 16.1 & 19.5 & $-$0.8 & 10100 & 0.59 & 0.6 & 8.4\\
9 & 3251748915515143296 & GD\,56 & 62.75904 & $-$3.97294 &  & 4.9 & 6.8 & 40.0 & 3.3 & 56.2 & 0.268 & 15.7 & 17.1 & 0.2 & 14900 & 0.61 & 0.1 & 10.7\\
10 & 4653404070862114176 & MCT\,0420$-$7310 & 64.90760 & $-$73.06236 & * & 2.4 & 3.9 & 3.5 & 1.3 & 59.6 & 0.150 & 15.7 & 17.7 & $-$0.1 & 17790 & 0.58 & 0.0 & 9.6\\
11 & 271992414775824640 & KPD\,0420+5203 & 66.06547 & 52.16962 & * & 3.1 & 4.9 & 17.4 & 2.2 & 40.1 & 0.876 & 15.2 & 17.6 & $-$0.3 & 22170 & 0.66 & 0.9 & 8.9\\
12 & 203931163247581184 & GD\,61 & 69.66406 & 41.15899 & 1 & 1.1 & 1.3 & 1.1 & 0.9 & 21.7 & 4.625 & 14.9 & 18.1 & $-$0.5 & 16940 & 0.63 & 0.9 & \\
13 & 706076062068973952 & GALEX\,J084244.5+294815 & 130.68559 & 29.80441 & 0 &  &  & 0.0 &  & 4.0 & 13.083 & 18.5 & 20.4 & $-$0.2 & 13120 & 0.47 & 0.0 & \\
14 & 1029081452683108480 & PG\,0843+517 & 131.75953 & 51.48149 &  & 0.9 & 1.5 & 0.9 & 0.6 & 17.9 & 0.000 & 16.3 & 18.9 & $-$0.3 & 21210 & 0.51 & 0.2 & 7.4\\
15 & 692795546450243456 & GALEX\,J085028.3+271956 & 132.61811 & 27.33230 &  & 0.4 & 0.5 & 0.5 & 0.2 & 6.7 & 0.000 & 17.5 & 19.9 & $-$0.6 & 18900 & 0.58 & 0.0 & 0.0\\
16 & 719422660057614080 & GALEX\,J090525.5+394339 & 136.35620 & 39.72756 & 0 &  &  &  &  &  & 0.269 & 18.0 &  & 0.1 & 12970 & 0.34 & 0.0 & \\
17 & 618139424181751552 & GALEX\,J092801.1+133219 & 142.00460 & 13.53869 &  & 0.7 & 0.9 & 0.7 & 0.0 & 4.5 & 0.000 & 18.1 & 20.3 & 0.1 & 22500 & 0.63 & 0.0 & 5.9\\
18 & 3888723386196630784 & PG\,1015+161 & 154.51595 & 15.86627 &  & 0.8 & 1.0 & 0.8 & 1.0 & 18.6 & 0.032 & 15.8 & 18.6 & $-$0.3 & 18370 & 0.60 & 0.0 & 7.4\\
19 & 804169064957527552 & PG\,1018+411 & 155.48128 & 40.83747 & 1 & 1.4 & 2.2 & 1.4 & 0.2 & 8.7 & 4.317 & 16.6 & 19.6 & $-$0.4 & 20750 & 0.63 & 0.0 & 8.0\\
20 & 3862040300575181184 & PG\,1031+063 & 158.52221 & 6.04658 & 2 & 0.9 & 0.9 & 0.9 & 0.0 & 7.0 & 0.000 & 16.5 & 19.8 & $-$1.0 & 19600 & 0.54 & 0.2 & 3.0\\
21 & 3808488483665268736 & GALEX\,J105807.8+020648 & 164.53280 & 2.11356 & 0 & 0.3 &  & 0.2 &  & 4.6 & 0.000 & 17.0 & 20.3 & $-$0.1 & 17350 & 0.52 & 0.0 & 5.4\\
22 & 3810933247769901696 & GD\,133 & 169.80166 & 2.34251 & 2 & 2.3 & 2.2 & 2.8 & 0.6 & 48.8 & 0.008 & 14.7 & 17.3 & $-$0.2 & 12150 & 0.63 & 1.1 & \\
23 & 4019789359821201536 & US\,2966 & 176.99441 & 28.53223 & * & 1.9 & 3.1 & 2.6 & 0.9 & 10.1 & 0.245 & 17.5 & 19.4 & 0.1 & 11990 & 0.58 & 0.3 & 8.3\\
24 & 3796414192429498880 & LBQS\,1145+0145 & 177.14012 & 1.48317 &  & 0.3 & 0.5 & 0.4 &  & 5.2 & 0.000 & 17.3 & 20.1 & $-$0.1 & 16030 & 0.66 & 5.4 & 6.8\\
25 & 3571559292842744960 & EC\,11507$-$1519 & 178.31358 & $-$15.61014 &  & 4.6 & 6.6 & 4.3 & 3.2 & 40.5 & 0.046 & 16.0 & 17.7 & 0.2 & 11770 & 0.62 & 0.1 & 9.4\\
26 & 3908649148233327232 & GALEX\,J122150.8+124512 & 185.46171 & 12.75370 & 1 & 1.0 & 1.7 & 1.0 & 0.4 & 8.7 & 1.687 & 18.3 & 19.5 & $-$0.1 & 12390 & 0.69 & 1.3 & \\
27 & 3583181371265430656 & PG\,1225$-$079 & 186.94733 & $-$8.24388 &  & 1.8 & 2.1 & 1.7 & 2.0 & 28.4 & 0.030 & 14.7 & 18.0 & $-$0.7 & 11720 & 0.77 & 0.0 & 8.4\\
28 & 1571584539980588544 & SBSS\,1232+563 & 188.63615 & 56.11195 &  & 1.9 & 2.5 & 1.8 & 0.5 & 8.7 & 0.258 & 18.1 & 19.8 & $-$0.0 & 12570 & 0.70 & 13.2 & 7.2\\
29 & 4011097617323921920 & GALEX\,J123843.9+291004 & 189.68301 & 29.16762 &  & 0.5 & 1.1 & 0.5 &  & 3.8 & 0.000 & 18.9 & 20.8 & $-$0.1 & 14820 & 0.86 & 0.0 & 2.6\\
30 & 1188075433968561792 & PG\,1454+173 & 224.17141 & 17.07087 &  & 1.4 & 2.0 & 1.4 &  & 5.6 & 0.647 & 17.3 & 20.2 & $-$0.2 & 34150 & 0.76 & 0.0 & \\
31 & 1281989124439286912 & EGGR\,298 & 224.52704 & 29.62487 & 2 & 2.4 & 2.2 & 3.9 & 0.5 & 15.0 & 0.044 & 15.2 & 19.1 & $-$0.2 & 7340 & 0.58 & 0.0 & \\
32 & 6332172027974304512 & PG\,1457$-$086 & 224.97079 & $-$8.82490 & 2 & 1.3 & 1.2 & 1.4 & 0.4 & 10.9 & 0.014 & 15.9 & 19.2 & $-$1.0 & 18330 & 0.51 & 2.1 & 6.6\\
33 & 1288812212565231232 & GALEX\,J150701.9+324545 & 226.75835 & 32.76250 & 3 & 1.4 & 1.3 & 1.4 & 0.7 & 6.4 & 1.864 & 18.5 & 20.2 & $-$0.2 & 6960 & 0.53 & 0.1 & 0.1\\
34 & 1595298501827000960 & SBSS\,1536+520 & 234.35719 & 51.85745 & 2 & 2.5 & 2.1 & 3.9 & 1.3 & 30.6 & 0.000 & 17.4 & 18.6 & 0.1 & 18180 & 0.51 & 0.0 & 3.6\\
35 & 1641326979142898048 & \,KX\,Dra & 235.43704 & 64.89805 &  & 1.9 & 2.5 & 3.2 & 1.6 & 65.7 & 0.188 & 15.6 & 17.8 & 0.2 & 11600 & 0.63 & 10.2 & 10.5\\
36 & 1196531988354226560 & KUV\,15519+1730 & 238.53757 & 17.35672 &  & 0.1 & 0.1 & 0.1 &  & 3.7 & 0.652 & 17.4 & 20.7 & $-$0.8 & 15630 & 0.58 & 0.0 & \\
37 & 4454599238843776128 & GALEX\,J155720.8+091625 & 239.33658 & 9.27355 & 0 & 0.0 &  & 0.0 &  & 3.7 & 0.000 & 18.8 & 20.7 & 0.2 & 25260 & 0.51 & 1.6 & 3.3\\
38 & 1336442472164656000 & GD\,362 & 262.89299 & 37.08909 & 3 & 3.2 & 1.3 & 3.3 & 1.1 & 31.4 & 66.096 & 16.0 & 18.3 & 0.1 & 10480 & 0.70 & 0.0 & 10.2\\
39 & 4287654959563143168 & GALEX\,J193156.8+011745 & 292.98722 & 1.29559 & 3 & 0.5 & 0.4 & 0.5 & 0.6 & 28.2 & 5.375 & 14.5 & 16.8 & $-$0.2 & 21540 & 0.62 & 0.7 & \\
40 & 6462911897617050240 & LAWD\,84 & 319.90222 & $-$55.83735 & 2 & 24.8 & 1.6 & 23.3 & 1.6 & 50.7 & 0.062 & 14.2 & 17.3 & $-$0.1 & 9520 & 0.58 & 0.0 & 10.1\\
41 & 1742342784582615936 & HS\,2132+0941 & 323.71172 & 9.92229 &  & 0.8 & 1.2 & 0.8 & 0.6 & 12.2 & 0.054 & 16.0 & 19.2 & $-$0.6 & 13210 & 0.59 & 0.0 & 8.0\\
42 & 6592315723192176896 & L\,570$-$26 & 325.48985 & $-$33.00828 & 2 & 4.5 & 1.6 & 4.9 & 1.5 & 55.9 & 0.007 & 14.2 & 17.1 & $-$0.6 & 7550 & 0.71 & 0.3 & 10.2\\
43 & 2683271785860305280 & Gaia\,DR2\,2683271785860305280 & 331.04273 & 2.55887 &  & 1.2 & 1.7 & 1.1 & 0.1 & 6.5 & 0.000 & 17.3 & 19.9 & $-$0.1 & 15480 & 0.67 & 0.8 & 2.6\\
44 & 2727904257071365760 & GALEX\,J220934.8+122337 & 332.39521 & 12.39349 &  & 1.5 & 2.4 & 1.4 & 0.7 & 10.8 & 1.150 & 17.4 & 19.3 & $-$0.1 & 14450 & 0.54 & 0.0 & \\
45 & 2595728287804350720 & PHL\,5103 & 336.07265 & $-$16.26310 &  & 1.6 & 2.3 & 3.4 & 0.7 & 19.5 & 0.663 & 15.9 & 18.5 & $-$0.4 & 9900 & 0.67 & 0.0 & 8.2\\
46 & 2660358032257156736 & G\,29$-$38 & 352.19849 & 5.24840 & 2 & 23.7 & 5.9 & 23.9 & 7.0 & 46.8 & 0.049 & 13.1 & 16.4 & 0.1 & 11520 & 0.63 & 8.2 & 11.5\\
47 & 2762605088857836288 & PG\,2328+108 & 352.67362 & 11.03513 &  & 1.3 & 2.1 & 1.5 & 0.9 & 12.3 & 0.673 & 15.7 & 19.1 & $-$0.8 & 17870 & 0.46 & 0.0 & 8.3\\
\enddata
\tablecomments{Description of columns: Flag legend: 1=likely crowded ($G_{\rm RP}$ Excess$>$1.415), 2=likely alternating (\chisqalt$<$\chisqone, see Section~\ref{sec:methods}), 3=both likely crowded and likely alternating, and 0=insufficient $W1$ data to analyze. Asterisks denote new near-infrared variables. We define $G_{\rm RP}$ Excess as the ratio of the summed flux in the Gaia $RP$ bandpass in a 30\arcsec\ cone around the target white dwarf to target white dwarf's $RP$ flux (Section~\ref{sec:methods}). S/N$_{W1}$ is the median signal-to-noise ratio of our $W1$ light curves. All magnitudes are reported here on the AB scale. catWISE $W1-W2$ colors are also on the AB scale. \teff and $M_{\rm WD}$ are derived from Gaia photometry assuming a hydrogen-dominated atmosphere \citep{GentileFusillo2021gaia}. EFF={\sc excess\_flux\_error} \citep{GentileFusillo2021gaia}. FoM is the Figure of Merit metric computed from the Gaia-AllWISE neighbor crossmatch analysis \citep{Marrese2017,Marrese2019,Marrese2022}.}
\end{deluxetable}
\end{longrotatetable}

\onecolumngrid
\section{LDT+DeVeny and SOAR+Goodman Follow-up Spectroscopy}\label{sec:soar_appendix}

Table~\ref{tab:spectra} shows an observing log of all of the near-infrared spectra we publish in Figure~\ref{fig:spectra}. A description of our observing setups is found in Section~\ref{sec:followup}.

\begin{deluxetable}{rrccccccc}[!h]
\tabletypesize{\scriptsize}
\tablecaption{Log of Observations at LDT and SOAR \label{tab:spectra}}
\tablehead{
        \colhead{Gaia DR3 ID}&
        \colhead{WDJ Name}&
        \colhead{Instrument+Facility}&
        \colhead{UTC Date}&
        \colhead{$G_{\rm RP}$}&
        \colhead{Exp. Time}&
        \colhead{S/N}&
        \colhead{Spectral Type} &
        \colhead{Ca~II Detected?}\\
        \colhead{}&
        \colhead{}&
        \colhead{}&
        \colhead{}&
        \colhead{(mag)}&
        \colhead{(sec.)}&
        \colhead{}&
        \colhead{}&
        \colhead{}
        }
\startdata
4907742035447425024 & WD\,J0058$-$5638 & SOAR+Goodman & 2023-08-13 & 16.9 & 7200 & 38.5 & DZ [1] & Absorption\\
5078074837769743744 & WD\,J0302$-$2301 & SOAR+Goodman & 2023-08-13 & 15.9 & 5400 & 61.5 & DA [2] & No\\
276192136878054656 & WD\,J0406+5431 & LDT+DeVeny & 2024-02-05 & 15.5 & 7200 & 32.9 & DA [2] & No\\
4653404070862114176 & WD\,J0419$-$7303 & SOAR+Goodman & 2023-08-13 & 15.7 & 1800 & 38.9 & DA [2] & No\\
271992414775824640 & WD\,J0424+5210 & LDT+DeVeny & 2023-10-06 & 15.2 & 7200 & 47.6& DA [2] & No\\
720353602808918016 & WD\,J0857+4016 & LDT+DeVeny & 2024-02-05 & 15.0 & 7200 & 35.4 & DA [2] & No\\
6306842338087328512 & WD\,J1500$-$1603 & SOAR+Goodman & 2023-08-18 & 16.9 & 6600 & 21.7& DA [1] & No\\
\enddata
\tablecomments{We calculate the tabulated signal-to-noise ratios (S/N) by taking the average of the ratio of the continuum normalized flux to its standard deviation at the wavelengths where we do not see features from the calcium triplet. References for the tabulated spectral types are as follows: [1] \citet{Vincent2024}, [2] \citet{McCookSion1999}.}
\end{deluxetable}

\section{Light Curves}\label{sec:light_curves}

\begin{figure}[t!]
\centering
\includegraphics[width=\textwidth]{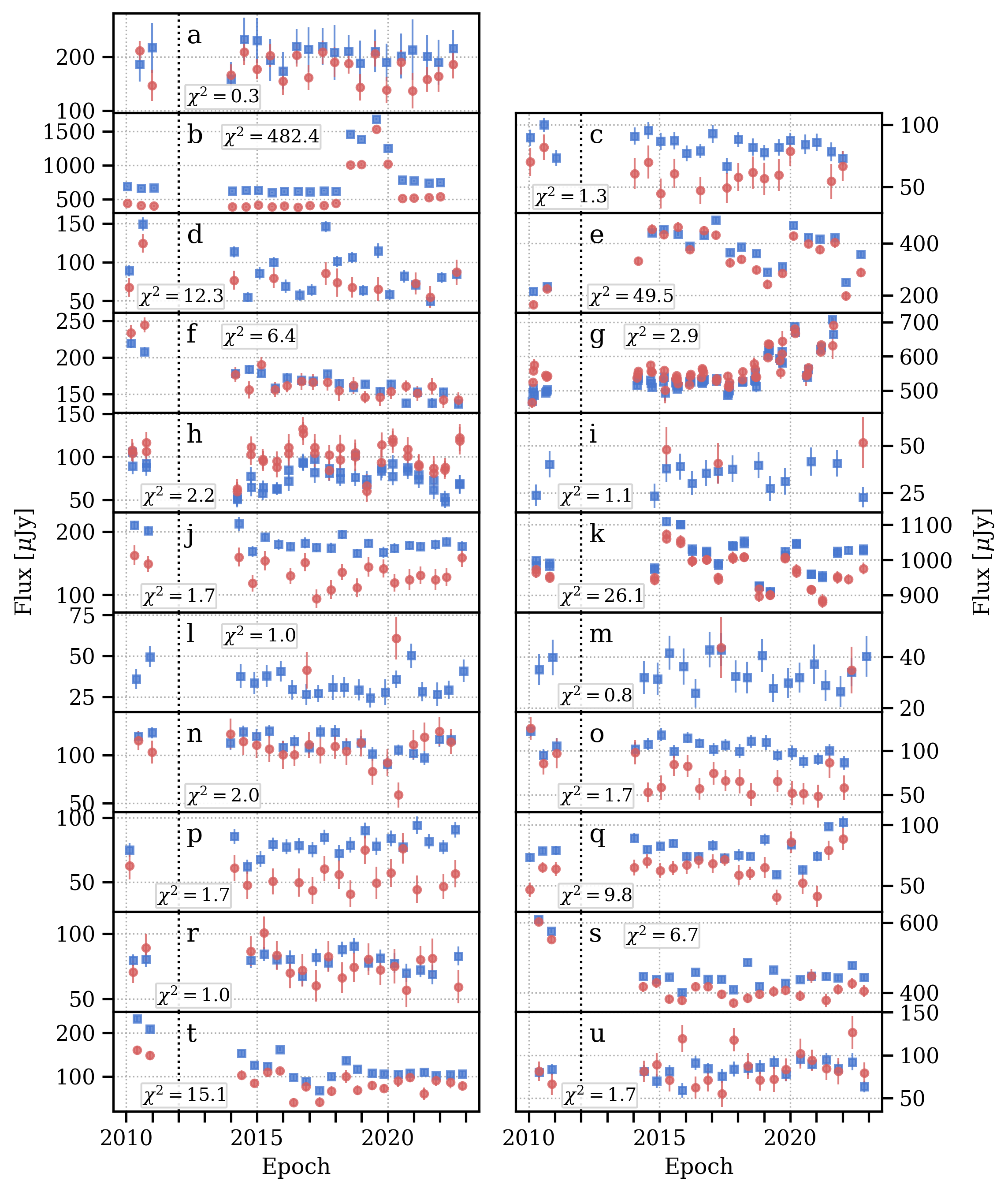}
\caption{Our WISE light curves for all 21 published white dwarfs known to harbor Ca~II gas disks \citep{GentileFusillo2021gas}, plotted following the convention in Figure~\ref{fig:4_modes}. Objects are shown in ascending order of right ascension. See Table~\ref{tab:known_gas_disks} for full information on this sub-sample.
\label{fig:all_gas_lcs}}
\end{figure}

Here were show our light curves of all 21 published Ca~II gas disks in our original input in Figure~\ref{fig:all_gas_lcs}. We show the first half of the 47 non-Ca~II disks in Figure~\ref{fig:all_dust_lcs_1} and the second half in Figure~\ref{fig:all_dust_lcs_2}.

\begin{figure}[t!]
\centering
\includegraphics[width=\textwidth]{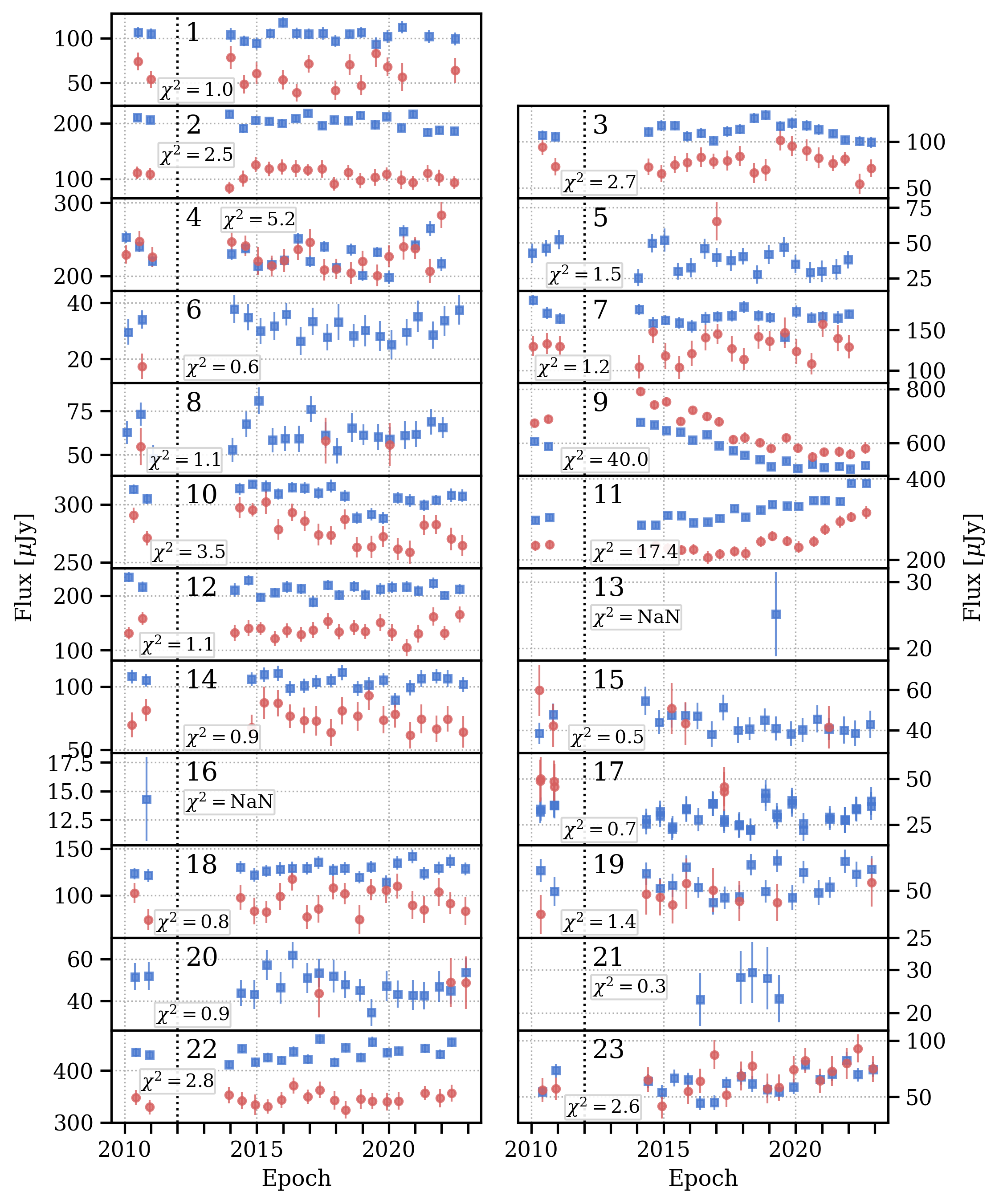}
\caption{Our WISE light curves for the first 23 of the non-Ca~II disk systems included in our input catalog, plotted following the convention in Figure~\ref{fig:4_modes}. Objects are shown in ascending order of right ascension. See Table~\ref{tab:known_dust_disks} for full information on this sub-sample.
\label{fig:all_dust_lcs_1}}
\end{figure}

\begin{figure}[t!]
\centering
\includegraphics[width=\textwidth]{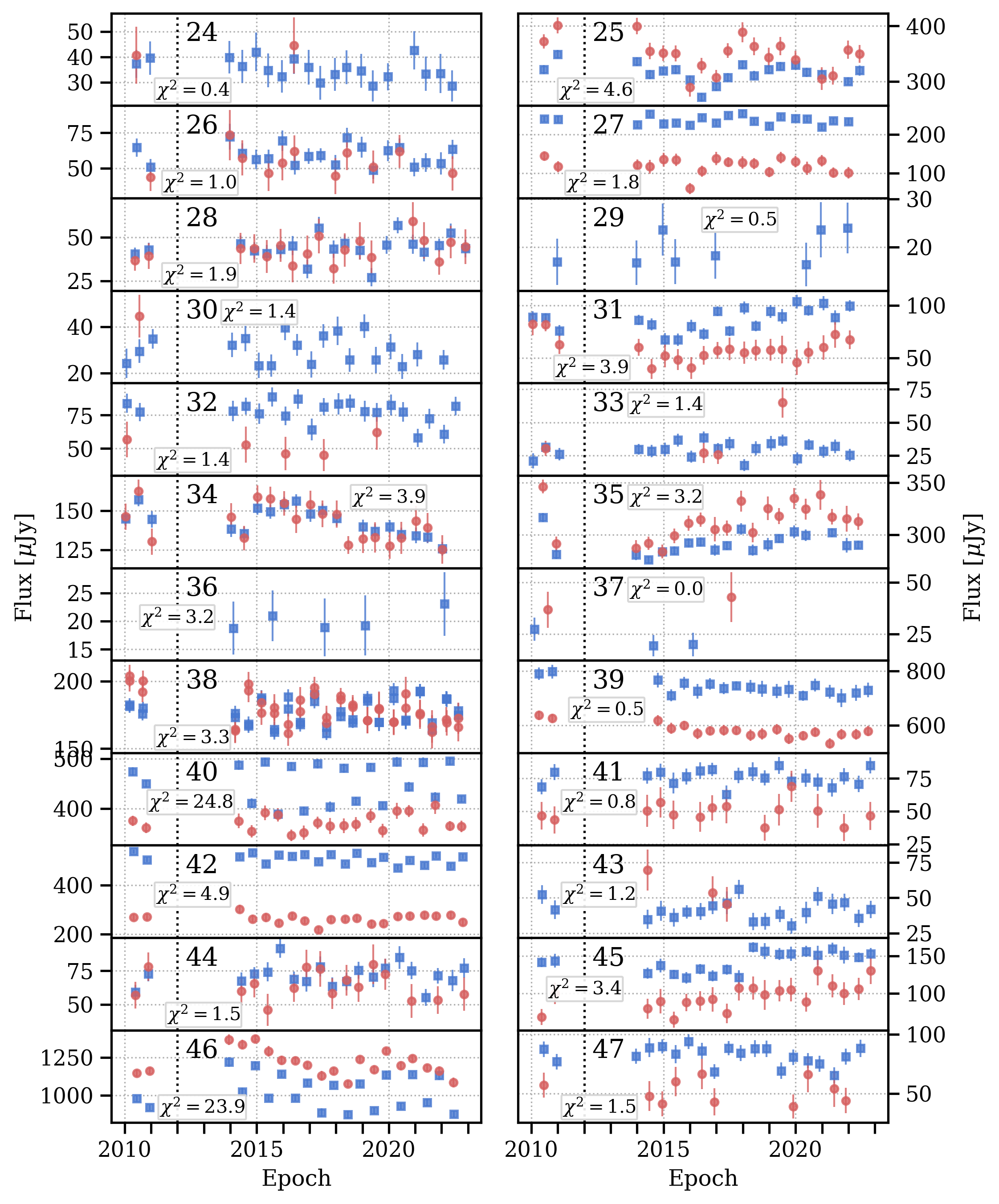}
\caption{Continuation of Figure~\ref{fig:all_dust_lcs_1}, showing the remaining 24 non-Ca~II disks. These light curves are plotted following the convention in Figure~\ref{fig:4_modes}. Objects are shown in ascending order of right ascension. See Table~\ref{tab:known_dust_disks} for full information on this sub-sample.
\label{fig:all_dust_lcs_2}}
\end{figure}

\newpage

\bibliography{references}{}

\begin{thebibliography}{}
\expandafter\ifx\csname natexlab\endcsname\relax\def\natexlab#1{#1}\fi
\providecommand{\url}[1]{\href{#1}{#1}}
\providecommand{\dodoi}[1]{doi:~\href{http://doi.org/#1}{\nolinkurl{#1}}}
\providecommand{\doeprint}[1]{\href{http://ascl.net/#1}{\nolinkurl{http://ascl.net/#1}}}
\providecommand{\doarXiv}[1]{\href{https://arxiv.org/abs/#1}{\nolinkurl{https://arxiv.org/abs/#1}}}

\bibitem[{{Astropy Collaboration} {et~al.}(2013){Astropy Collaboration},
  {Robitaille}, {Tollerud}, {Greenfield}, {Droettboom}, {Bray}, {Aldcroft},
  {Davis}, {Ginsburg}, {Price-Whelan}, {Kerzendorf}, {Conley}, {Crighton},
  {Barbary}, {Muna}, {Ferguson}, {Grollier}, {Parikh}, {Nair}, {Unther},
  {Deil}, {Woillez}, {Conseil}, {Kramer}, {Turner}, {Singer}, {Fox}, {Weaver},
  {Zabalza}, {Edwards}, {Azalee Bostroem}, {Burke}, {Casey}, {Crawford},
  {Dencheva}, {Ely}, {Jenness}, {Labrie}, {Lim}, {Pierfederici}, {Pontzen},
  {Ptak}, {Refsdal}, {Servillat}, \& {Streicher}}]{astropy2013}
{Astropy Collaboration}, {Robitaille}, T.~P., {Tollerud}, E.~J., {et~al.} 2013,
  \aap, 558, A33, \dodoi{10.1051/0004-6361/201322068}

\bibitem[{{Astropy Collaboration} {et~al.}(2018){Astropy Collaboration},
  {Price-Whelan}, {Sip{\H{o}}cz}, {G{\"u}nther}, {Lim}, {Crawford}, {Conseil},
  {Shupe}, {Craig}, {Dencheva}, {Ginsburg}, {Vand erPlas}, {Bradley},
  {P{\'e}rez-Su{\'a}rez}, {de Val-Borro}, {Aldcroft}, {Cruz}, {Robitaille},
  {Tollerud}, {Ardelean}, {Babej}, {Bach}, {Bachetti}, {Bakanov}, {Bamford},
  {Barentsen}, {Barmby}, {Baumbach}, {Berry}, {Biscani}, {Boquien}, {Bostroem},
  {Bouma}, {Brammer}, {Bray}, {Breytenbach}, {Buddelmeijer}, {Burke},
  {Calderone}, {Cano Rodr{\'\i}guez}, {Cara}, {Cardoso}, {Cheedella}, {Copin},
  {Corrales}, {Crichton}, {D'Avella}, {Deil}, {Depagne}, {Dietrich}, {Donath},
  {Droettboom}, {Earl}, {Erben}, {Fabbro}, {Ferreira}, {Finethy}, {Fox},
  {Garrison}, {Gibbons}, {Goldstein}, {Gommers}, {Greco}, {Greenfield},
  {Groener}, {Grollier}, {Hagen}, {Hirst}, {Homeier}, {Horton}, {Hosseinzadeh},
  {Hu}, {Hunkeler}, {Ivezi{\'c}}, {Jain}, {Jenness}, {Kanarek}, {Kendrew},
  {Kern}, {Kerzendorf}, {Khvalko}, {King}, {Kirkby}, {Kulkarni}, {Kumar},
  {Lee}, {Lenz}, {Littlefair}, {Ma}, {Macleod}, {Mastropietro}, {McCully},
  {Montagnac}, {Morris}, {Mueller}, {Mumford}, {Muna}, {Murphy}, {Nelson},
  {Nguyen}, {Ninan}, {N{\"o}the}, {Ogaz}, {Oh}, {Parejko}, {Parley}, {Pascual},
  {Patil}, {Patil}, {Plunkett}, {Prochaska}, {Rastogi}, {Reddy Janga},
  {Sabater}, {Sakurikar}, {Seifert}, {Sherbert}, {Sherwood-Taylor}, {Shih},
  {Sick}, {Silbiger}, {Singanamalla}, {Singer}, {Sladen}, {Sooley},
  {Sornarajah}, {Streicher}, {Teuben}, {Thomas}, {Tremblay}, {Turner},
  {Terr{\'o}n}, {van Kerkwijk}, {de la Vega}, {Watkins}, {Weaver}, {Whitmore},
  {Woillez}, {Zabalza}, \& {Astropy Contributors}}]{astropy2018}
{Astropy Collaboration}, {Price-Whelan}, A.~M., {Sip{\H{o}}cz}, B.~M., {et~al.}
  2018, \aj, 156, 123, \dodoi{10.3847/1538-3881/aabc4f}

\bibitem[{{Barber} {et~al.}(2016){Barber}, {Belardi}, {Kilic}, \&
  {Gianninas}}]{Barber2016}
{Barber}, S.~D., {Belardi}, C., {Kilic}, M., \& {Gianninas}, A. 2016, \mnras,
  459, 1415, \dodoi{10.1093/mnras/stw683}

\bibitem[{{Bateson} {et~al.}(1991){Bateson}, {McIntosh}, \&
  {Brunt}}]{1991PVSS...17...38B}
{Bateson}, F.~M., {McIntosh}, R., \& {Brunt}, D. 1991, Royal Astronomical
  Society of New Zealand Publications of Variable Star Section, 17, 38

\bibitem[{{Bergeron} {et~al.}(2001){Bergeron}, {Leggett}, \&
  {Ruiz}}]{Bergeron2001}
{Bergeron}, P., {Leggett}, S.~K., \& {Ruiz}, M.~T. 2001, \apjs, 133, 413,
  \dodoi{10.1086/320356}

\bibitem[{{Bergeron} {et~al.}(1997){Bergeron}, {Ruiz}, \&
  {Leggett}}]{Bergeron1997}
{Bergeron}, P., {Ruiz}, M.~T., \& {Leggett}, S.~K. 1997, \apjs, 108, 339,
  \dodoi{10.1086/312955}

\bibitem[{{Bergeron} {et~al.}(2011){Bergeron}, {Wesemael}, {Dufour},
  {Beauchamp}, {Hunter}, {Saffer}, {Gianninas}, {Ruiz}, {Limoges}, {Dufour},
  {Fontaine}, \& {Liebert}}]{Bergeron2011}
{Bergeron}, P., {Wesemael}, F., {Dufour}, P., {et~al.} 2011, \apj, 737, 28,
  \dodoi{10.1088/0004-637X/737/1/28}

\bibitem[{{Bergeron} {et~al.}(2021){Bergeron}, {Wesemael}, {Fontaine},
  {Lamontagne}, {Demers}, {B{\'e}dard}, {Gingras}, {Blouin}, {Irwin}, \&
  {Kepler}}]{Bergeron2021}
{Bergeron}, P., {Wesemael}, F., {Fontaine}, G., {et~al.} 2021, \aj, 162, 188,
  \dodoi{10.3847/1538-3881/ac22b1}

\bibitem[{{Bida} {et~al.}(2014){Bida}, {Dunham}, {Massey}, \&
  {Roe}}]{DeVeny2014}
{Bida}, T.~A., {Dunham}, E.~W., {Massey}, P., \& {Roe}, H.~G. 2014, in Society
  of Photo-Optical Instrumentation Engineers (SPIE) Conference Series, Vol.
  9147, Ground-based and Airborne Instrumentation for Astronomy V, ed. S.~K.
  {Ramsay}, I.~S. {McLean}, \& H.~{Takami}, 91472N, \dodoi{10.1117/12.2056872}

\bibitem[{{Brinkworth} {et~al.}(2005){Brinkworth}, {Marsh}, {Morales-Rueda},
  {Maxted}, {Burleigh}, \& {Good}}]{Brinkworth2005}
{Brinkworth}, C.~S., {Marsh}, T.~R., {Morales-Rueda}, L., {et~al.} 2005,
  \mnras, 357, 333, \dodoi{10.1111/j.1365-2966.2005.08649.x}

\bibitem[{{Brouwers} {et~al.}(2022){Brouwers}, {Bonsor}, \&
  {Malamud}}]{Brouwers2022}
{Brouwers}, M.~G., {Bonsor}, A., \& {Malamud}, U. 2022, \mnras, 509, 2404,
  \dodoi{10.1093/mnras/stab3009}

\bibitem[{{Burbidge} \& {Strittmatter}(1971)}]{1971ApJ...170L..39B}
{Burbidge}, E.~M., \& {Strittmatter}, P.~A. 1971, \apjl, 170, L39,
  \dodoi{10.1086/180836}

\bibitem[{{Caron} {et~al.}(2023){Caron}, {Bergeron}, {Blouin}, \&
  {Leggett}}]{Caron2023}
{Caron}, A., {Bergeron}, P., {Blouin}, S., \& {Leggett}, S.~K. 2023, \mnras,
  519, 4529, \dodoi{10.1093/mnras/stac3733}

\bibitem[{{Clemens} {et~al.}(2004){Clemens}, {Crain}, \&
  {Anderson}}]{Clemens2004}
{Clemens}, J.~C., {Crain}, J.~A., \& {Anderson}, R. 2004, in Society of
  Photo-Optical Instrumentation Engineers (SPIE) Conference Series, Vol. 5492,
  Ground-based Instrumentation for Astronomy, ed. A.~F.~M. {Moorwood} \&
  M.~{Iye}, 331--340, \dodoi{10.1117/12.550069}

\bibitem[{{Coutu} {et~al.}(2019){Coutu}, {Dufour}, {Bergeron}, {Blouin},
  {Loranger}, {Allard}, \& {Dunlap}}]{Coutu2019}
{Coutu}, S., {Dufour}, P., {Bergeron}, P., {et~al.} 2019, \apj, 885, 74,
  \dodoi{10.3847/1538-4357/ab46b9}

\bibitem[{{Debes} {et~al.}(2012){Debes}, {Kilic}, {Faedi}, {Shkolnik},
  {Lopez-Morales}, {Weinberger}, {Slesnick}, \& {West}}]{Debes2012b}
{Debes}, J.~H., {Kilic}, M., {Faedi}, F., {et~al.} 2012, \apj, 754, 59,
  \dodoi{10.1088/0004-637X/754/1/59}

\bibitem[{{Dennihy} {et~al.}(2017){Dennihy}, {Clemens}, {Debes}, {Dunlap},
  {Kilkenny}, {O'Brien}, \& {Fuchs}}]{Dennihy2017}
{Dennihy}, E., {Clemens}, J.~C., {Debes}, J.~H., {et~al.} 2017, \apj, 849, 77,
  \dodoi{10.3847/1538-4357/aa8ef2}

\bibitem[{{Dennihy} {et~al.}(2018){Dennihy}, {Clemens}, {Dunlap}, {Fanale},
  {Fuchs}, \& {Hermes}}]{Dennihy2018}
{Dennihy}, E., {Clemens}, J.~C., {Dunlap}, B.~H., {et~al.} 2018, \apj, 854, 40,
  \dodoi{10.3847/1538-4357/aaa89b}

\bibitem[{{Dennihy} {et~al.}(2020{\natexlab{a}}){Dennihy}, {Farihi}, {Gentile
  Fusillo}, \& {Debes}}]{Dennihy2020a}
{Dennihy}, E., {Farihi}, J., {Gentile Fusillo}, N.~P., \& {Debes}, J.~H.
  2020{\natexlab{a}}, \apj, 891, 97, \dodoi{10.3847/1538-4357/ab7249}

\bibitem[{{Dennihy} {et~al.}(2020{\natexlab{b}}){Dennihy}, {Xu}, {Lai},
  {Bonsor}, {Clemens}, {Dufour}, {G{\"a}nsicke}, {Gentile Fusillo}, {Hardy},
  {Hegedus}, {Hermes}, {Kaiser}, {Kissler-Patig}, {Klein}, {Manser}, \&
  {Reding}}]{Dennihy2020b}
{Dennihy}, E., {Xu}, S., {Lai}, S., {et~al.} 2020{\natexlab{b}}, \apj, 905, 5,
  \dodoi{10.3847/1538-4357/abc339}

\bibitem[{{Dominik} \& {Decin}(2003)}]{Dominik2003}
{Dominik}, C., \& {Decin}, G. 2003, \apj, 598, 626, \dodoi{10.1086/379169}

\bibitem[{{Dor{\'e}} {et~al.}(2018){Dor{\'e}}, {Werner}, {Ashby}, {Bleem},
  {Bock}, {Burt}, {Capak}, {Chang}, {Chaves-Montero}, {Chen}, {Civano},
  {Cleeves}, {Cooray}, {Crill}, {Crossfield}, {Cushing}, {de la Torre},
  {DiMatteo}, {Dvory}, {Dvorkin}, {Espaillat}, {Ferraro}, {Finkbeiner},
  {Greene}, {Hewitt}, {Hogg}, {Huffenberger}, {Jun}, {Ilbert}, {Jeong},
  {Johnson}, {Kim}, {Kirkpatrick}, {Kowalski}, {Korngut}, {Li}, {Lisse},
  {MacGregor}, {Mamajek}, {Mauskopf}, {Melnick}, {M{\'e}nard}, {Neyrinck},
  {{\"O}berg}, {Pisani}, {Rocca}, {Salvato}, {Schaan}, {Scoville}, {Song},
  {Stevens}, {Tenneti}, {Teplitz}, {Tolls}, {Unwin}, {Urry}, {Wandelt},
  {Williams}, {Wilner}, {Windhorst}, {Wolk}, {Yorke}, \& {Zemcov}}]{Dore2018}
{Dor{\'e}}, O., {Werner}, M.~W., {Ashby}, M. L.~N., {et~al.} 2018, arXiv
  e-prints, arXiv:1805.05489, \dodoi{10.48550/arXiv.1805.05489}

\bibitem[{{Dufour} {et~al.}(2017){Dufour}, {Blouin}, {Coutu},
  {Fortin-Archambault}, {Thibeault}, {Bergeron}, \& {Fontaine}}]{MWDD2017}
{Dufour}, P., {Blouin}, S., {Coutu}, S., {et~al.} 2017, in Astronomical Society
  of the Pacific Conference Series, Vol. 509, 20th European White Dwarf
  Workshop, ed. P.~E. {Tremblay}, B.~{Gaensicke}, \& T.~{Marsh}, 3.
\newblock \doarXiv{1610.00986}

\bibitem[{{Eisenhardt} {et~al.}(2020){Eisenhardt}, {Marocco}, {Fowler},
  {Meisner}, {Kirkpatrick}, {Garcia}, {Jarrett}, {Koontz}, {Marchese},
  {Stanford}, {Caselden}, {Cushing}, {Cutri}, {Faherty}, {Gelino}, {Gonzalez},
  {Mainzer}, {Mobasher}, {Schlegel}, {Stern}, {Teplitz}, \&
  {Wright}}]{Eisenhardt2020}
{Eisenhardt}, P. R.~M., {Marocco}, F., {Fowler}, J.~W., {et~al.} 2020, \apjs,
  247, 69, \dodoi{10.3847/1538-4365/ab7f2a}

\bibitem[{{Eyer} {et~al.}(2023){Eyer}, {Audard}, {Holl}, {Rimoldini},
  {Carnerero}, {Clementini}, {De Ridder}, {Distefano}, {Evans}, {Gavras},
  {Gomel}, {Lebzelter}, {Marton}, {Mowlavi}, {Panahi}, {Ripepi}, {Wyrzykowski},
  {Nienartowicz}, {Jevardat de Fombelle}, {Lecoeur-Taibi}, {Rohrbasser},
  {Riello}, {Garc{\'\i}a-Lario}, {Lanzafame}, {Mazeh}, {Raiteri}, {Zucker},
  {{\'A}brah{\'a}m}, {Aerts}, {Aguado}, {Anderson}, {Bashi}, {Binnenfeld},
  {Faigler}, {Garofalo}, {Karbevska}, {K{\'o}sp{\'a}l}, {Kruszy{\'n}ska},
  {Kun}, {Lanza}, {Leccia}, {Marconi}, {Messina}, {Molinaro}, {Moln{\'a}r},
  {Muraveva}, {Musella}, {Nagy}, {Pagano}, {Palaversa}, {Plachy}, {Pr{\v{s}}a},
  {Rybicki}, {Shahaf}, {Szabados}, {Szegedi-Elek}, {Trabucchi}, {Barblan},
  {Grenon}, {Roelens}, \& {S{\"u}veges}}]{Eyer2023}
{Eyer}, L., {Audard}, M., {Holl}, B., {et~al.} 2023, \aap, 674, A13,
  \dodoi{10.1051/0004-6361/202244242}

\bibitem[{{Farihi}(2016)}]{Farihi2016}
{Farihi}, J. 2016, \nar, 71, 9, \dodoi{10.1016/j.newar.2016.03.001}

\bibitem[{{Farihi} {et~al.}(2005){Farihi}, {Becklin}, \&
  {Zuckerman}}]{Farihi2005}
{Farihi}, J., {Becklin}, E.~E., \& {Zuckerman}, B. 2005, \apjs, 161, 394,
  \dodoi{10.1086/444362}

\bibitem[{{Farihi} {et~al.}(2012){Farihi}, {G{\"a}nsicke}, {Steele}, {Girven},
  {Burleigh}, {Breedt}, \& {Koester}}]{Farihi2012}
{Farihi}, J., {G{\"a}nsicke}, B.~T., {Steele}, P.~R., {et~al.} 2012, \mnras,
  421, 1635, \dodoi{10.1111/j.1365-2966.2012.20421.x}

\bibitem[{{Farihi} {et~al.}(2018){Farihi}, {van Lieshout}, {Cauley}, {Dennihy},
  {Su}, {Kenyon}, {Wilson}, {Toloza}, {G{\"a}nsicke}, {von Hippel}, {Redfield},
  {Debes}, {Xu}, {Rogers}, {Bonsor}, {Swan}, {Pala}, \& {Reach}}]{Farihi2018}
{Farihi}, J., {van Lieshout}, R., {Cauley}, P.~W., {et~al.} 2018, \mnras, 481,
  2601, \dodoi{10.1093/mnras/sty2331}

\bibitem[{{Fontaine} {et~al.}(2003){Fontaine}, {Bergeron}, {Bill{\`e}res}, \&
  {Charpinet}}]{Fontaine2003}
{Fontaine}, G., {Bergeron}, P., {Bill{\`e}res}, M., \& {Charpinet}, S. 2003,
  \apj, 591, 1184, \dodoi{10.1086/375490}

\bibitem[{{G{\"a}nsicke}(2011)}]{Gaensicke2011}
{G{\"a}nsicke}, B.~T. 2011, in AIP Conf. Proc. 1331, Planetary Systems Beyond
  the Main Sequence, ed. S. Schuh, H. Drechsel, \& U. Heber (Melville, NY:
  AIP), 211

\bibitem[{{G{\"a}nsicke} {et~al.}(2008){G{\"a}nsicke}, {Koester}, {Marsh},
  {Rebassa-Mansergas}, \& {Southworth}}]{Gaensicke2008}
{G{\"a}nsicke}, B.~T., {Koester}, D., {Marsh}, T.~R., {Rebassa-Mansergas}, A.,
  \& {Southworth}, J. 2008, \mnras, 391, L103,
  \dodoi{10.1111/j.1745-3933.2008.00565.x}

\bibitem[{{G{\"a}nsicke} {et~al.}(2007){G{\"a}nsicke}, {Marsh}, \&
  {Southworth}}]{Gaensicke2007}
{G{\"a}nsicke}, B.~T., {Marsh}, T.~R., \& {Southworth}, J. 2007, \mnras, 380,
  L35, \dodoi{10.1111/j.1745-3933.2007.00343.x}

\bibitem[{{G{\"a}nsicke} {et~al.}(2006){G{\"a}nsicke}, {Marsh}, {Southworth},
  \& {Rebassa-Mansergas}}]{Gaensicke2006}
{G{\"a}nsicke}, B.~T., {Marsh}, T.~R., {Southworth}, J., \&
  {Rebassa-Mansergas}, A. 2006, Science, 314, 1908,
  \dodoi{10.1126/science.1135033}

\bibitem[{{G{\"a}nsicke} {et~al.}(2019){G{\"a}nsicke}, {Schreiber}, {Toloza},
  {Gentile Fusillo}, {Koester}, \& {Manser}}]{Gaensicke2019}
{G{\"a}nsicke}, B.~T., {Schreiber}, M.~R., {Toloza}, O., {et~al.} 2019, \nat,
  576, 61, \dodoi{10.1038/s41586-019-1789-8}

\bibitem[{{Gentile Fusillo} {et~al.}(2021{\natexlab{a}}){Gentile Fusillo},
  {Manser}, {G{\"a}nsicke}, {Toloza}, {Koester}, {Dennihy}, {Brown}, {Farihi},
  {Hollands}, {Hoskin}, {Izquierdo}, {Kinnear}, {Marsh},
  {Santamar{\'\i}a-Miranda}, {Pala}, {Redfield}, {Rodr{\'\i}guez-Gil},
  {Schreiber}, {Veras}, \& {Wilson}}]{GentileFusillo2021gas}
{Gentile Fusillo}, N.~P., {Manser}, C.~J., {G{\"a}nsicke}, B.~T., {et~al.}
  2021{\natexlab{a}}, \mnras, 504, 2707, \dodoi{10.1093/mnras/stab992}

\bibitem[{{Gentile Fusillo} {et~al.}(2021{\natexlab{b}}){Gentile Fusillo},
  {Tremblay}, {Cukanovaite}, {Vorontseva}, {Lallement}, {Hollands},
  {G{\"a}nsicke}, {Burdge}, {McCleery}, \& {Jordan}}]{GentileFusillo2021gaia}
{Gentile Fusillo}, N.~P., {Tremblay}, P.~E., {Cukanovaite}, E., {et~al.}
  2021{\natexlab{b}}, \mnras, 508, 3877, \dodoi{10.1093/mnras/stab2672}

\bibitem[{{Gianninas} {et~al.}(2006){Gianninas}, {Bergeron}, \&
  {Fontaine}}]{Gianninas2006}
{Gianninas}, A., {Bergeron}, P., \& {Fontaine}, G. 2006, \aj, 132, 831,
  \dodoi{10.1086/506516}

\bibitem[{{Gianninas} {et~al.}(2011){Gianninas}, {Bergeron}, \&
  {Ruiz}}]{Gianninas2011}
{Gianninas}, A., {Bergeron}, P., \& {Ruiz}, M.~T. 2011, \apj, 743, 138,
  \dodoi{10.1088/0004-637X/743/2/138}

\bibitem[{{Ginsburg} {et~al.}(2019){Ginsburg}, {Sip{\H o}cz}, {Brasseur},
  {Cowperthwaite}, {Craig}, {Deil}, {Guillochon}, {Guzman}, {Liedtke}, {Lian
  Lim}, {Lockhart}, {Mommert}, {Morris}, {Norman}, {Parikh}, {Persson},
  {Robitaille}, {Segovia}, {Singer}, {Tollerud}, {de Val-Borro}, {Valtchanov},
  {Woillez}, {The Astroquery collaboration}, \& {a subset of the astropy
  collaboration}}]{Astroquery_Ginsburg2019}
{Ginsburg}, A., {Sip{\H o}cz}, B.~M., {Brasseur}, C.~E., {et~al.} 2019, \aj,
  157, 98, \dodoi{10.3847/1538-3881/aafc33}

\bibitem[{{Girven} {et~al.}(2011){Girven}, {G{\"a}nsicke}, {Steeghs}, \&
  {Koester}}]{Girven2011}
{Girven}, J., {G{\"a}nsicke}, B.~T., {Steeghs}, D., \& {Koester}, D. 2011,
  \mnras, 417, 1210, \dodoi{10.1111/j.1365-2966.2011.19337.x}

\bibitem[{Guidry {et~al.}(2021)Guidry, Vanderbosch, Hermes, Barlow, Lopez,
  Boudreaux, Corcoran, Bell, Montgomery, Heintz, Castanheira, Reding, Dunlap,
  Winget, Winget, \& Kuehne}]{Guidry2021}
Guidry, J.~A., Vanderbosch, Z.~P., Hermes, J.~J., {et~al.} 2021, The
  Astrophysical Journal, 912, 125, \dodoi{10.3847/1538-4357/abee68}

\bibitem[{{Guo} {et~al.}(2015){Guo}, {Zhao}, {Tziamtzis}, {Liu}, {Li}, {Zhang},
  {Hou}, \& {Wang}}]{Guo2015}
{Guo}, J., {Zhao}, J., {Tziamtzis}, A., {et~al.} 2015, \mnras, 454, 2787,
  \dodoi{10.1093/mnras/stv2104}

\bibitem[{Harris {et~al.}(2020)Harris, Millman, van~der Walt, Gommers,
  Virtanen, Cournapeau, Wieser, Taylor, Berg, Smith, Kern, Picus, Hoyer, van
  Kerkwijk, Brett, Haldane, del R{'{\i}}o, Wiebe, Peterson,
  G{'{e}}rard-Marchant, Sheppard, Reddy, Weckesser, Abbasi, Gohlke, \&
  Oliphant}]{numpy2020}
Harris, C.~R., Millman, K.~J., van~der Walt, S.~J., {et~al.} 2020, Nature, 585,
  357, \dodoi{10.1038/s41586-020-2649-2}

\bibitem[{{Harrison} {et~al.}(2009){Harrison}, {Bornak}, {Howell}, {Mason},
  {Szkody}, \& {McGurk}}]{2009AJ....137.4061H}
{Harrison}, T.~E., {Bornak}, J., {Howell}, S.~B., {et~al.} 2009, \aj, 137,
  4061, \dodoi{10.1088/0004-6256/137/4/4061}

\bibitem[{{Hart} {et~al.}(2023){Hart}, {Shappee}, {Hey}, {Kochanek}, {Stanek},
  {Lim}, {Dobbs}, {Tucker}, {Jayasinghe}, {Beacom}, {Boright}, {Holoien},
  {Ong}, {Prieto}, {Thompson}, \& {Will}}]{Hart2023}
{Hart}, K., {Shappee}, B.~J., {Hey}, D., {et~al.} 2023, arXiv e-prints,
  arXiv:2304.03791, \dodoi{10.48550/arXiv.2304.03791}

\bibitem[{{Higgins} \& {Bell}(2023)}]{Higgins2023}
{Higgins}, M.~E., \& {Bell}, K.~J. 2023, \aj, 165, 141,
  \dodoi{10.3847/1538-3881/acb20c}

\bibitem[{{Hoard} {et~al.}(2013){Hoard}, {Debes}, {Wachter}, {Leisawitz}, \&
  {Cohen}}]{Hoard2013}
{Hoard}, D.~W., {Debes}, J.~H., {Wachter}, S., {Leisawitz}, D.~T., \& {Cohen},
  M. 2013, \apj, 770, 21, \dodoi{10.1088/0004-637X/770/1/21}

\bibitem[{{Howarth}(1978)}]{1978JBAA...88..292H}
{Howarth}, I.~D. 1978, Journal of the British Astronomical Association, 88, 292

\bibitem[{Hunter(2007)}]{Matplotlib2007}
Hunter, J.~D. 2007, Computing in Science \& Engineering, 9, 90,
  \dodoi{10.1109/MCSE.2007.55}

\bibitem[{{Jarrett} {et~al.}(2011){Jarrett}, {Cohen}, {Masci}, {Wright},
  {Stern}, {Benford}, {Blain}, {Carey}, {Cutri}, {Eisenhardt}, {Lonsdale},
  {Mainzer}, {Marsh}, {Padgett}, {Petty}, {Ressler}, {Skrutskie}, {Stanford},
  {Surace}, {Tsai}, {Wheelock}, \& {Yan}}]{Jarrett2011}
{Jarrett}, T.~H., {Cohen}, M., {Masci}, F., {et~al.} 2011, \apj, 735, 112,
  \dodoi{10.1088/0004-637X/735/2/112}

\bibitem[{{Jura}(2003)}]{Jura2003}
{Jura}, M. 2003, \apjl, 584, L91, \dodoi{10.1086/374036}

\bibitem[{{Kanaan} {et~al.}(1992){Kanaan}, {Kepler}, {Giovannini}, \&
  {Diaz}}]{Kanaan1992}
{Kanaan}, A., {Kepler}, S.~O., {Giovannini}, O., \& {Diaz}, M. 1992, \apjl,
  390, L89, \dodoi{10.1086/186379}

\bibitem[{{Kato} {et~al.}(2000){Kato}, {Hanson}, {Poyner}, {Muyllaert},
  {Reszelski}, \& {Dubovsky}}]{2000IBVS.4932....1K}
{Kato}, T., {Hanson}, G., {Poyner}, G., {et~al.} 2000, Information Bulletin on
  Variable Stars, 4932, 1

\bibitem[{{Kawka} \& {Vennes}(2012)}]{KawkaVennes2012}
{Kawka}, A., \& {Vennes}, S. 2012, \mnras, 425, 1394,
  \dodoi{10.1111/j.1365-2966.2012.21574.x}

\bibitem[{{Kilic} {et~al.}(2020){Kilic}, {Bergeron}, {Kosakowski}, {Brown},
  {Ag{\"u}eros}, \& {Blouin}}]{Kilic2020}
{Kilic}, M., {Bergeron}, P., {Kosakowski}, A., {et~al.} 2020, \apj, 898, 84,
  \dodoi{10.3847/1538-4357/ab9b8d}

\bibitem[{{Kilic} {et~al.}(2012){Kilic}, {Patterson}, {Barber}, {Leggett}, \&
  {Dufour}}]{Kilic2012}
{Kilic}, M., {Patterson}, A.~J., {Barber}, S., {Leggett}, S.~K., \& {Dufour},
  P. 2012, \mnras, 419, L59, \dodoi{10.1111/j.1745-3933.2011.01177.x}

\bibitem[{{Kilkenny} {et~al.}(2016){Kilkenny}, {Worters}, {O'Donoghue}, {Koen},
  {Koen}, {Hambly}, {MacGillivray}, \& {Stobie}}]{Kilkenny2016}
{Kilkenny}, D., {Worters}, H.~L., {O'Donoghue}, D., {et~al.} 2016, \mnras, 459,
  4343, \dodoi{10.1093/mnras/stw916}

\bibitem[{{Kleinman} {et~al.}(2013){Kleinman}, {Kepler}, {Koester}, {Pelisoli},
  {Pe{\c{c}}anha}, {Nitta}, {Costa}, {Krzesinski}, {Dufour}, {Lachapelle},
  {Bergeron}, {Yip}, {Harris}, {Eisenstein}, {Althaus}, \&
  {C{\'o}rsico}}]{Kleinman2013}
{Kleinman}, S.~J., {Kepler}, S.~O., {Koester}, D., {et~al.} 2013, \apjs, 204,
  5, \dodoi{10.1088/0067-0049/204/1/5}

\bibitem[{{Kochanek} {et~al.}(2017){Kochanek}, {Shappee}, {Stanek}, {Holoien},
  {Thompson}, {Prieto}, {Dong}, {Shields}, {Will}, {Britt}, {Perzanowski}, \&
  {Pojma{\'n}ski}}]{Koachanek2017}
{Kochanek}, C.~S., {Shappee}, B.~J., {Stanek}, K.~Z., {et~al.} 2017, \pasp,
  129, 104502, \dodoi{10.1088/1538-3873/aa80d9}

\bibitem[{{Koester} {et~al.}(2014){Koester}, {G{\"a}nsicke}, \&
  {Farihi}}]{Koester2014}
{Koester}, D., {G{\"a}nsicke}, B.~T., \& {Farihi}, J. 2014, \aap, 566, A34,
  \dodoi{10.1051/0004-6361/201423691}

\bibitem[{{Koester} {et~al.}(2005){Koester}, {Rollenhagen}, {Napiwotzki},
  {Voss}, {Christlieb}, {Homeier}, \& {Reimers}}]{Koester2005}
{Koester}, D., {Rollenhagen}, K., {Napiwotzki}, R., {et~al.} 2005, \aap, 432,
  1025, \dodoi{10.1051/0004-6361:20041927}

\bibitem[{{Koester} {et~al.}(1982){Koester}, {Weidemann}, \&
  {Zeidler}}]{Koester1982}
{Koester}, D., {Weidemann}, V., \& {Zeidler}, E.~M. 1982, \aap, 116, 147

\bibitem[{{Lang}(2014)}]{Lang2014}
{Lang}, D. 2014, \aj, 147, 108, \dodoi{10.1088/0004-6256/147/5/108}

\bibitem[{{Limoges} {et~al.}(2015){Limoges}, {Bergeron}, \&
  {L{\'e}pine}}]{Limoges2015}
{Limoges}, M.~M., {Bergeron}, P., \& {L{\'e}pine}, S. 2015, \apjs, 219, 19,
  \dodoi{10.1088/0067-0049/219/2/19}

\bibitem[{{Limoges} {et~al.}(2013){Limoges}, {L{\'e}pine}, \&
  {Bergeron}}]{Limoges2013}
{Limoges}, M.~M., {L{\'e}pine}, S., \& {Bergeron}, P. 2013, \aj, 145, 136,
  \dodoi{10.1088/0004-6256/145/5/136}

\bibitem[{{Mainzer} {et~al.}(2011){Mainzer}, {Bauer}, {Grav}, {Masiero},
  {Cutri}, {Dailey}, {Eisenhardt}, {McMillan}, {Wright}, {Walker}, {Jedicke},
  {Spahr}, {Tholen}, {Alles}, {Beck}, {Brandenburg}, {Conrow}, {Evans},
  {Fowler}, {Jarrett}, {Marsh}, {Masci}, {McCallon}, {Wheelock}, {Wittman},
  {Wyatt}, {DeBaun}, {Elliott}, {Elsbury}, {Gautier}, {Gomillion}, {Leisawitz},
  {Maleszewski}, {Micheli}, \& {Wilkins}}]{Mainzer2011}
{Mainzer}, A., {Bauer}, J., {Grav}, T., {et~al.} 2011, \apj, 731, 53,
  \dodoi{10.1088/0004-637X/731/1/53}

\bibitem[{{Mainzer} {et~al.}(2014){Mainzer}, {Bauer}, {Cutri}, {Grav},
  {Masiero}, {Beck}, {Clarkson}, {Conrow}, {Dailey}, {Eisenhardt}, {Fabinsky},
  {Fajardo-Acosta}, {Fowler}, {Gelino}, {Grillmair}, {Heinrichsen}, {Kendall},
  {Kirkpatrick}, {Liu}, {Masci}, {McCallon}, {Nugent}, {Papin}, {Rice},
  {Royer}, {Ryan}, {Sevilla}, {Sonnett}, {Stevenson}, {Thompson}, {Wheelock},
  {Wiemer}, {Wittman}, {Wright}, \& {Yan}}]{Mainzer2014}
{Mainzer}, A., {Bauer}, J., {Cutri}, R.~M., {et~al.} 2014, \apj, 792, 30,
  \dodoi{10.1088/0004-637X/792/1/30}

\bibitem[{{Mainzer} {et~al.}(2023){Mainzer}, {Masiero}, {Abell}, {Bauer},
  {Bottke}, {Buratti}, {Carey}, {Cotto-Figueroa}, {Cutri}, {Dahlen},
  {Eisenhardt}, {Fernandez}, {Furfaro}, {Grav}, {Hoffman}, {Kelley}, {Kim},
  {Kirkpatrick}, {Lawler}, {Lilly}, {Liu}, {Marocco}, {Marsh}, {Masci},
  {McMurtry}, {Pourrahmani}, {Reinhart}, {Ressler}, {Satpathy}, {Schambeau},
  {Sonnett}, {Spahr}, {Surace}, {Vaquero}, {Wright}, {Zengilowski}, \& {NEO
  Surveyor Mission Team}}]{Mainzer2023}
{Mainzer}, A.~K., {Masiero}, J.~R., {Abell}, P.~A., {et~al.} 2023, \psj, 4,
  224, \dodoi{10.3847/PSJ/ad0468}

\bibitem[{{Malamud} {et~al.}(2021){Malamud}, {Grishin}, \&
  {Brouwers}}]{Malamud2021}
{Malamud}, U., {Grishin}, E., \& {Brouwers}, M. 2021, \mnras, 501, 3806,
  \dodoi{10.1093/mnras/staa3940}

\bibitem[{{Manser} {et~al.}(2020){Manser}, {G{\"a}nsicke}, {Gentile Fusillo},
  {Ashley}, {Breedt}, {Hollands}, {Izquierdo}, \& {Pelisoli}}]{Manser2020}
{Manser}, C.~J., {G{\"a}nsicke}, B.~T., {Gentile Fusillo}, N.~P., {et~al.}
  2020, \mnras, 493, 2127, \dodoi{10.1093/mnras/staa359}

\bibitem[{{Marrese} {et~al.}(2019){Marrese}, {Marinoni}, {Fabrizio}, \&
  {Altavilla}}]{Marrese2019}
{Marrese}, P.~M., {Marinoni}, S., {Fabrizio}, M., \& {Altavilla}, G. 2019,
  \aap, 621, A144, \dodoi{10.1051/0004-6361/201834142}

\bibitem[{{Marrese} {et~al.}(2022){Marrese}, {Marinoni}, {Fabrizio}, \&
  {Altavilla}}]{Marrese2022}
---. 2022, {Gaia DR3 documentation Chapter 15: Cross-match with external
  catalogues}, Gaia DR3 documentation, European Space Agency; Gaia Data
  Processing and Analysis Consortium.

\bibitem[{{Marrese} {et~al.}(2017){Marrese}, {Marinoni}, {Fabrizio}, \&
  {Giuffrida}}]{Marrese2017}
{Marrese}, P.~M., {Marinoni}, S., {Fabrizio}, M., \& {Giuffrida}, G. 2017,
  \aap, 607, A105, \dodoi{10.1051/0004-6361/201730965}

\bibitem[{{Masterson} {et~al.}(2024){Masterson}, {De}, {Panagiotou}, {Kara},
  {Arcavi}, {Eilers}, {Frostig}, {Gezari}, {Grotova}, {Liu}, {Malyali},
  {Meisner}, {Merloni}, {Newsome}, {Rau}, {Simcoe}, \& {van
  Velzen}}]{Masterson2024}
{Masterson}, M., {De}, K., {Panagiotou}, C., {et~al.} 2024, \apj, 961, 211,
  \dodoi{10.3847/1538-4357/ad18bb}

\bibitem[{{McCleery} {et~al.}(2020){McCleery}, {Tremblay}, {Gentile Fusillo},
  {Hollands}, {G{\"a}nsicke}, {Izquierdo}, {Toonen}, {Cunningham}, \&
  {Rebassa-Mansergas}}]{McCleery2020}
{McCleery}, J., {Tremblay}, P.-E., {Gentile Fusillo}, N.~P., {et~al.} 2020,
  \mnras, 499, 1890, \dodoi{10.1093/mnras/staa2030}

\bibitem[{{McCook} \& {Sion}(1999)}]{McCookSion1999}
{McCook}, G.~P., \& {Sion}, E.~M. 1999, \apjs, 121, 1, \dodoi{10.1086/313186}

\bibitem[{{Meinunger}(1984)}]{1984IBVS.2483....1M}
{Meinunger}, L. 1984, Information Bulletin on Variable Stars, 2483, 1

\bibitem[{{Meisner} {et~al.}(2023){Meisner}, {Caselden}, {Schlafly}, \&
  {Kiwy}}]{Meisner2023}
{Meisner}, A.~M., {Caselden}, D., {Schlafly}, E.~F., \& {Kiwy}, F. 2023, \aj,
  165, 36, \dodoi{10.3847/1538-3881/aca2ab}

\bibitem[{{Meisner} {et~al.}(2019){Meisner}, {Lang}, {Schlafly}, \&
  {Schlegel}}]{Meisner2019}
{Meisner}, A.~M., {Lang}, D., {Schlafly}, E.~F., \& {Schlegel}, D.~J. 2019,
  \pasp, 131, 124504, \dodoi{10.1088/1538-3873/ab3df4}

\bibitem[{{Meisner} {et~al.}(2017{\natexlab{a}}){Meisner}, {Lang}, \&
  {Schlegel}}]{Meisner2017a}
{Meisner}, A.~M., {Lang}, D., \& {Schlegel}, D.~J. 2017{\natexlab{a}}, \aj,
  153, 38, \dodoi{10.3847/1538-3881/153/1/38}

\bibitem[{{Meisner} {et~al.}(2017{\natexlab{b}}){Meisner}, {Lang}, \&
  {Schlegel}}]{Meisner2017b}
---. 2017{\natexlab{b}}, \aj, 154, 161, \dodoi{10.3847/1538-3881/aa894e}

\bibitem[{{Meisner} {et~al.}(2018){Meisner}, {Lang}, \&
  {Schlegel}}]{Meisner2018b}
{Meisner}, A.~M., {Lang}, D.~A., \& {Schlegel}, D.~J. 2018, Research Notes of
  the American Astronomical Society, 2, 202, \dodoi{10.3847/2515-5172/aaecd5}

\bibitem[{{Melis} {et~al.}(2020){Melis}, {Klein}, {Doyle}, {Weinberger},
  {Zuckerman}, \& {Dufour}}]{Melis2020}
{Melis}, C., {Klein}, B., {Doyle}, A.~E., {et~al.} 2020, \apj, 905, 56,
  \dodoi{10.3847/1538-4357/abbdfa}

\bibitem[{{Melis} {et~al.}(2012){Melis}, {Dufour}, {Farihi}, {Bochanski},
  {Burgasser}, {Parsons}, {G{\"a}nsicke}, {Koester}, \& {Swift}}]{Melis2012}
{Melis}, C., {Dufour}, P., {Farihi}, J., {et~al.} 2012, \apjl, 751, L4,
  \dodoi{10.1088/2041-8205/751/1/L4}

\bibitem[{{Nather} {et~al.}(1981){Nather}, {Robinson}, \&
  {Stover}}]{Nather1981}
{Nather}, R.~E., {Robinson}, E.~L., \& {Stover}, R.~J. 1981, \apj, 244, 269,
  \dodoi{10.1086/158704}

\bibitem[{Newville {et~al.}(2014)Newville, Stensitzki, Allen, \&
  Ingargiola}]{newville_matthew_2014_11813}
Newville, M., Stensitzki, T., Allen, D.~B., \& Ingargiola, A. 2014, {LMFIT:
  Non-Linear Least-Square Minimization and Curve-Fitting for Python}, 0.8.0,
  Zenodo, \dodoi{10.5281/zenodo.11813}

\bibitem[{{O'Brien} {et~al.}(2023){O'Brien}, {Tremblay}, {Gentile Fusillo},
  {Hollands}, {G{\"a}nsicke}, {Koester}, {Pelisoli}, {Cukanovaite},
  {Cunningham}, {Doyle}, {Elms}, {Farihi}, {Hermes}, {Holberg}, {Jordan},
  {Klein}, {Kleinman}, {Manser}, {De Martino}, {Marsh}, {McCleery}, {Melis},
  {Nitta}, {Parsons}, {Raddi}, {Rebassa-Mansergas}, {Schreiber}, {Silvotti},
  {Steeghs}, {Toloza}, {Toonen}, {Torres}, {Weinberger}, \&
  {Zuckerman}}]{OBrien2023}
{O'Brien}, M.~W., {Tremblay}, P.~E., {Gentile Fusillo}, N.~P., {et~al.} 2023,
  \mnras, 518, 3055, \dodoi{10.1093/mnras/stac3303}

\bibitem[{{O'Brien} {et~al.}(2024){O'Brien}, {Tremblay}, {Klein}, {Koester},
  {Melis}, {B{\'e}dard}, {Cukanovaite}, {Cunningham}, {Doyle}, {G{\"a}nsicke},
  {Gentile Fusillo}, {Hollands}, {McCleery}, {Pelisoli}, {Toonen},
  {Weinberger}, \& {Zuckerman}}]{OBrien2024}
{O'Brien}, M.~W., {Tremblay}, P.~E., {Klein}, B.~L., {et~al.} 2024, \mnras,
  527, 8687, \dodoi{10.1093/mnras/stad3773}

\bibitem[{{O'Donoghue} {et~al.}(2013){O'Donoghue}, {Kilkenny}, {Koen},
  {Hambly}, {MacGillivray}, \& {Stobie}}]{ODonoghue2013}
{O'Donoghue}, D., {Kilkenny}, D., {Koen}, C., {et~al.} 2013, \mnras, 431, 240,
  \dodoi{10.1093/mnras/stt158}

\bibitem[{{Ould Rouis} {et~al.}(2024){Ould Rouis}, {Hermes}, \&
  {Gaensicke}}]{OuldRouis2024ESSV}
{Ould Rouis}, L.~B., {Hermes}, J., \& {Gaensicke}, B. 2024, in AAS/Division for
  Extreme Solar Systems Abstracts, Vol.~56, AAS/Division for Extreme Solar
  Systems Abstracts, 612.02

\bibitem[{pandas~development team(2020)}]{pandas2020}
pandas~development team, T. 2020, pandas-dev/pandas: Pandas, latest,  Zenodo,
  \dodoi{10.5281/zenodo.3509134}

\bibitem[{{Petrosky} {et~al.}(2021){Petrosky}, {Hwang}, {Zakamska}, {Chandra},
  \& {Hill}}]{Petrosky2021}
{Petrosky}, E., {Hwang}, H.-C., {Zakamska}, N.~L., {Chandra}, V., \& {Hill},
  M.~J. 2021, \mnras, 503, 3975, \dodoi{10.1093/mnras/stab592}

\bibitem[{{Prochaska} {et~al.}(2020){Prochaska}, {Hennawi}, {Westfall},
  {Cooke}, {Wang}, {Hsyu}, {Davies}, {Farina}, \&
  {Pelliccia}}]{pypeit:joss_arXiv}
{Prochaska}, J., {Hennawi}, J., {Westfall}, K., {et~al.} 2020, The Journal of
  Open Source Software, 5, 2308, \dodoi{10.21105/joss.02308}

\bibitem[{{Pr{\v{s}}a} {et~al.}(2022){Pr{\v{s}}a}, {Kochoska}, {Conroy},
  {Eisner}, {Hey}, {IJspeert}, {Kruse}, {Fleming}, {Johnston}, {Kristiansen},
  {LaCourse}, {Mortensen}, {Pepper}, {Stassun}, {Torres}, {Abdul-Masih},
  {Chakraborty}, {Gagliano}, {Guo}, {Hambleton}, {Hong}, {Jacobs}, {Jones},
  {Kostov}, {Lee}, {Omohundro}, {Orosz}, {Page}, {Powell}, {Rappaport}, {Reed},
  {Schnittman}, {Schwengeler}, {Shporer}, {Terentev}, {Vanderburg}, {Welsh},
  {Caldwell}, {Doty}, {Jenkins}, {Latham}, {Ricker}, {Seager}, {Schlieder},
  {Shiao}, {Vanderspek}, \& {Winn}}]{Prsa2022}
{Pr{\v{s}}a}, A., {Kochoska}, A., {Conroy}, K.~E., {et~al.} 2022, \apjs, 258,
  16, \dodoi{10.3847/1538-4365/ac324a}

\bibitem[{{Raddi} {et~al.}(2017){Raddi}, {Gentile Fusillo}, {Pala}, {Hermes},
  {G{\"a}nsicke}, {Chote}, {Hollands}, {Henden}, {Catal{\'a}n}, {Geier},
  {Koester}, {Munari}, {Napiwotzki}, \& {Tremblay}}]{Raddi2017}
{Raddi}, R., {Gentile Fusillo}, N.~P., {Pala}, A.~F., {et~al.} 2017, \mnras,
  472, 4173, \dodoi{10.1093/mnras/stx2243}

\bibitem[{{Rebassa-Mansergas} {et~al.}(2019){Rebassa-Mansergas}, {Solano},
  {Xu}, {Rodrigo}, {Jim{\'e}nez-Esteban}, \& {Torres}}]{Rebassa-Mansergas2019}
{Rebassa-Mansergas}, A., {Solano}, E., {Xu}, S., {et~al.} 2019, \mnras, 489,
  3990, \dodoi{10.1093/mnras/stz2423}

\bibitem[{{Ricker} {et~al.}(2015){Ricker}, {Winn}, {Vanderspek}, {Latham},
  {Bakos}, {Bean}, {Berta-Thompson}, {Brown}, {Buchhave}, {Butler}, {Butler},
  {Chaplin}, {Charbonneau}, {Christensen-Dalsgaard}, {Clampin}, {Deming},
  {Doty}, {De Lee}, {Dressing}, {Dunham}, {Endl}, {Fressin}, {Ge}, {Henning},
  {Holman}, {Howard}, {Ida}, {Jenkins}, {Jernigan}, {Johnson}, {Kaltenegger},
  {Kawai}, {Kjeldsen}, {Laughlin}, {Levine}, {Lin}, {Lissauer}, {MacQueen},
  {Marcy}, {McCullough}, {Morton}, {Narita}, {Paegert}, {Palle}, {Pepe},
  {Pepper}, {Quirrenbach}, {Rinehart}, {Sasselov}, {Sato}, {Seager},
  {Sozzetti}, {Stassun}, {Sullivan}, {Szentgyorgyi}, {Torres}, {Udry}, \&
  {Villasenor}}]{Ricker2015}
{Ricker}, G.~R., {Winn}, J.~N., {Vanderspek}, R., {et~al.} 2015, Journal of
  Astronomical Telescopes, Instruments, and Systems, 1, 014003,
  \dodoi{10.1117/1.JATIS.1.1.014003}

\bibitem[{{Rocchetto} {et~al.}(2015){Rocchetto}, {Farihi}, {G{\"a}nsicke}, \&
  {Bergfors}}]{Rocchetto2015}
{Rocchetto}, M., {Farihi}, J., {G{\"a}nsicke}, B.~T., \& {Bergfors}, C. 2015,
  \mnras, 449, 574, \dodoi{10.1093/mnras/stv282}

\bibitem[{{Rodriguez} {et~al.}(2023{\natexlab{a}}){Rodriguez}, {Kulkarni},
  {Prince}, {Szkody}, {Burdge}, {Caiazzo}, {van Roestel}, {Vanderbosch},
  {El-Badry}, {Bellm}, {G{\"a}nsicke}, {Graham}, {Mahabal}, {Masci},
  {Mr{\'o}z}, {Riddle}, \& {Rusholme}}]{Rodriguez2023a}
{Rodriguez}, A.~C., {Kulkarni}, S.~R., {Prince}, T.~A., {et~al.}
  2023{\natexlab{a}}, \apj, 945, 141, \dodoi{10.3847/1538-4357/acbb6f}

\bibitem[{{Rodriguez} {et~al.}(2023{\natexlab{b}}){Rodriguez}, {Galiullin},
  {Gilfanov}, {Kulkarni}, {Khamitov}, {Bikmaev}, {van Roestel}, {Yungelson},
  {El-Badry}, {Sunayev}, {Prince}, {Buntov}, {Caiazzo}, {Drake}, {Gorbachev},
  {Graham}, {Gumerov}, {Irtuganov}, {Laher}, {Masci}, {Medvedev}, {Purdum},
  {Sakhibullin}, {Sklyanov}, {Smith}, {Szkody}, \&
  {Vanderbosch}}]{Rodriguez2023b}
{Rodriguez}, A.~C., {Galiullin}, I., {Gilfanov}, M., {et~al.}
  2023{\natexlab{b}}, \apj, 954, 63, \dodoi{10.3847/1538-4357/ace698}

\bibitem[{{Rogers} {et~al.}(2020){Rogers}, {Xu}, {Bonsor}, {Hodgkin}, {Su},
  {von Hippel}, \& {Jura}}]{Rogers2020}
{Rogers}, L.~K., {Xu}, S., {Bonsor}, A., {et~al.} 2020, \mnras, 494, 2861,
  \dodoi{10.1093/mnras/staa873}

\bibitem[{{Rogers} {et~al.}(2024{\natexlab{a}}){Rogers}, {Bonsor}, {Xu},
  {Dufour}, {Klein}, {Buchan}, {Hodgkin}, {Hardy}, {Kissler-Patig}, {Melis},
  {Weinberger}, \& {Zuckerman}}]{Rogers2024a}
{Rogers}, L.~K., {Bonsor}, A., {Xu}, S., {et~al.} 2024{\natexlab{a}}, \mnras,
  527, 6038, \dodoi{10.1093/mnras/stad3557}

\bibitem[{{Rogers} {et~al.}(2024{\natexlab{b}}){Rogers}, {Bonsor}, {Xu},
  {Buchan}, {Dufour}, {Klein}, {Hodgkin}, {Kissler-Patig}, {Melis}, {Walton},
  \& {Weinberger}}]{Rogers2024b}
---. 2024{\natexlab{b}}, arXiv e-prints, arXiv:2406.11470.
\newblock \doarXiv{2406.11470}

\bibitem[{{Romero} {et~al.}(2019){Romero}, {Amaral}, {Klippel}, {Sanmartim},
  {Fraga}, {Ourique}, {Pelisoli}, {Lauffer}, {Kepler}, \&
  {Koester}}]{Romero2019}
{Romero}, A.~D., {Amaral}, L.~A., {Klippel}, T., {et~al.} 2019, \mnras, 490,
  1803, \dodoi{10.1093/mnras/stz2571}

\bibitem[{{Romero} {et~al.}(2022){Romero}, {Kepler}, {Hermes}, {Amaral},
  {Uzundag}, {Bogn{\'a}r}, {Bell}, {VanWyngarden}, {Baran}, {Pelisoli},
  {Oliveira}, {Koester}, {Klippel}, {Fraga}, {Bradley}, {Vu{\v{c}}kovi{\'c}},
  {Heintz}, {Reding}, {Kaiser}, \& {Charpinet}}]{Romero2022}
{Romero}, A.~D., {Kepler}, S.~O., {Hermes}, J.~J., {et~al.} 2022, \mnras, 511,
  1574, \dodoi{10.1093/mnras/stac093}

\bibitem[{{Sayres} {et~al.}(2012){Sayres}, {Subasavage}, {Bergeron}, {Dufour},
  {Davenport}, {AlSayyad}, \& {Tofflemire}}]{Sayres2012}
{Sayres}, C., {Subasavage}, J.~P., {Bergeron}, P., {et~al.} 2012, \aj, 143,
  103, \dodoi{10.1088/0004-6256/143/4/103}

\bibitem[{{Schlafly} {et~al.}(2019){Schlafly}, {Meisner}, \&
  {Green}}]{Schlafly2019}
{Schlafly}, E.~F., {Meisner}, A.~M., \& {Green}, G.~M. 2019, \apjs, 240, 30,
  \dodoi{10.3847/1538-4365/aafbea}

\bibitem[{{Shappee} {et~al.}(2014){Shappee}, {Prieto}, {Grupe}, {Kochanek},
  {Stanek}, {De Rosa}, {Mathur}, {Zu}, {Peterson}, {Pogge}, {Komossa}, {Im},
  {Jencson}, {Holoien}, {Basu}, {Beacom}, {Szczygie{\l}}, {Brimacombe},
  {Adams}, {Campillay}, {Choi}, {Contreras}, {Dietrich}, {Dubberley},
  {Elphick}, {Foale}, {Giustini}, {Gonzalez}, {Hawkins}, {Howell}, {Hsiao},
  {Koss}, {Leighly}, {Morrell}, {Mudd}, {Mullins}, {Nugent}, {Parrent},
  {Phillips}, {Pojmanski}, {Rosing}, {Ross}, {Sand}, {Terndrup}, {Valenti},
  {Walker}, \& {Yoon}}]{Shappee2014}
{Shappee}, B.~J., {Prieto}, J.~L., {Grupe}, D., {et~al.} 2014, \apj, 788, 48,
  \dodoi{10.1088/0004-637X/788/1/48}

\bibitem[{{Sion} {et~al.}(2014){Sion}, {Holberg}, {Oswalt}, {McCook},
  {Wasatonic}, \& {Myszka}}]{Sion2014}
{Sion}, E.~M., {Holberg}, J.~B., {Oswalt}, T.~D., {et~al.} 2014, \aj, 147, 129,
  \dodoi{10.1088/0004-6256/147/6/129}

\bibitem[{{Steele} {et~al.}(2021){Steele}, {Debes}, {Xu}, {Yeh}, \&
  {Dufour}}]{Steele2021}
{Steele}, A., {Debes}, J., {Xu}, S., {Yeh}, S., \& {Dufour}, P. 2021, \apj,
  911, 25, \dodoi{10.3847/1538-4357/abc262}

\bibitem[{{Subasavage} {et~al.}(2022){Subasavage}, {Bergeron}, {L{\'e}pine}, \&
  {Blouin}}]{Subasavage2022}
{Subasavage}, J., {Bergeron}, P., {L{\'e}pine}, S., \& {Blouin}, S. 2022,
  Montreal White Dwarf Database

\bibitem[{{Swan} {et~al.}(2024){Swan}, {Farihi}, {Su}, \& {Desch}}]{Swan2024}
{Swan}, A., {Farihi}, J., {Su}, K. Y.~L., \& {Desch}, S.~J. 2024, \mnras, 529,
  L41, \dodoi{10.1093/mnrasl/slad198}

\bibitem[{{Swan} {et~al.}(2019){Swan}, {Farihi}, \& {Wilson}}]{Swan2019}
{Swan}, A., {Farihi}, J., \& {Wilson}, T.~G. 2019, \mnras, 484, L109,
  \dodoi{10.1093/mnrasl/slz014}

\bibitem[{{Swan} {et~al.}(2020){Swan}, {Farihi}, {Wilson}, \&
  {Parsons}}]{Swan2020}
{Swan}, A., {Farihi}, J., {Wilson}, T.~G., \& {Parsons}, S.~G. 2020, \mnras,
  496, 5233, \dodoi{10.1093/mnras/staa1688}

\bibitem[{{Swan} {et~al.}(2021){Swan}, {Kenyon}, {Farihi}, {Dennihy},
  {G{\"a}nsicke}, {Hermes}, {Melis}, \& {von Hippel}}]{Swan2021}
{Swan}, A., {Kenyon}, S.~J., {Farihi}, J., {et~al.} 2021, \mnras, 506, 432,
  \dodoi{10.1093/mnras/stab1738}

\bibitem[{{Taylor}(2005)}]{Taylor2005_TOPCAT}
{Taylor}, M.~B. 2005, in Astronomical Society of the Pacific Conference Series,
  Vol. 347, Astronomical Data Analysis Software and Systems XIV, ed.
  P.~{Shopbell}, M.~{Britton}, \& R.~{Ebert}, 29

\bibitem[{{Tran} {et~al.}(2024){Tran}, {De}, \& {Hillenbrand}}]{Tran2024}
{Tran}, V., {De}, K., \& {Hillenbrand}, L. 2024, \mnras,
  \dodoi{10.1093/mnras/stae953}

\bibitem[{{Tremblay} {et~al.}(2020){Tremblay}, {Hollands}, {Gentile Fusillo},
  {McCleery}, {Izquierdo}, {G{\"a}nsicke}, {Cukanovaite}, {Koester}, {Brown},
  {Charpinet}, {Cunningham}, {Farihi}, {Giammichele}, {van Grootel}, {Hermes},
  {Hoskin}, {Jordan}, {Kepler}, {Kleinman}, {Manser}, {Marsh}, {de Martino},
  {Nitta}, {Parsons}, {Pelisoli}, {Raddi}, {Rebassa-Mansergas}, {Ren},
  {Schreiber}, {Silvotti}, {Toloza}, {Toonen}, \& {Torres}}]{Tremblay2020}
{Tremblay}, P.~E., {Hollands}, M.~A., {Gentile Fusillo}, N.~P., {et~al.} 2020,
  \mnras, 497, 130, \dodoi{10.1093/mnras/staa1892}

\bibitem[{{Udalski}(1990)}]{1990AJ....100..226U}
{Udalski}, A. 1990, \aj, 100, 226, \dodoi{10.1086/115508}

\bibitem[{{Valyavin} {et~al.}(2011){Valyavin}, {Antonyuk}, {Plachinda},
  {Clark}, {Wade}, {Fox Machado}, {Alvarez}, {Lopez}, {Hiriart}, {Han}, {Jeon},
  {Bagnulo}, {Zharikov}, {Zurita}, {Mujica}, {Shulyak}, \&
  {Burlakova}}]{Valyavin2011}
{Valyavin}, G., {Antonyuk}, K., {Plachinda}, S., {et~al.} 2011, \apj, 734, 17,
  \dodoi{10.1088/0004-637X/734/1/17}

\bibitem[{{Vauclair} {et~al.}(1987){Vauclair}, {Chevreton}, \&
  {Dolez}}]{Vauclair1987}
{Vauclair}, G., {Chevreton}, M., \& {Dolez}, N. 1987, \aap, 175, L13

\bibitem[{{Vauclair} {et~al.}(1997){Vauclair}, {Dolez}, {Fu}, \&
  {Chevreton}}]{Vauclair1997}
{Vauclair}, G., {Dolez}, N., {Fu}, J.~N., \& {Chevreton}, M. 1997, \aap, 322,
  155

\bibitem[{{Vauclair} {et~al.}(2000){Vauclair}, {Dolez}, {Fu}, {Homeier},
  {Roques}, {Chevreton}, \& {Koester}}]{Vauclair2000}
{Vauclair}, G., {Dolez}, N., {Fu}, J.~N., {et~al.} 2000, Baltic Astronomy, 9,
  133, \dodoi{10.1515/astro-2000-0120}

\bibitem[{{Veras}(2021)}]{Veras2021}
{Veras}, D. 2021, in Oxford Research Encyclopedia of Planetary Science, 1,
  \dodoi{10.1093/acrefore/9780190647926.013.238}

\bibitem[{{Vincent} {et~al.}(2024){Vincent}, {Barstow}, {Jordan}, {Mander},
  {Bergeron}, \& {Dufour}}]{Vincent2024}
{Vincent}, O., {Barstow}, M.~A., {Jordan}, S., {et~al.} 2024, \aap, 682, A5,
  \dodoi{10.1051/0004-6361/202347694}

\bibitem[{{Vincent} {et~al.}(2020){Vincent}, {Bergeron}, \&
  {Lafreni{\`e}re}}]{Vincent2020}
{Vincent}, O., {Bergeron}, P., \& {Lafreni{\`e}re}, D. 2020, \aj, 160, 252,
  \dodoi{10.3847/1538-3881/abbe20}

\bibitem[{Virtanen {et~al.}(2020)Virtanen, Gommers, Oliphant, Haberland, Reddy,
  Cournapeau, Burovski, Peterson, Weckesser, Bright, {van der Walt}, Brett,
  Wilson, Millman, Mayorov, Nelson, Jones, Kern, Larson, Carey, Polat, Feng,
  Moore, {VanderPlas}, Laxalde, Perktold, Cimrman, Henriksen, Quintero, Harris,
  Archibald, Ribeiro, Pedregosa, {van Mulbregt}, \& {SciPy 1.0
  Contributors}}]{2020SciPy-NMeth}
Virtanen, P., Gommers, R., Oliphant, T.~E., {et~al.} 2020, Nature Methods, 17,
  261, \dodoi{10.1038/s41592-019-0686-2}

\bibitem[{{{\v{S}}imon}(2000)}]{Simon2000}
{{\v{S}}imon}, V. 2000, \aap, 360, 627

\bibitem[{{Wang} {et~al.}(2023){Wang}, {Zhang}, {Wang}, {Zhang}, {Fang}, {Gu},
  {Guo}, \& {Jiang}}]{Wang2023}
{Wang}, L., {Zhang}, X., {Wang}, J., {et~al.} 2023, \apj, 944, 23,
  \dodoi{10.3847/1538-4357/acaf5a}

\bibitem[{{Wang} {et~al.}(2019){Wang}, {Jiang}, {Ge}, {Cutri}, {Jiang},
  {Sheng}, {Zhou}, {Bauer}, {Mainzer}, \& {Wright}}]{Wang2019}
{Wang}, T.-g., {Jiang}, N., {Ge}, J., {et~al.} 2019, \apjl, 886, L5,
  \dodoi{10.3847/2041-8213/ab53ed}

\bibitem[{{Warner}(1974)}]{1974MNRAS.168..235W}
{Warner}, B. 1974, \mnras, 168, 235, \dodoi{10.1093/mnras/168.1.235}

\bibitem[{Waskom(2021)}]{Waskom2021}
Waskom, M.~L. 2021, Journal of Open Source Software, 6, 3021,
  \dodoi{10.21105/joss.03021}

\bibitem[{{Wilson} {et~al.}(2014){Wilson}, {G{\"a}nsicke}, {Koester}, {Raddi},
  {Breedt}, {Southworth}, \& {Parsons}}]{Wilson2014}
{Wilson}, D.~J., {G{\"a}nsicke}, B.~T., {Koester}, D., {et~al.} 2014, \mnras,
  445, 1878, \dodoi{10.1093/mnras/stu1876}

\bibitem[{{Wilson} {et~al.}(2019){Wilson}, {Farihi}, {G{\"a}nsicke}, \&
  {Swan}}]{Wilson2019}
{Wilson}, T.~G., {Farihi}, J., {G{\"a}nsicke}, B.~T., \& {Swan}, A. 2019,
  \mnras, 487, 133, \dodoi{10.1093/mnras/stz1050}

\bibitem[{{Wright} {et~al.}(2010){Wright}, {Eisenhardt}, {Mainzer}, {Ressler},
  {Cutri}, {Jarrett}, {Kirkpatrick}, {Padgett}, {McMillan}, {Skrutskie},
  {Stanford}, {Cohen}, {Walker}, {Mather}, {Leisawitz}, {Gautier}, {McLean},
  {Benford}, {Lonsdale}, {Blain}, {Mendez}, {Irace}, {Duval}, {Liu}, {Royer},
  {Heinrichsen}, {Howard}, {Shannon}, {Kendall}, {Walsh}, {Larsen}, {Cardon},
  {Schick}, {Schwalm}, {Abid}, {Fabinsky}, {Naes}, \& {Tsai}}]{Wright2010}
{Wright}, E.~L., {Eisenhardt}, P. R.~M., {Mainzer}, A.~K., {et~al.} 2010, \aj,
  140, 1868, \dodoi{10.1088/0004-6256/140/6/1868}

\bibitem[{{Xu} \& {Bonsor}(2021)}]{XuBonsor2021}
{Xu}, S., \& {Bonsor}, A. 2021, Elements, 17, 241,
  \dodoi{10.2138/gselements.17.4.241}

\bibitem[{{Xu} \& {Jura}(2014)}]{Xu&Jura2014}
{Xu}, S., \& {Jura}, M. 2014, \apjl, 792, L39,
  \dodoi{10.1088/2041-8205/792/2/L39}

\bibitem[{{Xu} {et~al.}(2016){Xu}, {Jura}, {Dufour}, \& {Zuckerman}}]{Xu2016}
{Xu}, S., {Jura}, M., {Dufour}, P., \& {Zuckerman}, B. 2016, \apjl, 816, L22,
  \dodoi{10.3847/2041-8205/816/2/L22}

\bibitem[{{Xu} {et~al.}(2014){Xu}, {Jura}, {Koester}, {Klein}, \&
  {Zuckerman}}]{Xu2014}
{Xu}, S., {Jura}, M., {Koester}, D., {Klein}, B., \& {Zuckerman}, B. 2014,
  \apj, 783, 79, \dodoi{10.1088/0004-637X/783/2/79}

\bibitem[{{Xu} {et~al.}(2018){Xu}, {Su}, {Rogers}, {Bonsor}, {Olofsson},
  {Veras}, {van Lieshout}, {Dufour}, {Green}, {Schlawin}, {Farihi}, {Wilson},
  {Wilson}, \& {G{\"a}nsicke}}]{Xu2018b}
{Xu}, S., {Su}, K. Y.~L., {Rogers}, L.~K., {et~al.} 2018, \apj, 866, 108,
  \dodoi{10.3847/1538-4357/aadcfe}

\bibitem[{{Zhang} {et~al.}(2013){Zhang}, {Deng}, {Liu}, {L{\'e}pine},
  {Newberg}, {Carlin}, {Carrell}, {Yang}, {Gao}, {Xu}, {Li}, {Zhang}, {Zhao},
  {Luo}, {Bai}, {Yuan}, \& {Jin}}]{Zhang2013}
{Zhang}, Y.-Y., {Deng}, L.-C., {Liu}, C., {et~al.} 2013, \aj, 146, 34,
  \dodoi{10.1088/0004-6256/146/2/34}

\bibitem[{{Zuckerman} {et~al.}(2003){Zuckerman}, {Koester}, {Reid}, \&
  {H{\"u}nsch}}]{Zuckerman2003}
{Zuckerman}, B., {Koester}, D., {Reid}, I.~N., \& {H{\"u}nsch}, M. 2003, \apj,
  596, 477, \dodoi{10.1086/377492}

\bibitem[{{Zuckerman} {et~al.}(2010){Zuckerman}, {Melis}, {Klein}, {Koester},
  \& {Jura}}]{Zuckerman2010}
{Zuckerman}, B., {Melis}, C., {Klein}, B., {Koester}, D., \& {Jura}, M. 2010,
  \apj, 722, 725, \dodoi{10.1088/0004-637X/722/1/725}

\end{thebibliography}
\bibliographystyle{aasjournal}

\end{CJK*}
\end{document}